BEN-GURION UNIVERSITY OF THE NEGEV

FACULTY OF ENGINEERING SCIENCES

SCHOOL OF ELECTRICAL AND COMPUTER ENGINEERING

# ANALYSIS OF LINEAR TIME-VARYING & PERIODIC SYSTEMS

THESIS SUBMITTED IN PARTIAL FULFILLMENT OF THE REQUIREMENTS FOR THE MSc. DEGREE

By: Oren Fivel

Supervised by:

Professor Izchak Lewkowicz

**BEN-GURION UNIVERSITY OF THE NEGEV**

**FACULTY OF ENGINEERING SCIENCES**

**SCHOOL OF ELECTRICAL AND COMPUTER ENGINEERING**

# ANALYSIS OF LINEAR TIME-VARYING & PERIODIC SYSTEMS

THESIS SUBMITTED IN PARTIAL FULFILLMENT OF THE REQUIREMENTS FOR THE MSc. DEGREE

By: Oren Fivel

Supervised by:

Professor Izchak Lewkowicz

Author: Oren Fivel ……………… Date: 15-Jul-21

Supervisor: Professor Izchak Lewkowicz ……………… Date: 15-Jul-21

Chairman of Graduate Studies Committee:

Name: Dr. Kobi Cohen ……………… Date: 02/08/2021

July 2021

# Abstract

This thesis applies *Floquet theory* to analyze linear periodic time-varying (LPTV) systems, which are represented by a system of ordinary differential equations (ODEs) that depend on a time variable $t$ and have a matrix of coefficients with period $T > 0$ (or equivalently, with frequency $\omega = \frac{2\pi}{T}$).

According to *Floquet theory*, the transition matrix of an LPTV system represented by a square periodic-function matrix $A(t) = A(t + T)$ can be expressed as the product of a square periodic function matrix $P(t) = P(t + T)$ and an exponentiated square matrix of the form $Rt$, where $R$ is a constant matrix (independent of $t$). Despite the validity of *Floquet theory*, it is difficult to find an analytical closed form for the matrices $P(t)$ and $R$ when the transition matrix $\Phi_A(t, t_0)$ is unknown. In essence, it is difficult to find an analytical solution for an LPTV system (i.e., a closed form for its transition matrix).

The objective of this work is to characterize a sufficiently large family of periodic matrices $A(t)$ with a finite number of harmonics such that the periodic part $P(t)$ of the solution also has a finite number of harmonics. In addition, for each family, an effective procedure is required to find the matrices $P(t)$ and $R$.

The research method involves studying periodic matrices $A(t)$ with frequency $\omega$ as a free parameter (i.e., not a fixed number) to obtain a sufficiently large family of LPTV systems so that their transition matrices are defined by the frequency $\omega$. In this work, we focus on cases in which the Fourier coefficients of $A(t)$ are finite polynomials in $\omega$, $R$ is a finite polynomial in $\omega$, and the Fourier coefficients of $P(t)$ do not depend on $\omega$.

The research results show that for a given family of periodic matrices $A(t)$, we can compare the powers of $\omega$ that multiply the harmonics (i.e., $\omega$ is part of the coefficients multiplying the cosine [sine] factors in even [odd] representations or of the exponential factors in complex representations) to determine the matrices

July 2021

$P(t)$ and $R$. In addition, the results lead to relations between LPTV systems at frequency $\omega$ and the associated linear time-invariant system, which is defined by having zero frequency ($\omega = 0$). A buy-product of using the frequency $\omega$ as a free parameter is that allows the stability of the LPTV system to be determined based on how $R$ depends on the frequency $\omega$.

***Keywords:*** Floquet theory; Linear ordinary differential equations; LTV systems, Periodic systems; Fourier series; Matrix equation; Eigen decomposition; Frequency-domain analysis; Power-coefficient comparison; Finite number of harmonics

July 2021

# Thesis Overview

This thesis analyzes linear periodic time-varying (LPTV) systems, which are defined by linear ordinary differential equations (ODEs) with a matrix of periodic functions $A(t)$, in turn defined by a frequency parameter $\omega$. This work is organized as follows:

1. **CHAPTER 1** introduces the topic, discusses the background of LPTV systems and *Floquet theory*, summarizes previous work by other authors, and suggests how to analyze and solve an LPTV system based on its frequency $\omega$, which is a free parameter (i.e., not fixed).

2. **CHAPTER 2** implements the cosine-sine Fourier series of LPTV systems that converts the LPTV system's ODE into a matrix algebraic equation that is similar to an eigen decomposition problem.

3. **CHAPTER 3** shows that the factorization as per *Floquet theory* is not unique by using a matrix similarity transformation and shifting the trace of LPTV system's matrix by a scalar function (more precisely, shifting to zero).

4. **CHAPTER 4** presents examples of LPTV systems and solutions with finite and infinite harmonics and outlines how the frequency $\omega$ affects the solution.

5. **CHAPTER 5** outlines our suggestion for comparing powers of $\omega$ to solve LPTV systems, which is incorporated with the information obtained from the associated linear time-invariant (LTI) system defined by using $\omega = 0$ in the LPTV system.

6. **CHAPTER 6** summarizes this work, discusses its scientific contributions, and presents avenues for future research.

The appendices provide additional information, such as mathematical background, an overview of previous work, and suggestions for future work. The appendices in this work are organized as follows:

A. **APPENDIX A** outlines the notion of Fourier series for matrices (exponential form and cosine-sine form) and properties thereof.

B. **APPENDIX B** applies exponential Fourier series analysis to LPTV systems and compares this approach with the cosine-sine form described in **CHAPTER 2**.

C. **APPENDIX C** outlines complex LPTV systems and how they can be converted to real LPTV systems based on the isomorphism $a + ib \leftrightarrow \begin{bmatrix} a & -b \\ b & a \end{bmatrix}$. In addition, this appendix outlines how to decompose the state vector into even and odd parts and, furthermore, how to rewrite the LPTV system matrix using even and odd matrix blocks. In this appendix we also introduce Split complex numbers (also known as hyperbolic complex numbers) in order to generalize the notion of the even-odd decomposition.

D. **APPENDIX D** outlines the main results of (Wang, 2017), which generalize the notion eigen decomposition of linear time-varying systems (this is a continuation of the literature survey of Section 1.4.2).

E. **APPENDIX E** outlines the main results of (Wereley, 1991), which suggest implementing LPTV systems by using generalized tools to analyze and control LTI systems (e.g., transfer functions and zeros and poles in the $s$ and $z$ domains). This is a continuation of the literature survey of Section 1.4.6.

# Acknowledgments

I would like to thank my supervisor Professor Izchak Lewkowicz, with whom I worked for two intensive years. Professor Lewkowicz has been a supportive tutor and has helped me tremendously during my research. I thank him for teaching me to think outside the box, to be persistent, and to strive to perfect my performance. I also thank him for improving my learning skills and for sharing outstanding ideas with me throughout my studies.

I also want to thank Mr. Yair Harpaz for being a great colleague and friend. I thank him for his collaboration, research work, and ideas. Yair is always happy to help everyone, at any time, anywhere.

With great appreciation, I thank my dear love Adi for her tremendous loving support, patience, and care. I want to thank my parents and sister as well for their support and care and for their contributions, each in his or her area of expertise.

July 2021

# Table of Contents

# Figure List

# Table List

# Acronyms and Abbreviations

| | |
|---|---|
| iif | if and only if |
| LHS | left-hand side |
| LPTV | linear periodic time varying |
| LTI | linear time invariant |
| LTV | linear time varying |
| ODE | ordinary differential equation |
| RHS | right-hand side |

# CHAPTER 1. Introduction

This thesis applies *Floquet theory* (Floquet, 1883) to solve linear periodic time-varying (LPTV) systems, which may be represented by systems of ordinary differential equations (ODEs). An LPTV system is a particular case of a linear time-varying (LTV) system, therefore. Herein, we outline some equations and definitions regarding to LTV system as basic tools for LPTV system.

This chapter is organized as follows: Section 1.1 outlines the basics definition and results regarding to LTV systems. Section 1.2 presents the LPTV system as a particular case of the LTV system, presents the *Floquet theory* and outlines some basic results related to LPTV systems and *Floquet theory*. In Section 1.3, we highlight some applications of LPTV systems, and we show, as an example for an LPTV system, a classic result of *Hill equation* (and its special cases: *Mathieu equation* and *Meissner equation*). In Section 1.4, we highlight some additional recent results from the literature regarding to LPTV systems and *Floquet theory*. In Section 1.5 we highlight the objective of the research, the suggested approach and scope of Work.

## 1.1 Linear Time-Varying System: Background

Let $\mathbb{T} \subseteq \mathbb{R}$ be the time domain of the independent variable $t$, i.e., $\mathbb{T} = \{t \in \mathbb{R}: t \geq t_0 \geq 0\}$, with the initial time denoted $t_0$. Suppose $x: \mathbb{T} \to \mathbb{R}^n$ (called the *state vector* or simply the *state*) is a solution to the following LTV system of differential equations:

$$\dot{x} = A(t)x \,, \tag{1.1}$$

where $A: \mathbb{T} \to \mathbb{R}^{n\times n}$ is some square matrix (called the *system matrix*) that is independent of the variable $x$ (i.e., the system is linear). The solution $x(t)$ is then represented as

$$x(t) = \Phi_A(t, t_0)x(t_0)\,, \tag{1.2}$$

where $t_0$ is the initial time, $x(t_0) \in \mathbb{R}^n$ is the initial condition for $x(t)$, and $\Phi_A: \mathbb{T} \times \mathbb{T} \to \mathbb{R}^{n\times n}$ is the *transition matrix* of $A(t)$ that solves the following differential equation:

$$\begin{gathered} \dot{\Phi}_A(t, t_0) = A(t)\Phi_A(t, t_0), \\ \forall\, \tau \in \mathbb{T}, \Phi_A(\tau, \tau) = I_n\,, \end{gathered} \tag{1.3}$$

where $I_n$ is the $n \times n$ identity matrix, and $\dot{\Phi}_A(t, t_0) = \frac{\partial \Phi_A(t,t_0)}{\partial t}$. Hereinafter, to simplify notation, we denote functions of $t$ as elements of real matrices; e.g., $x: \mathbb{T} \to \mathbb{R}^n$ is denoted $x(t) \in \mathbb{R}^n$, $A: \mathbb{T} \to \mathbb{R}^{n\times n}$ is denoted $A(t) \in \mathbb{R}^{n\times n}$, and $\Phi_A: \mathbb{T} \times \mathbb{T} \to \mathbb{R}^{n\times n}$ is denoted $\Phi_A(t, t_0) \in \mathbb{R}^{n\times n}$. The transition matrix can be factorized by the following

$$\Phi_A(t, t_0) = W(t)W^{-1}(t_0)\,, \tag{1.4}$$

Where $W(t) \in \mathbb{R}^{n\times n}$ is a nonsingular matrix $W(t)$ is called the *fundamental matrix*, and by Eq. (1.3) it obtains the following equation.

$$\dot{W}(t) = A(t)W(t). \tag{1.5}$$

The initial condition $W(t_0)$ may be any invertible square matrix.

In general, explicitly obtaining a transition matrix $\Phi_A(t,t_0)$ is difficult. Moreover, unlike in e.g., the LTI case, it is not always valid to exponentiate the antiderivative of $A(t)$ to obtain the transition matrix, i.e., $\Phi_A(t,t_0) \neq e^{\int_{t_0}^{t} A(\tau)d\tau}$ in general. Referring to (Rugh, 1996) Property 4.2, if an LTV system matrix and its anti-derivative multiplicatively commute, then the transition matrix is obtained by exponentiating this anti-derivative (similar to the scalar ODE case), i.e.,

$$A(t)\int_{t_0}^{t} A(\tau)\,d\tau = \int_{t_0}^{t} A(\tau)\,d\tau A(t) \Rightarrow \Phi_A(t,t_0) = e^{\int_{t_0}^{t} A(\tau)d\tau}, \forall\, t_0, t. \qquad (1.6)$$

Two special cases of Eq (1.6) allow for solutions:

1. If $A(t) \equiv A$ is a constant matrix, then $\Phi_A(t,t_0) = e^{(t-t_0)A}$. This case is referred to as a linear time-invariant (LTI) system.
2. If $A(t)$ is a diagonal matrix [i.e., $A_{ij}(t) = 0, \forall\, i \neq j$], then $\Phi_A(t,t_0) = e^{\int_{t_0}^{t} A(\tau)d\tau}$ is also a diagonal matrix such that $[\Phi_A]_{ii}(t,t_0) = e^{\int_{t_0}^{t} A_{ii}(\tau)d\tau}$.

Case 1 is a special case of Eq (1.6) because $A(t) \equiv A$ and $\int_{t_0}^{t} A(\tau)d\tau \equiv (t-t_0)A$, so $A(t)$ and $\int_{t_0}^{t} A(\tau)d\tau$ trivially commute because any matrix commutes with itself and with a scalar multiple of itself. Case 2 is a special case of Eq (1.6), because both $A(t)$ and $\int_{t_0}^{t} A(\tau)d\tau$ are diagonal square matrices and, thus, trivially commute. This result emphasizes that a transition matrix of an LTV system, $\Phi_A(t,t_0)$, is difficult to obtain in general, and usually $\Phi_A(t,t_0) \neq e^{\int_{t_0}^{t} A(\tau)d\tau}$. At the following example we demonstrate a case that $\Phi_A(t,t_0) \neq e^{\int_{t_0}^{t} A(\tau)d\tau}$.

**<u>Example 1.1</u>: Cauchy–Euler Equation (non-periodic example)**

Consider the following LTV system matrix:

$$\begin{bmatrix} \dot{x}_1 \\ \dot{x}_2 \end{bmatrix} = \begin{bmatrix} 0 & 1 \\ 6t^{-2} & 0 \end{bmatrix} \begin{bmatrix} x_1 \\ x_2 \end{bmatrix}, t \geq t_0 \geq 1, \tag{1.7}$$

where $A(t) = \begin{bmatrix} 0 & 1 \\ 6t^{-2} & 0 \end{bmatrix}$, $x = \begin{bmatrix} x_1 \\ x_2 \end{bmatrix}$. Based on a solution in which $x_1$ is a power of $t$ [e.g., $x_1(t) = ct^r$], the transition matrix $\Phi_A(t, t_0)$ obtained by calculating of the fundamental matrix $W(t)$ and its inverse $W^{-1}(t_0)$ [Eq. (1.4)] with

$$W(t) = \begin{bmatrix} t^{-2} & t^3 \\ -2t^{-3} & 3t^2 \end{bmatrix}, \quad W^{-1}(t_0) = \frac{1}{5}\begin{bmatrix} 3t_0^2 & -t_0^3 \\ 2t_0^{-3} & t_0^{-2} \end{bmatrix}. \tag{1.8}$$

After multiplying out $W(t)W^{-1}(t_0)$, we arrive at:

$$\Phi_A(t, t_0) = \frac{1}{5(t^2 t_0^3)} \begin{bmatrix} 2t^5 + 3t_0^5 & t^5 - t_0^5 \\ 6t^5 - 6t_0^5 & 3t^5 + 2t_0^5 \end{bmatrix} \tag{1.9}$$

In contrast, $e^{\int_{t_0}^{t} A(\tau)d\tau}$ is given by:

$$e^{\int_{t_0}^{t} A(\tau)d\tau} = \begin{bmatrix} \cosh(S) & \sqrt{\dfrac{tt_0}{6}}\sinh(S) \\ \sqrt{\dfrac{6}{tt_0}}\sinh(S) & \cosh(S) \end{bmatrix}, \quad S = \sqrt{\frac{6}{tt_0}}(t - t_0), \tag{1.10}$$

so that $\Phi_A(t, t_0) \neq e^{\int_{t_0}^{t} A(\tau)d\tau}$.

□

Case 2 can be generalized to the case when $A(t)$ is diagonalizable by a constant diagonalizing matrix $U$ and a diagonal matrix $B(t)$ such that

$$B(t) = U^{-1}A(t)U. \tag{1.11}$$ [1]

[1] Two square matrices $A, B \in \mathbb{C}^{n\times n}$ are called "similar" (denoted $A \sim B$) if exist a non-singular matrix $U \in \mathbb{C}^{n\times n}$ such that $B = U^{-1}AU$. $U$ denotes a "similarity transformation matrix" (or a "change of basis matrix").

The proof is straightforward by Eq. (1.6) and the commutative property of the diagonal matrix $B(t)$ with its antiderivative $\int_{t_0}^{t} B(\tau)d\tau$. For a general matrix $B(t)$ (that is not necessarily diagonal), by inserting Eq. (1.11) into Eq. (1.3), we can obtain the following identity

$$\Phi_A(t, t_0) = U\Phi_B(t, t_0)U^{-1}, \tag{1.12}$$

and therefore, if $B(t)$ is diagonal we have $\Phi_A(t, t_0) = e^{\int_{t_0}^{t} A(\tau)d\tau} = U\Phi_B(t, t_0)U^{-1}$, with $[\Phi_B(t, t_0)]_{ij} = e^{\int_{t_0}^{t} B_{ii}(\tau)d\tau}\delta_{ij}$ where $\delta_{ij}$ is the Kronecker delta function ($\delta_{ij} = 0$ for $i \neq j$, $\delta_{ij} = 1$ for $i = j$). The notion of Eqs. (1.11)-(1.12) can be generalized to a time dependent matrix $U(t)$ by the following [see (Rugh, 1996) Property 4.13]

$$\begin{aligned} B(t) &= U^{-1}(t)\left(A(t)U(t) - \dot{U}(t)\right), \\ \Phi_A(t, t_0) &= U(t)\Phi_B(t, t_0)U^{-1}(t_0). \end{aligned} \qquad (1.13)[2]$$

The proof is detailed in (Rugh, 1996) by supposing $x(t)$ be a solution of to an LPTV ODE with a system matrix $A(t)$ and a transition matrix $\Phi_A(t, t_0)$. Then, perform a coordinate change by $\tilde{x}(t)$ such that $x(t) = U(t)\tilde{x}(t)$, and by computing the derivative of this coordinate change, we have new LPTV ODE in terms of $\tilde{x}(t)$, so, by some matrix algebra, we can construct $B(t)$ in terms of $A(t)$ and $U(t)$, and also construct $\Phi_A(t, t_0)$ by $\Phi_B(t, t_0)$ [if known] and $U(t)$. Even

[2] Eq. (1.13) can be written equivalently by the following two forms:

(a) $A(t) = \left[\dot{U}(t) + U(t)B(t)\right]U^{-1}(t)$ , $\Phi_B(t, t_0) = U^{-1}(t)\Phi_A(t, t_0)U(t_0)$; or

(b) $A(t) = \widetilde{U}(t)^{-1}\left[B(t)\widetilde{U}(t) - \dot{\widetilde{U}}(t)\right]$, $\Phi_B(t, t_0) = \widetilde{U}(t)\Phi_A(t, t_0)\widetilde{U}^{-1}(t_0)$ with $\widetilde{U}(t) = U^{-1}(t)$.

Options (a) and (b) are equivalent via some matrix calculus [e.g., $U(t)\widetilde{U}(t) = I \Rightarrow \dot{U}U^{-1} = -\widetilde{U}^{-1}\dot{\widetilde{U}}$ ]. Option (a) presents the matrix transformation by an opposite direction of matrix product corresponding to a transformation matrix $U(t)$, while option (b) keeps the structure of Eq. (1.13) but using a matrix transformation $U^{-1}(t)$ instead of $U(t)$.

though, $A(t)$ may not satisfy the commutative condition in Eq. (1.6), we may have a new LPTV system matrix $B(t)$ that does satisfy this commutative condition so that $\Phi_B(t,t_0) = e^{\int_{t_0}^{t} B(\tau)d\tau}$, and thus $\Phi_A(t,t_0)$ can be obtained by (1.13). We use this notion of similarity in Section 1.2 [Eqs. (1.15)-(1.20)] and in Section 3.3.

An LTI system can be generated by a local linearization of an autonomous nonlinear system $\dot{x} = f(x)$ adjacent to equilibrium points. In other words, although an autonomous nonlinear system is difficult to solve directly, several initial conditions and initial times $t_0$ can be set to shift the solution without deforming it. On the other hand, linearizing a general nonlinear system $\dot{x} = f(x,t)$ adjacent to equilibrium points would create an LTV system, which remains difficult to solve. Moreover, it is impossible to set several initial conditions with several initial times $t_0$ because the system is time dependent (changing the initial time $t_0$ would change the solution).

In this work, we focus on a periodic system matrix so that $A(t) = A(t+T)$ for some $T > 0$ [$A(t)$ is called a $T$-periodic function of $t$]. On the next section, we outline the definition of an LPTV system, *Floquet Theory* and some basic results related to LPTV systems and *Floquet Theory*.

## 1.2 Linear Periodic Time-Varying System and Floquet Theory: Basic Results

**Definition 1.2**: An LPTV system is defined as a linear system of ODEs (namely, an LTV system) that can be represented by a coefficient matrix of $T$-periodic functions, i.e.,

$$\begin{gathered} \dot{x} = A(t)x, \quad x(t_0) = x_0, \\ A(t) = A(t+T) \in \mathbb{R}^{n\times n}, \quad T = \frac{2\pi}{\omega} > 0. \end{gathered} \tag{1.14}$$

□

**<u>Theorem 1.3</u>: (***Floquet theory***)** Suppose $x(t) \in \mathbb{R}^n$ is a solution to the ODE of the following LPTV system defined by the $T$-periodic matrix $A(t)$. Then, according to *Floquet theory* [e.g., see (Rugh, 1996) property 5.11], the transition matrix of $A(t)$ can be decomposed into

$$\Phi_A(t,t_0) = P(t)e^{(t-t_0)R}P^{-1}(t_0), \tag{1.15}$$

where $R$ is a constant matrix for all $t$ and $P(t)$ is a non-singular $T$-periodic matrix [i.e., $P(t) = P(t+T)\ \forall\, t$].

□

Note that the product $P(t)e^{tR} \equiv W(t)$ is a fundamental matrix [Eq. (1.5)] of the LPTV system matrix $A(t)$. By plugging Eq. (1.15) into Eq. (1.3), we obtain the following matrix differential equation:

$$A(t)P(t) = \dot{P}(t) + P(t)R. \tag{1.16}$$

Two equivalent equations for Eq. (1.16) are given by solving for $A(t)$ and for $R$:

$$A(t) = \left[\dot{P}(t) + P(t)R\right]P^{-1}(t) \tag{1.17}$$

and

$$R = P^{-1}(t)\left[A(t)P(t) - \dot{P}(t)\right]. \tag{1.18}$$

Eq. (1.17) stipulates that, given the matrices $P(t)$ and $R$, we can generate an LPTV system matrix $A(t)$. Eq. (1.17) is a special case of LTV matrix transformation in Eq. (1.13) [see option (a) in footnote[2] with transformation matrix $U(t) \coloneqq P(t)$ and system matrix $B(t) \coloneqq R$]. However, the goal is to solve in the opposite direction; i.e., given a matrix $A(t)$, we need to find the matrices $P(t)$ and $R$ that construct the transition matrix of $A(t)$ given by Eq. (1.15). In this work, we use Eq. (1.17) to produce an LPTV system matrix $A(t)$ for use as a tool to explore the properties and solutions of LPTV systems.

Eq. (1.18) says that, for a given LPTV system matrix $A(t)$, a periodic matrix $P(t)$ inserted into the RHS of Eq. (1.18) should produce a constant matrix on the LHS if $P(t)$ is correct. Eq. (1.18) is a useful tool to double-check solutions and to calculate the constant matrix $R$ when information is obtained regarding the structure of $R$. Eqs. (1.18) and (1.15) [in this order] are special case of Eq. (1.13) with a transformation matrix $U(t) \coloneqq P(t)$ [used in the coordinates change Eq. (1.19) below] and a new system's matrix $B(t) \coloneqq R$ [present the LTI system in Eq. (1.20) below] such that $\Phi_B(t, t_0) \coloneqq e^{(t-t_0)R}$.

Form Eq. (1.18), a relation between LPTV and LTI Systems can be obtained as per *Lyapunov reducibility theorem* [see Chapter 2, Section 2.4 of (Yakubovich & Starzhinskii, 1975)], which considers the following coordinates change:

$$x = P(t)\tilde{x}. \tag{1.19}$$

Plugging this equation into Eq. (1.14) and using Eqs. (1.13) and (1.18) produces an LTI system,

$$\dot{\tilde{x}} = R\tilde{x} \tag{1.20}$$

that connects the LPTV system represented by the system matrix $A(t)$ to its LTI system represented by the system matrix $R$.

Unfortunately, neither $P(t)$ nor $R$ is known. If we try an incorrect periodic part $P(t)$, Eq. (1.20) becomes a new LPTV system with a new periodic system matrix defined by the RHS of Eq. (1.18). Furthermore, we would like to find a case where we can obtain information from $A(t)$ regarding the structure and values of $P(t)$ and $R$ by mathematical manipulation.

**Lemma 1.4:** The transition matrix $\Phi_A(t, t_0)$ of an LPTV system is simultaneously $T$-periodic in both time inputs $(t, t_0)$ such that [(Wereley, 1991) Eqs. (2.32)]:

$$\Phi_A(t + T, t_0 + T) = \Phi_A(t, t_0). \quad (1.21)$$

**Proof:** In Wereley's PhD work, a proof is bases on *Peano-Baker* series [(Wereley, 1991) Eqs. (2.33)-(2.35)], that is a presentation of the transition matrix $\Phi_A(t, t_0)$ by an infinite sum of multiple integrals on $A(t)$. Herein, we proof this lemma straightforward from *Floquet theory*, by using the periodicity of $P(t)$ and the fact that $R$ is constant in time:

$$\begin{aligned}\Phi_A(t + T, t_0 + T) &= P(t + T)e^{((t+T)-(t_0+T))R}P^{-1}(t_0 + T)\\ &= P(t)e^{(t+T-t_0-T)R}P^{-1}(t_0)\\ &= P(t)e^{(t-t_0)R}P^{-1}(t_0)\\ &= \Phi_A(t, t_0).\end{aligned}$$

■

Inspired by this proof, we have the following result [see (Wereley, 1991) Eq. (2.59)]

$$\Phi_A(kT, 0) = [\Phi_A(T, 0)]^k, \forall k \in \mathbb{Z}. \quad (1.22)$$

In the literature the matrix $\Phi_A(T, 0)$ denotes the *Monodromy Matrix* and usually defined by $\Phi_A(T, 0) = e^{TR}$ with $P(0) = I_n$ [e.g., (Chicone, 2006) page 210]. This definition for *Monodromy Matrix* leads to another result, addressed to a factorization of a known Transition Matrix of a $T$-periodic LPTV system using matrix logarithm. This may cause complex solutions for the constant matrix $R$ and $T$-periodic matrix $P(t)$ [see e.g.: (Yakubovich & Starzhinskii, 1975) Sections 2.2; (Rugh, 1996) Property 5.11; (Chicone, 2006) Theorems 2.82-2.83], by the following setup. Suppose $\Phi_A(t,0)$ is a known Transition Matrix corresponding to an LPTV system's $T$-periodic matrix $A(t)$. The following factorization of $\Phi_A(t,0)$ is required:

$$\Phi_A(t,0) = P(t)e^{Rt}, \quad \Phi_A(0,0) = P(0) = I_n \, , \tag{1.23}$$

A (possibly) complex solution can be obtained for a constant matrix $R$ (using matrix logarithm) and a $T$-periodic matrix $P(t)$ by the following

$$R_T = \frac{1}{T}\ln[\Phi_A(T,0)] \Rightarrow P_T(t) = \Phi_A(t,0)e^{-R_T t} \tag{1.24}$$

A subscript $T$ denotes the case when $P(t)$ is $T$-periodic. Because $\Phi_A(2T,0) = \left(\Phi_A(T,0)\right)^2$ [see Eq. (1.22)], a real solution can be obtained by assuming $P(t)$ to be $2T$-periodic instead as follows

$$R_{2T} = \frac{1}{2T}\ln[\Phi_A(2T,0)] \Rightarrow P_{2T}(t) = \Phi_A(t,0)e^{-R_{2T} t} \tag{1.25}$$

These results may affect the number of harmonics of the periodic part of the solution $P(t)$ (see section 4.2 below), e.g., first harmonic term in $A(t)$ is $\omega_1 = \frac{2\pi}{T} = \omega$, but in $P(t)$ is $\omega_1 = \frac{2\pi}{2T} = \frac{\omega}{2}$ and not $\omega$. Despite these results, Montagnier et al. emphasize in their work (Montagnier, Paige, & Spiteri, 2003) that it is important to obtain real factorizations from computations on a single period. A real floquet factorization of an LPTV Transition Matrix is discussed in Section 1.4.4 with regards to Montagnier et al. and also to (Yakubovich & Starzhinskii, 1975) Section 2.3.

Due to Eqs. (1.23)-(1.24), other basic results are shown in the literature corresponding to the eigen decomposition of a *Monodromy Matrix* $\Phi_A(T,0) = e^{TR_T}$. An eigenvalue of $\Phi_A(T,0) = e^{TR_T}$, denoted by a complex number $\rho$, is called a *characteristic multiplier* (or *Floquet multiplier*). A complex number $\mu$ is called a *characteristic exponent* (or *Floquet exponent*) if $e^{\mu T}$ is a *characteristic multiplier* [$\mu$ is defined up to modulo $i\frac{2\pi}{T}$ because $e^{\mu T} = e^{\left(\mu + i\frac{2\pi}{T}k\right)T}$, so μ is an eigenvalue of $R_T$ up to modulo $i\frac{2\pi}{T}$]. Note that any linear combination of the vector functions with the form $e^{\mu_i t}p_i(t)$, where $p_i(t)$ is a $T$-periodic (and possibly complex) vector corresponding to a characteristic exponent $\mu_i$, is a non-trivial solution (not the zero function) to the LPTV ODE in Eq. (1.1) up to initial conditions [see e.g., (Chicone, 2006) Theorem 2.96]. For **Lemma 1.5** below, we refer (Kelley & Peterson, 2010) Theorem 2.74:

**Lemma 1.5:** Let $A(t)$ be an LPTV system's $T$-periodic matrix satisfying Eq. (1.14), $\Phi_A(t,0)$ be a transition matrix satisfying Eqs. (1.23)-(1.24) and $\rho_1 = e^{\mu_1 T}, \dots, \rho_n = e^{\mu_n T}$ be the corresponding characteristic multipliers. Then:

$$\rho_1\rho_2 \cdots \rho_n = e^{\mathrm{trace}\left\{\int_{t_0}^{t_0+T} A(t)dt\right\}} \tag{1.26}$$

This lemma is proved by (Kelley & Peterson, 2010) [Theorem 2.74], and guided by (Chicone, 2006) in exercise 2.87, using *Jacobi–Liouville formula* [see (Rugh, 1996), Property 4.9]

$$e^{\int_{t_0}^{t} \mathrm{trace}\{A(\tau)\}d\tau} = \det\{\Phi_A(t,t_0)\}. \tag{1.27}$$

In that proof the initial time was zero ($t_0 = 0$).

□

Inspired by this lemma, *Jacobi–Liouville formula* and Exercise 2.87 of (Chicone, 2006), we have that any real matrix $R$ that is a constant part of a floquet factorization of $\Phi_A(t, t_0)$ [Eq. (1.14)], i.e.,

$$\text{trace}\{R\} = \frac{1}{T}\text{trace}\left\{\int_{t_0}^{t_0+T} A(t)dt\right\}. \tag{1.28}$$

For simplicity denote the average of $A(t)$ by $A_o$ (Lewkowicz, 2020) such that

$$A_o = \frac{1}{T}\int_{t_0}^{t_0+T} A(t)dt. \tag{1.29}$$

Then, Eq. (1.28) is simplified to

$$\text{trace}\{R\} = \text{trace}\{A_o\}. \tag{1.30}$$

The notion of a transition matrix of an LPTV system might appear straightforward enough because its structure is known. However, in practice, it is difficult to find an analytical closed-form solution of an LPTV system, or, equivalently, for the matrices $P(t)$ and $R$. Moreover, no procedure is available in the literature for solving the ODE of a general LPTV system (or even for specific families thereof), or, equivalently, for finding the matrices $P(t)$ and $R$ when the transition matrix $\Phi_A(t, t_0)$ is unknown.

## 1.3 LPTV Systems Applications and Physical Examples

LPTV systems are used to describe the motion of lunar perigee (Hill, 1886), (Hill, 1878), vibrations of stretched elliptical membranes (Mathieu, 1868), the motion of side rods of a locomotive (Meissner, 1918), elliptical waveguides (McLachlan, 1947), (Pillay & Kumar, 2017)**,** the motion of gravitationally stabilized Earth-pointing satellites (Schechter, 1964), quadrupole mass spectrometry (Dawson, 1976), the rolling motion of ships (Jovanoski & Robinson, 2009), micromechanical tuning fork gyroscope dynamics (King, 1989), pendulum dynamics (Seyranian & Seyranian, 2006) and (Rugh, 1996) Example 5.21, helicopter rotors (Friedmann, 1986), wind turbines (Stol, Balas, & Bir, 2002), multistage DC-DC converters (Li, Guo, Ren, Zhang, & Zhang, 2017), etc. An LPTV system can be obtained by linearizing a nonlinear system that has periodic components (e.g., the vertical motion of a pendulum undergoes periodic motion, where the pendulum's angle with the vertical axis and its angular velocity is the output).

In the following example we highlight the classic result of an LPTV system, the *Hill equation* and its special subcases: *Mathieu equation* and *Meissner equation.*

**<u>Example 1.6</u>: Hill Equation**

The *Hill equation*[3] is a second-order linear ODE reduced to first order. Consider the following second-order ODE:

$$\frac{d}{dt}\begin{pmatrix} y \\ \dot{y} \end{pmatrix} = \begin{bmatrix} 0 & 1 \\ 2q\psi(t) - a & 0 \end{bmatrix}\begin{pmatrix} y \\ \dot{y} \end{pmatrix}, \tag{1.31}$$

where $\psi(t)$ is a scalar $T$-periodic function with amplitude equals to 1 and mean equals to 0, $a$ and $q$ are constant parameters, and $y = y(t)$ is the single variable, which, together with its derivative $\dot{y}$, constructs the state space $x = [y, \dot{y}]^T$. The Hill equation is taken from the work on a lunar system of George William Hill (Hill, 1886), (Hill, 1878) [see e.g., (Richards, 1983), Eq. (2.5)].

[3] A damped Hill equation occurs when $[A(t)]_{2,2} \not\equiv 0$ is a $T$-periodic function.

A special case of the Hill equation is called the *Mathieu equation* and is defined by using $\psi(t) = \cos(\omega t)$. Typically, the frequency $\omega = 2\ \mathrm{rad/s}$ or, equivalently, the time period $T = \pi$ sec. The Mathieu equation was originally derived in the context of vibrations of stretched elliptical membranes (Mathieu, 1868) [see e.g., (Richards, 1983), Eqs. (2.6) and (6.1)]. Another special case of the Hill equation is called the *Meissner equation* and is defined by $\psi(t) = \mathrm{sgn}\{\cos(\omega t)\}$. This equation was derived by Meissner while studying the motion of the side rods of a locomotive (Meissner, 1918). The function $\mathrm{sgn}\{\cos(\omega t)\}$ generates a $\frac{2\pi}{\omega}$-periodic square function that changes its sign from +1 over the first half of a period to $-1$ over the second half [see e.g., (Richards, 1983), section 3.2].

We demonstrate *Hill equation* (and *Mathieu equation*) on the following two examples. The first example is a vertical motion of a pendulum undergoes periodic motion, illustrated below:

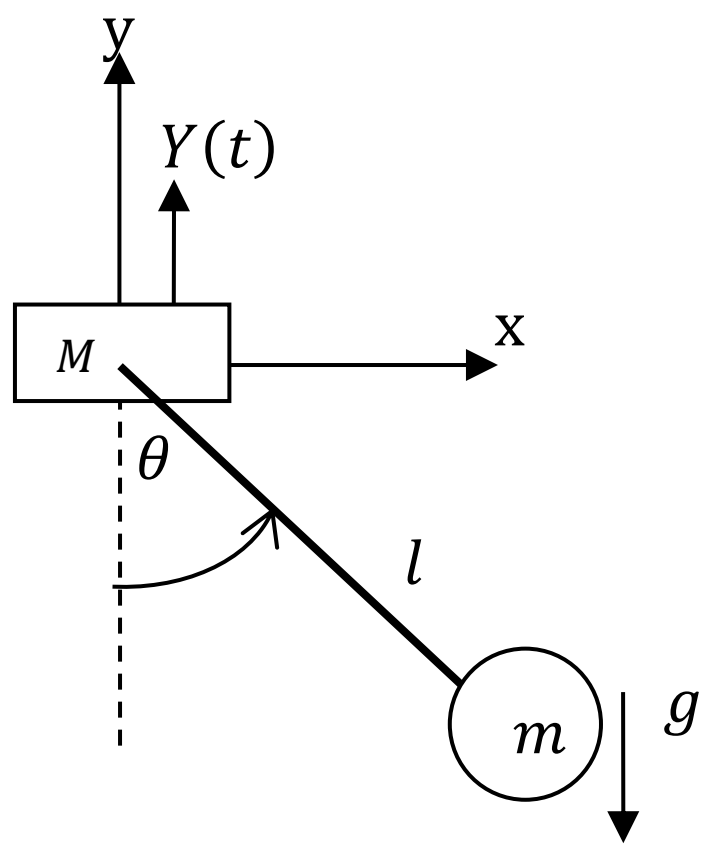

***Figure 1-1*** *Vertical pendulum undergoes periodic motion.*

The pendulum system consists of a rigid and massless rod with length $l$, that connects between a bob with mass $m$ and a pivot point with mass $M$. The vertical displacement of the pivot is given by the T-periodic scalar function $Y(t)$. $\theta$ is the degree of freedom tin the system, the angle between the rod and the negative y-axis. $g$ is the acceleration of the gravity at the negative y-axis direction. The equation of motion in $\theta$ direction can be derived e.g., by Lagrange equation and we have

$$ml^2\ddot{\theta} + \left(mgl + ml\ddot{Y}\right)\sin(\theta) = 0,$$

and for the state vector $\left[\theta, \dot{\theta}\right]^T$, the following *Hill equation* is derived adjacent the stable equilibrium point $\left[\theta, \dot{\theta}\right]^T = [0,0]$ as follows:

$$\frac{d}{dt}\begin{pmatrix}\theta\\ \dot{\theta}\end{pmatrix} = \begin{bmatrix}\dot{\theta}\\ \left(-\frac{\ddot{Y}(t)}{l} - \frac{g}{l}\right)\sin\theta\end{bmatrix} \overset{(a)}{\approx} \begin{bmatrix}\dot{\theta}\\ \left(-\frac{\ddot{Y}(t)}{l} - \frac{g}{l}\right)\theta\end{bmatrix} = \begin{bmatrix}0 & 1\\ -\frac{1}{l}\ddot{Y}(t) - \frac{g}{l} & 0\end{bmatrix}\begin{pmatrix}\theta\\ \dot{\theta}\end{pmatrix}.$$

The approximation in (a) is obtained due to $\sin\theta \approx \theta$ for $|\theta| \ll 1$. In this setup, refer to the notation in *Hill equation* in Eq. (1.31), $a = \frac{g}{l}$, $q = -\frac{1}{2l}$ and $\psi(t) = \ddot{Y}(t)$. In order to have a *Hill equation* for an inverted pendulum (adjacent the unstable equilibrium point $\left[\theta, \dot{\theta}\right]^T = [\pi, 0]$), we can define a new angle $\tilde{\theta}$ such that $\tilde{\theta} = \theta - \pi$, and to derive the *Hill equation* WRT $\tilde{\theta}$ (adjacent the unstable equilibrium point $\left[\tilde{\theta}, \dot{\tilde{\theta}}\right]^T = [0,0]$). Using the identity $\sin\left(\tilde{\theta} + \pi\right) = -\sin\tilde{\theta}$, we have

$$\frac{d}{dt}\begin{pmatrix}\tilde{\theta}\\ \dot{\tilde{\theta}}\end{pmatrix} = \begin{bmatrix}0 & 1\\ \frac{1}{l}\ddot{Y}(t) + \frac{g}{l} & 0\end{bmatrix}\begin{pmatrix}\tilde{\theta}\\ \dot{\tilde{\theta}}\end{pmatrix}.$$

*Mathieu equation* is obtained if we set the displacement function $Y(t) = Y_0 \cos(\omega t)$ so that the acclamation $\ddot{Y}(t) = -Y_0\omega^2 \cos(\omega t)$ is induced in the system.

The second example is a degenerate parametric amplification in electric circuit applications [see (Richards, 1983) section 7.5.1]. consider the following electrical cicuit in the figure below:

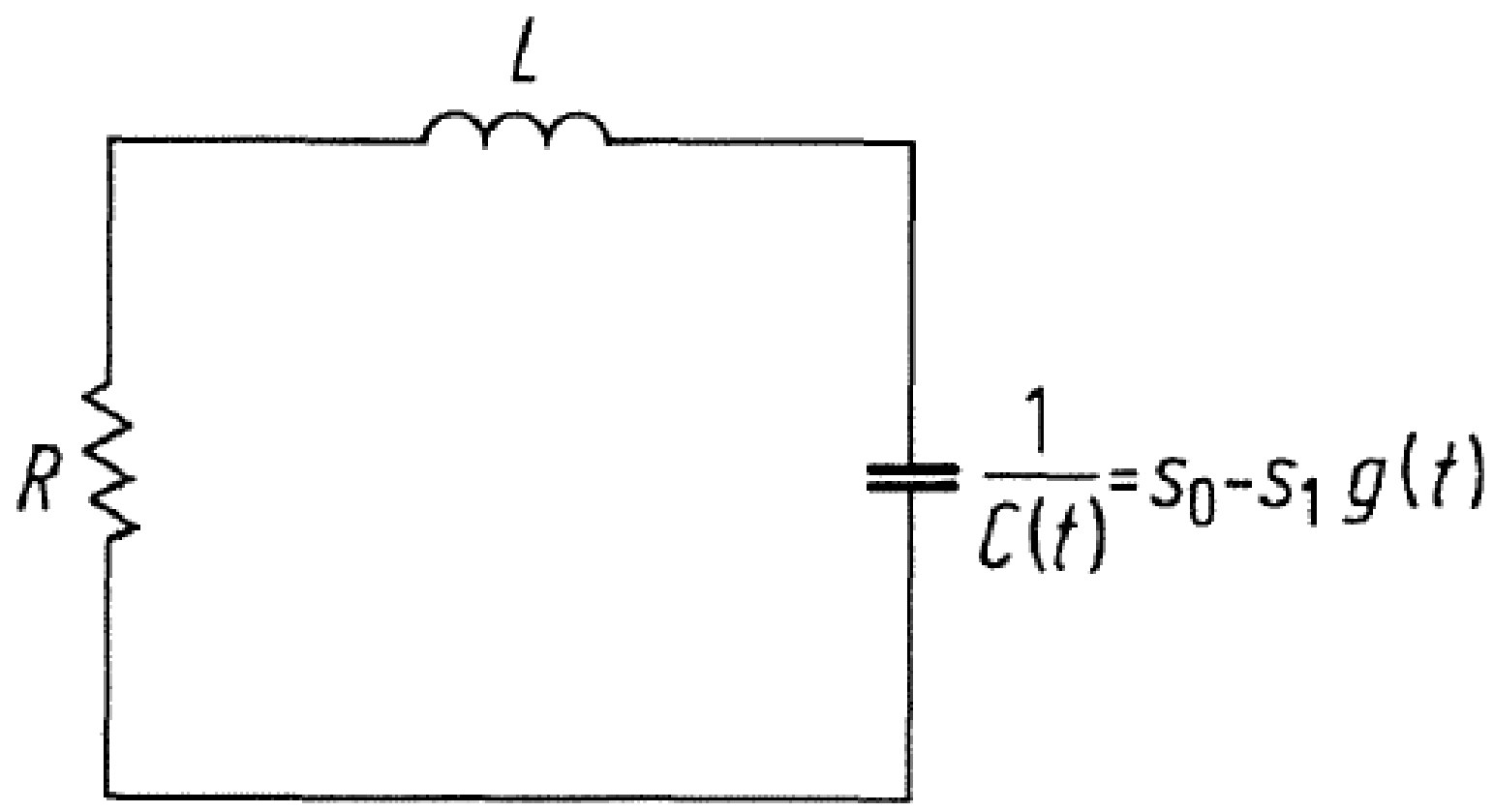

***Figure 1-2*** *Series RLC network with periodically varying capacitance. Source: (Richards, 1983) Fig. 7.5.1*

In this electrical cicuit, we have a series connection of RLC componects with constant resistor with resistance $R$, constant inductor with inductance $L$ and a time varying capacitor $C(t)$ with elastance $\frac{1}{C(t)} = s_0 - s_1 g(t)$, where $g(t)$ is T-periodic. Denote the state vector $[Q, \dot{Q}]^T$ where $Q$ is the electrical charge and $\dot{Q}$=$i$ is the curret of the circutry. Using the kirchhoff's voltage law, we construct the *Hill equation* WRT the state vector $[q, \dot{q}]^T$ as follows:

$$\frac{d}{dt}\begin{pmatrix} Q \\ \dot{Q} \end{pmatrix} = \begin{bmatrix} 0 & 1 \\ \frac{s_1}{L} g(t) - \frac{s_0}{L} & -\frac{R}{L} \end{bmatrix} \begin{pmatrix} Q \\ \dot{Q} \end{pmatrix}$$

This equation is a damped version of *Hill equation* due to the term $-\frac{R}{L}$ on the last entry in the matrix. refer to the notation in *Hill equation* in Eq. (1.31), $a = \frac{s_0}{L}$, $q =$ $\frac{s_1}{2L}$ and. $\psi(t) = g(t)$. A (damped) *Mathieu equation* is obtained when we set $g(t) = \cos(\omega t)$.

The general *Hill equation* (and *Mathieu equation*) is very difficult to solve analytically and usually requires summing infinite terms, but the *Meissner equation* is easy to solve piecewise by applying a technique for solving LTI systems. The purpose of the present study is to find a family of LPTV systems with a finite number of harmonics, and that is analytically solvable by applying a nontrivial technique (i.e., not a technique for solving LTI systems), such as exponentiating the antiderivative of an LPTV matrix.

□

## 1.4 Additional Recent Results

### 1.4.1 Stability Analysis of LPTV System

Whether a system is linear or nonlinear, its stability is one of the most significant properties to be analyzed. A necessary condition for quantitative exponential stability of an LTV system is outlined, e.g., in (Lewkowicz, 1999). An unforced LTI system is stable (asymptotically) if and only if all the eigenvalues of the system matrix $A$ have a negative real part. However, this statement is false for a more general LTV system and, in particular, for an LPTV system. Two counterexamples are given by (Markus & Yamabe, 1960, pp. 130-131) and (Rosenbrook, 1963, p. 73)[4]. In both examples, the LPTV system is represented by a 2×2 periodic matrix $A(t)$ with frequency and Fourier-coefficient matrices that are numerically known and have constant eigenvalues with negative real parts, and the solution includes terms that grow exponentially instead of decaying (i.e., are unstable).

[4] It may be referred to as the "Vinograd Example" (Vinograd, 1952) in the literature.

These counterexamples (or variations thereof) are discussed in other textbooks [see, e.g., (Amato, 2006) Example 2.2; (Bittanti & Colaneri, 2009) Example 1.1; (Chicone, 2006) Eq. (2.28); (Colonius & Kliemann, 2014) Chapter 6, pages 109–111; (Kelley & Peterson, 2010) Exercise 2.69 and similarly Exercise 2.66; (Khalil, 2002) Example 4.22 and Exercise 10.10(2); (Mathis, 1987) Example 5.13; (Rugh, 1996) Example 8.1]. These counterexamples are also used in some articles to demonstrate stability analyses and solutions of LTV systems and, in particular, of LPTV systems. Aggarwal and Infante (Aggarwal & Infante, 1968) parametrized the example from (Markus & Yamabe, 1960) by scaling the periodic part $a$ of the LPTV system. The stability condition is based on this parametrization factor $a$, which is obtained directly from the solution of the LPTV system rather than from the eigenvalues of the LPTV system matrix.

Numerous authors (Colaneri, 2005), (DaCunha, 2004), (Mullhaupt, Buccieri, & Bonvin, 2007), (Varbel, 2020), (Wu, 1974), (Yao, Liu, Sun, Balakrishnan, & Guo, 2012), analyze the stability of LPTV systems and refer to the examples in (Aggarwal & Infante, 1968), (Markus & Yamabe, 1960), or (Rosenbrook, 1963) without suggesting any solution. For example, by using a *logarithmic norm*[5], (Varbel, 2020) derived a new criterion for uniform exponential stability of an LPTV system without finding the constant matrix $R$. However, (Varbel, 2020) refers to a stability analysis without solving the ODE of the LPTV system itself.

[5] The *logarithmic norm* ,defined by $\mu[A] \stackrel{\text{def}}{=} \lim_{h\to 0^+} \frac{\|I_n - hA\| - 1}{h}$ , is called in some books a *measure of a matrix* [see e.g., (Desoer & Vidyasagar, 2009), chapter II section 8] or simply a *matrix measure* [see e.g., (Rugh, 1996), note 8.2, referring to (Coppel, 1965). More specifically, see in (Coppel, 1965): page 41 Eq. (1) and page 58 Theorem 3]. A *logarithmic norm* may be negative.

### 1.4.2 Dynamic Eigen Decomposition of LTV System

van der Kloet and Neerhoff (van der Kloet & Neerhoff, 2004a) (van der Kloet & Neerhoff, 2004b) suggest a generalization of dynamic eigenvalues and a procedure to diagonalize a general LTV system based on the Riccati equation, which is not a linear ODE and might be difficult to solve. This approach is demonstrated on an LPTV system (Aggarwal & Infante, 1968), (Rosenbrook, 1963). The article discusses the use of dynamic eigenvalues along with the Riccati equation and similar approaches (Wang, 2017), and the LPTV system addressed is the same as in (Aggarwal & Infante, 1968). However, instead of using the approach based on the Riccati equation, the LPTV system is solved by using an *auxiliary equation*, which is outlined in APPENDIX D. This requires that an appropriate matrix $G(t)$ be selected to find dynamic eigenvalues of the matrix $A(t) - G(t)$. In addition, $G(t)$ is required to be an LTV system with solutions consisting of the corresponding dynamic eigenvectors, which may make it difficult to guess $G(t)$ and verify whether it is difficult to implement (see APPENDIX D for more details).

### 1.4.3 Small-Perturbation Approach

Yakubovich and Starzhinskii,[ (Yakubovich & Starzhinskii, 1975) Chapter 4], use a small-parameter approach so that the LPTV system matrix $A(t)$ and its solution matrices $P(t)$ and $R$ can each be decomposed into a power series (infinite or finite) in some small parameter $\varepsilon > 0$ such that:

$$\begin{aligned} A(t) &= A_0 + \sum_{k=1}^{\infty} \varepsilon^k A_k(t), \\ P(t) &= P_0 + \sum_{k=1}^{\infty} \varepsilon^k P_k(t), \\ R &= \sum_{k=0}^{\infty} \varepsilon^k R_k, \end{aligned} \tag{1.32}$$

where the matrices $A_0$ and $P_0$ are not functions of $t.$ Note that for $\varepsilon = 0$, we have an LTI system [see, e.g., (Prokopenya, 2007) and (Sinha, Pandiyan, & Bibb, 1996)].

Herein, inspired from that small-perturbation approach, we suggest to present $R$ and the Fourier series coefficients of $A(t)$ as polynomials in the frequency $\omega$, but $P(t)$'s Fourier series coefficients to be independent on $\omega$. In our approach, the term $A_0$ is also $T$-periodic. Our approached was developed so that when $\omega \to 0$, the LPTV system approaches to be an LTI system [i.e., $A(t)$ approaches to be a constant matrix]. However, unlike the small-perturbation approach, our approach of polynomials in the frequency $\omega$ is develop, so that the obtained solution for $P(t)$ and $R$ is valid not only to small values of $\omega$, but for any $\omega$.

### 1.4.4 Real Floquet factorization of LPTV Transition Matrix

The work (Montagnier, Paige, & Spiteri, 2003) addresses to a factorization of a known Transition Matrix of a $T$-periodic LPTV system using matrix logarithm. This may cause complex solutions for the constant matrix $R$ and $T$-periodic matrix $P(t)$ [see Eqs. (1.23)-(1.25)]. Montagnier et al. emphasize in their work that it is important to obtain real factorizations from computations on a single period, and refer to the results in (Yakubovich & Starzhinskii, 1975) Section 2.3 to obtain such real factorizations [Theorem 2.3 in (Montagnier, Paige, & Spiteri, 2003)]. Suppose an arbitrary real matrix $W(t)$ is a Fundamental Solution [Eq. (1.5)] corresponding to an LPTV system's $T$-periodic matrix $A(t)$. The following factorization of $W(t)$ is required:

$$W(t) = P(t)e^{Rt}, \tag{1.33}$$

then, there exists a real matrix $Y$ such that there is a real factorization for Eq. (1.33), a real $2T$-periodic function matrix $P(t)$ and a real constant matrix $R$, satesfying the following

$$\begin{aligned} P(t+T) &= P(t)Y, \\ Y^2 &= I, RY = YR, \\ P(t+2T) &= P(t). \end{aligned} \tag{1.34}$$

When $W(t)$ is constructed, then the transition matrix $\Phi_A(t, t_0)$ is constructed by Eq. (1.4). Montagnier et al. characterize at Theorem 3.1 of their work all real Floquet factorizations $\Phi_A(t,0) = P(t)e^{Rt}$, with $P(t)$ being a $T$-periodic or $2T$-periodic such that $\Phi_A(0,0) = P(0) = I$, by applying Eqs. (1.33)-(1.34) on this transition matrix. According to (Montagnier, Paige, & Spiteri, 2003) Theorem 3.1, suppose that exist a real matrix $Y$ such that there is a real factorization for Eq. (1.33), a real function matrix $P(t)$ and a real constant matrix $R$, satesfying the following:

$$\begin{gathered} \Phi_A(t,0) = P(t)e^{Rt}, \quad \Phi_A(0,0) = P(0) = I_n, \\ Y\Phi_A(T,0) = e^{TR}, \ln\big(Y\Phi_A(T,0)\big) \in \mathbb{R}^{n\times n}, \end{gathered} \tag{1.35}$$

then,

$$P(t+T) = P(t)e^{tR}P(T)e^{-tR}, \tag{1.36}$$

and also, the choice of $Y$ affects $P(t)$ as follows:

$$\begin{gathered} P(T) = Y^{-1}; \\ P(t) = P(t+T)\ \forall t \iff Y = I_n; \\ P(t) = -P(t+T)\ \forall t \iff Y = -I_n; \\ P(t) = P(t+2T)\ \forall t \iff [\Phi_A(T,0)]^2 = [Y\Phi_A(T,0)]^2. \end{gathered} \tag{1.37}$$

These results may affect the number of harmonics of the periodic part of the solution $P(t)$ (see section 4.2 below). Moreover, this real Floquet factorization for the transition matrix $\Phi_A(t,0)$ is valid only when this transition matrix $\Phi_A(t,0)$ is known, which is not the case in general. However, this result motivates us to explore, when for a real $T$-periodic $A(t)$ (with a finite number of harmonics) we can obtain a real Floquet factorization matrix pair: a $T$-periodic $P(t)$ (with a finite number of harmonics with a possible condition $P(0) = I_n$) and a constant $R$ (perhaps with some real canonical form e.g., Jordan Matrix).

### 1.4.5 Alternative Factorization of LPTV Transition Matrix

Jikuya and Hodaka (Jikuya & Hodaka, 2009), (Jikuya & Hodaka, 2010) refer to (Montagnier, Paige, & Spiteri, 2003) and (Yakubovich & Starzhinskii, 1975) to emphasize the importance of having a real Floquet factorization. Jikuya and Hodaka suggest factorizing the solution of an LPTV system into a $T$-periodic function $P(t)$ and two matrix exponential functions $e^{Gt}, e^{Ft}$ with $F, G \in \mathbb{R}^{n\times n}$ such that:

$$\begin{gathered}\Phi_A(t,0) = P(t)e^{Gt}e^{Ft}, \quad \Phi_A(0,0) = P(0) = I_n\,, \\ e^{GT}F = Fe^{GT}, e^{GT}e^{Ft} = e^{Ft}e^{GT}, e^{2GT} = I_n.\end{gathered} \tag{1.38}$$

$F$ is chosen to characterize the stability of the system (such as the matrix $R$ given above), and $G$ is chosen so that $e^{Gt}$ is $T$-periodic. This suggestion assumes that the solution is known, although it is not easy to find. Since $e^{Gt}$ is $T$-periodic, we can write the matrix $G$ as a linear function of $\omega = \frac{2\pi}{T}$ with $G = G_1\omega$. If $F$ and $G$ commute, we have $e^{Gt}e^{Ft} = e^{(F+G)t} = e^{(F+G_1\omega)t}$, and therefore, the constant part of the floquet factorization is given by $R = F + G_1\omega$ namely a linear function in $\omega$. We generalize this notion by assuming $R$ to be any arbitrary polynomial function in $\omega$ (see the last paragraph in section 1.4.3).

### 1.4.6 Generalization of LTI-System Tools for LPTV Systems

Wereley suggests in his PhD work (Wereley, 1991) applying generalized tools to LPTV systems to analyze and control LTI systems (e.g., transfer functions, zeros and poles in $s$ and $z$ domains), based on complex exponential functions representation of Fourier series. Some examples from (Wereley, 1991) are outlined in APPENDIX E. Similar concept of generalized transfer functions is shown on (Gad & Nakhla, 2005). We are inspired by Wereley's work by converting the LPTV ODE in Eq. (1.16) to an algebraic matrix equation, but with respect to a cosine and sine representation of Fourier series instead of complex exponential functions.

## 1.5 Objective of the Research, Suggested Approach and Scope of Work

The objective of this work is to find a large family of periodic matrices, $A(t)$, so that with an efficient procedure, we can find their suitable matrices $P(t)$ and $R$, and to obtain the transition matrix $\Phi_A(t, t_0)$. Additional goal is to find this efficient procedure. This work is inspired by my supervisor's working draft (Lewkowicz, 2020) . This work is limited to the case that both $A(t)$ and $P(t)$ have finite numbers of harmonics (motivated at section 4.2) with the following Fourier series representations[6]:

[6] We limit our work to a real LPTV system matrix $A(t)$, so we would like to find a real matrix pair $P(t)$ and $R$. Therefore, we use a cosine-sine Fourier series to avoid complex matrix computations. In addition, we extend the Fourier series to negative indices (see APPENDIX A for details).

$$A(t)=\sum_{l=-L}^{L}\left(\frac{A_l^{\text{even}}}{2}\cos(l\omega t)+\frac{A_l^{\text{odd}}}{2}\sin(l\omega t)\right), \tag{1.39}$$

$$P(t)=\sum_{k=-p}^{p}\left(\frac{P_k^{\text{even}}}{2}\cos(k\omega t)+\frac{P_k^{\text{odd}}}{2}\sin(k\omega t)\right), \tag{1.40}$$

In order to find to obtain the suitable $P(t)$ and $R$, we need to solve the above matrix ODE in Eq. (1.16).

The key idea in this work is to address cases where in the Fourier series expansion of each $A(t)$ and $P(t)$, there is a finite number of harmonics. Then to exploit this property to solve the matrix ODE in Eq. (1.16), using a comparison of the Fourier series matrix coefficients and rearrangement into an algebraic matrix equation with finite dimensions. This notion is discussed at CHAPTER 2 for a cosine-sine form of the Fourier series (and at APPENDIX B for the exponential form), so that by Eq. (2.22) below we have

$$\widetilde{\mathbf{A}}\widetilde{\mathbf{P}}=\widetilde{\mathbf{P}}R, \tag{1.41}$$

Where $\widetilde{\mathbf{A}}$ is a block matrix contains the Fourier series the known Fourier series matrix coefficients of $A(t)$ [see Eqs. (2.22)-(2.25)], and $\widetilde{\mathbf{P}}$ is a block matrix contains the unknown Fourier series matrix coefficients of $P(t)$ [see Eq. (2.23)]. We require a finite number of harmonics both in $A(t)$ and $P(t)$, in order to construct the matrices $\widetilde{\mathbf{A}}$ and $\widetilde{\mathbf{P}}$ in Eq. (1.41) to be with finite dimensions. Solving Eq. (1.41) with infinite will be extremely difficult.

This algebraic equation has a similar form of an eigen decomposition setup with the matrix $R$ plays the role of the eigenvalues' matrix, and the block matrix $\widetilde{\mathbf{P}}$ plays the role of the eigenvectors' matrix. Hence, a similar approach to this eigen decomposition is required to be developed, e.g., finding a nontrivial solution for $\widetilde{\mathbf{P}}$ (i.e., not the zero solution) with its corresponding eigen pair $R$.

In order to enhance the approach of solving the algebraic Eq. (1.41), we suggest to treat the frequency $\omega$ as a free parameter and not a fixed number (motivated at section 4.3), so that we can, e.g., perform a comparison of the powers of $\omega$. Herein, this work focuses on cases in which the coefficients of the Fourier series of $A(t)$ are finite polynomials in $\omega$,[7] $R$ is a finite polynomial in $\omega$, but the coefficients of the Fourier series of $P(t)$ do not depend on $\omega$, such that:

$$\frac{A_l^{\text{even}}}{2} = \sum_{r=0}^{N}\left(\omega^r \frac{{A^{\{r\}}}_l^{\text{even}}}{2}\right), \qquad \frac{A_l^{\text{odd}}}{2} = \sum_{r=0}^{N}\left(\omega^r \frac{{A^{\{r\}}}_l^{\text{odd}}}{2}\right), \tag{1.42}$$

$$R = \sum_{r=0}^{N} \omega^r R^{\{r\}}\,, \tag{1.43}$$

where ${A^{\{r\}}}_l^{\text{even}}$ and ${A^{\{r\}}}_l^{\text{odd}}$ are the coefficients of $\omega^r$ in the Fourier series (and likewise for $R^{\{r\}}$). This setup for $A(t)$, $P(t)$, and $R$ serves to compare two coefficients to find $P(t)$ and $R$: Fourier coefficients and powers of $\omega$. The Fourier coefficients of $P(t)$ must not depend on $\omega$ to assure that the terms ${A^{\{r\}}}_l^{\text{even}}$ and ${A^{\{r\}}}_l^{\text{odd}}$ are polynomials in $\omega$ (i.e., without negative powers), and also to assure a continuity in the parameter $\omega$ in $A(t)$, $P(t)$, and $R$ [sometimes this condition is denoted $A(t|\omega)$, $P(t|\omega)$, and $R(\omega)$ to emphasize the dependency on $\omega$]. This notion is implemented at CHAPTER 5 so that Eq. (1.41) is reformed as follows:

$$\widetilde{\mathbf{A}}^{\{r\}}\widetilde{\mathbf{P}} = \widetilde{\mathbf{P}}R^{\{r\}}, \tag{1.44}$$

[7] Although this work is not limited to a small frequency parameter $\omega$ [in contrast to (Yakubovich & Starzhinskii, 1975) who define a small perturbation by $\varepsilon > 0$], this work discusses properties related to the LPTV system matrix $A(t)$ and its LTI version when $\omega = 0$, so that the matrix $A(t)$ is not periodic and its components become constant (i.e., $A(t|\omega = 0)$ is an LTI system).

Where $\widetilde{\mathbf{A}}^{\{r\}}$ and $R^{\{r\}}$ are the matrices corresponding to the comparison of $\omega^r$, for $r = 0,1, \dots N$. It seems that we need to solve for all $r = 0,1, \dots N$ in order to obtain all $R^{\{r\}}$. However, due to our setup for $P(t)$'s Fourier series coefficient to be independent on $\omega$, the solution for $\widetilde{\mathbf{P}}$ is universal for all $r = 0,1, \dots N$, i.e., it is sufficient to solve Eq. (1.44) for some $r \in \{0,1, \dots N\}$ [for example $r = 0$ where we can find a relation between $A(t|\omega = 0)$ and $R^{\{r\}}$], so that from $\widetilde{\mathbf{P}}$ we geneate our solution to $P(t)$ and solving for $R$.

# CHAPTER 2. Fourier Analysis of LPTV Systems

## 2.1 Overview

This chapter implements a Fourier analysis of LPTV systems [presented by the $T$-periodic matrix $A(t)$ ] and their $T$-periodic part $P(t)$ of the solution provided by *Floquet theory*. We develop the equation under the assumption that the LPTV system is real [i.e., $A(t) \in \mathbb{R}^{n\times n}$], so it is convenient to use the cosine-sine Fourier decomposition to find a pair of real matrices, $P(t),\ R \in \mathbb{R}^{n\times n}$, as a solution per *Floquet theory*. The cosine-sine Fourier decomposition gives matrices with real Fourier coefficients separated according to their parity (even or odd).

For comparison, APPENDIX B outlines a case that applies a complex Fourier decomposition to an LTPV system and its *Floquet theory* solution. APPENDIX C, outlines how a complex LPTV system can be converted to a real LPTV system based on the isomorphism $a + ib \leftrightarrow \begin{bmatrix} a & -b \\ b & a \end{bmatrix}$, and outlines how to decompose the state vector into even and odd parts and then rewrites the LPTV system matrix with even and odd matrix blocks.

## 2.2 Cosine-Sine Fourier Analysis of LPTV Systems

### 2.2.1 General

Suppose that the periodic matrices $P(t)$ and $R$ are solutions to Eq. (1.16) and are used to construct the transition matrix $\Phi_A(t, t_0)$. As per APPENDIX A, Eq. (A.6), we use the following cosine-sine Fourier decomposition for $A(t)$ and $P(t)$:

$$A(t) = \sum_{l=-\infty}^{\infty} \left( \frac{A_l^{\text{even}}}{2} \cos(l\omega t) + \frac{A_l^{\text{odd}}}{2} \sin(l\omega t) \right), \tag{2.1}$$

$$P(t) = \sum_{k=-\infty}^{\infty} \left( \frac{P_k^{\text{even}}}{2} \cos(k\omega t) + \frac{P_k^{\text{odd}}}{2} \sin(k\omega t) \right), \tag{2.2}$$

where $A_l^{\text{even}}$ ($A_l^{\text{odd}}$), and $P_l^{\text{even}}$ ($P_l^{\text{odd}}$) are real Fourier coefficients (in general, they may be complex) of matrices of $\cos(l\omega t)$ [ $\sin(l\omega t)$ ] in $A(t)$ and $P(t)$, respectively.

We need to decompose all terms in Eq. (1.16) into even functions [i.e., coefficients of $\cos(l\omega t)$ and 1] and odd functions [i.e., coefficients of $\sin(l\omega t)$], and likewise also decompose $A(t)$ and $P(t)$ themselves. The product $A(t)P(t)$ includes a linear combination of products of cosine and sine factors, so the following trigonometric identities will be useful to convert these products to sums:

$$\sin(a\omega t)\cos(b\omega t) = \frac{1}{2}\sin([a-b]\omega t) + \frac{1}{2}\sin([a+b]\omega t), \tag{2.3}$$

$$\cos(a\omega t)\cos(b\omega t) = \frac{1}{2}\cos([a-b]\omega t) + \frac{1}{2}\cos([a+b]\omega t), \tag{2.4}$$

$$\sin(a\omega t)\sin(b\omega t) = \frac{1}{2}\cos([a-b]\omega t) - \frac{1}{2}\cos([a+b]\omega t). \tag{2.5}$$

### 2.2.2 Even-Odd Decomposition

Any function $f(t)$ can be decomposed into a sum of an even function $f^{\text{even}}(t)$ and an odd function $f^{\text{odd}}(t)$:

$$f(t) = f^{\text{even}}(t) + f^{\text{odd}}(t)\,, \tag{2.6}$$

where

$$f^{\text{even}}(t) = \frac{f(t) + f(-t)}{2} \tag{2.7}$$

and

$$f^{\text{odd}}(t) = \frac{f(t) - f(-t)}{2}. \tag{2.8}$$

By using the above even-odd decomposition property, $A(t)$ is decomposed as follows:

$$A(t) = A^{\text{even}}(t) + A^{\text{odd}}(t), \tag{2.9}$$

where

$$\begin{aligned} A^{\text{even}}(t) &= \sum_{l=-\infty}^{\infty} \frac{A_l^{\text{even}}}{2} \cos(l\omega t), \\ A^{\text{odd}}(t) &= \sum_{l=-\infty}^{\infty} \frac{A_l^{\text{odd}}}{2} \sin(l\omega t), \end{aligned} \tag{2.10}$$

and likewise, for $P(t)$. Plugging the above even-odd decomposition of Eq. (2.9) [for both $A(t)$ and $P(t)$] into Eq. (1.16) gives:

$$\left(A^{\text{even}}(t) + A^{\text{odd}}(t)\right)\left(P^{\text{even}}(t) + P^{\text{odd}}(t)\right) = \frac{d}{dt}\{P^{\text{even}}(t) + P^{\text{odd}}(t)\} + \left(P^{\text{even}}(t) + P^{\text{odd}}(t)\right)R. \tag{2.11}$$

Note that the derivative of an even function is an odd function and vice versa. In addition, multiplication of even or odd functions gives even functions, and multiplication of even by odd functions (or vice versa) gives odd functions. Therefore, we obtain:

$$A^{\text{even}}(t)P^{\text{even}}(t) + A^{\text{odd}}(t)P^{\text{odd}}(t) = \frac{d}{dt}\{P^{\text{odd}}(t)\} + P^{\text{even}}(t)R \tag{2.12}$$

as even terms, and:

$$A^{\text{even}}(t)P^{\text{odd}}(t) + A^{\text{odd}}(t)P^{\text{even}}(t) = \frac{d}{dt}\{P^{\text{even}}(t)\} + P^{\text{odd}}(t)R \qquad (2.13)$$

as odd terms.

The terms of the form $A^{\text{even}}(t)P^{\text{even}}(t)$ and $A^{\text{odd}}(t)P^{\text{odd}}(t)$ on the LHS of Eqs. (2.12) and (2.13) are linear combinations of products of a pair of cosine and/or sine terms, so we use the trigonometric identities (2.3)-(2.5) to convert these products to sums. Re-indexing the result and using the even-odd properties from Eq. (A.5) gives:

$$A^{\text{even}}(t)P^{\text{even}}(t) + A^{\text{odd}}(t)P^{\text{odd}}(t) = \sum_{k=-\infty}^{\infty}\sum_{l=-\infty}^{\infty}\left[\frac{A_l^{\text{even}}}{2}\frac{P_{k-l}^{\text{even}}}{2} - \frac{A_l^{\text{odd}}}{2}\frac{P_{k-l}^{\text{odd}}}{2}\right]\cos(k\omega t)\,, \qquad (2.14)$$

$$A^{\text{even}}(t)P^{\text{odd}}(t) + A^{\text{odd}}(t)P^{\text{even}}(t) = \sum_{k=-\infty}^{\infty}\sum_{l=-\infty}^{\infty}\left(\frac{A_l^{\text{odd}}}{2}\frac{P_{k-l}^{\text{even}}}{2} + \frac{A_l^{\text{even}}}{2}\frac{P_{k-l}^{\text{odd}}}{2}\right)\sin(k\omega t)\,. \qquad (2.15)$$

**Proof of Eq. (2.14)**

$A^{\text{even}}(t)P^{\text{even}}(t) + A^{\text{odd}}(t)P^{\text{odd}}(t) =$

$$= \sum_{l=-\infty}^{\infty}\sum_{k=-\infty}^{\infty}\frac{A_l^{\text{even}}}{2}\frac{P_k^{\text{even}}}{2}\cos(l\omega t)\cos(k\omega t) + \sum_{l=-\infty}^{\infty}\sum_{k=-\infty}^{\infty}\frac{A_l^{\text{odd}}}{2}\frac{P_k^{\text{odd}}}{2}\sin(l\omega t)\sin(k\omega t)$$

$$+ \sum_{l=-\infty}^{\infty}\sum_{k=-\infty}^{\infty}\frac{A_l^{\text{even}}}{2}\frac{P_k^{\text{even}}}{2}\frac{1}{2}[\cos([l-k]\omega t) + \cos([l+k]\omega t)] +$$

$$+ \sum_{l=-\infty}^{\infty}\sum_{k=-\infty}^{\infty}\frac{A_l^{\text{odd}}}{2}\frac{P_k^{\text{odd}}}{2}\frac{1}{2}[\cos([l-k]\omega t) - \cos([l+k]\omega t)]$$

$$= \sum_{l=-\infty}^{\infty}\sum_{m=-\infty}^{\infty}\frac{A_l^{\text{even}}}{2}\frac{P_{l-m}^{\text{even}}}{2}\frac{1}{2}[\cos([m]\omega t)]\,|_{k=l-m} + \sum_{l=-\infty}^{\infty}\sum_{m=-\infty}^{\infty}\frac{A_l^{\text{even}}}{2}\frac{P_{m-l}^{\text{even}}}{2}\frac{1}{2}\cos([m]\omega t)\,|_{k=m-l}$$

$$+ \sum_{l=-\infty}^{\infty}\sum_{m=-\infty}^{\infty}\frac{A_l^{\text{odd}}}{2}\frac{P_{l-m}^{\text{odd}}}{2}\frac{1}{2}[\cos([m]\omega t)]\,|_{k=l-m} - \sum_{l=-\infty}^{\infty}\sum_{m=-\infty}^{\infty}\frac{A_l^{\text{odd}}}{2}\frac{P_{m-l}^{\text{odd}}}{2}\frac{1}{2}\cos([m]\omega t)|_{k=m-l}$$

$$= \sum_{l=-\infty}^{\infty}\sum_{m=-\infty}^{\infty}\frac{A_l^{\text{even}}}{2}\frac{P_{m-l}^{\text{even}}}{2}\frac{1}{2}[\cos([m]\omega t)]\,|_{k=l-m} + \sum_{l=-\infty}^{\infty}\sum_{m=-\infty}^{\infty}\frac{A_l^{\text{even}}}{2}\frac{P_{m-l}^{\text{even}}}{2}\frac{1}{2}\cos([m]\omega t)\,|_{k=m-l}$$

$$- \sum_{l=-\infty}^{\infty}\sum_{m=-\infty}^{\infty}\frac{A_l^{\text{odd}}}{2}\frac{P_{m-l}^{\text{odd}}}{2}\frac{1}{2}[\cos([m]\omega t)]\,|_{k=l-m} - \sum_{l=-\infty}^{\infty}\sum_{m=-\infty}^{\infty}\frac{A_l^{\text{odd}}}{2}\frac{P_{m-l}^{\text{odd}}}{2}\frac{1}{2}\cos([m]\omega t)|_{k=m-l}$$

$$= \sum_{k=-\infty}^{\infty}\sum_{l=-\infty}^{\infty}\left[\frac{A_l^{\text{even}}}{2}\frac{P_{k-l}^{\text{even}}}{2} - \frac{A_l^{\text{odd}}}{2}\frac{P_{k-l}^{\text{odd}}}{2}\right]\cos(k\omega t)$$

■

**Proof of Eq. (2.15)**

$$A^{\text{even}}(t)P^{\text{odd}}(t)+A^{\text{odd}}(t)P^{\text{even}}(t)=$$

$$=\sum_{l=-\infty}^{\infty}\sum_{k=-\infty}^{\infty}\frac{A_l^{\text{odd}}}{2}\frac{P_k^{\text{even}}}{2}\sin(l\omega t)\cos(k\omega t)\ +\sum_{l=-\infty}^{\infty}\sum_{k=-\infty}^{\infty}\frac{A_l^{\text{even}}}{2}\frac{P_k^{\text{odd}}}{2}\cos(l\omega t)\sin(k\omega t)$$

$$=\sum_{l=-\infty}^{\infty}\sum_{k=-\infty}^{\infty}\frac{A_l^{\text{odd}}}{2}\frac{P_k^{\text{even}}}{2}\sin(l\omega t)\cos(k\omega t)\ +\sum_{l=-\infty}^{\infty}\sum_{k=-\infty}^{\infty}\frac{A_l^{\text{even}}}{2}\frac{P_k^{\text{odd}}}{2}\sin(k\omega t)\cos(l\omega t)$$

$$=\sum_{l=-\infty}^{\infty}\sum_{k=-\infty}^{\infty}\frac{A_l^{\text{odd}}}{2}\frac{P_k^{\text{even}}}{2}\frac{1}{2}[\sin([l-k]\omega t)+\sin([l+k]\omega t)]\ +$$

$$+\sum_{l=-\infty}^{\infty}\sum_{k=-\infty}^{\infty}\frac{A_l^{\text{even}}}{2}\frac{P_k^{\text{odd}}}{2}\frac{1}{2}[\sin([k-l]\omega t)+\sin([k+l]\omega t)]$$

$$=\sum_{l=-\infty}^{\infty}\sum_{k=-\infty}^{\infty}\frac{A_l^{\text{odd}}}{2}\frac{P_k^{\text{even}}}{2}\frac{1}{2}\sin([l-k]\omega t)\ |_{k=l-m}+\sum_{l=-\infty}^{\infty}\sum_{k=-\infty}^{\infty}\frac{A_l^{\text{odd}}}{2}\frac{P_k^{\text{even}}}{2}\frac{1}{2}\sin([l+k]\omega t)\,|_{k=m-l}\ +$$

$$+\sum_{l=-\infty}^{\infty}\sum_{k=-\infty}^{\infty}\frac{A_l^{\text{even}}}{2}\frac{P_k^{\text{odd}}}{2}\frac{1}{2}\sin([k-l]\omega t)\,|_{k=l-m}\ +\sum_{l=-\infty}^{\infty}\sum_{k=-\infty}^{\infty}\frac{A_l^{\text{even}}}{2}\frac{P_k^{\text{odd}}}{2}\frac{1}{2}\sin([k+l]\omega t)\,|_{k=m-l}$$

$$=\sum_{l=-\infty}^{\infty}\sum_{m=-\infty}^{\infty}\frac{A_l^{\text{odd}}}{2}\frac{P_{l-m}^{\text{even}}}{2}\frac{1}{2}\sin([m]\omega t)\ +\sum_{l=-\infty}^{\infty}\sum_{m=-\infty}^{\infty}\frac{A_l^{\text{odd}}}{2}\frac{P_{m-l}^{\text{even}}}{2}\frac{1}{2}\sin([m]\omega t)$$

$$+\sum_{l=-\infty}^{\infty}\sum_{m=-\infty}^{\infty}\frac{A_l^{\text{even}}}{2}\frac{P_{l-m}^{\text{odd}}}{2}\frac{1}{2}\sin([-m]\omega t)\ +\sum_{l=-\infty}^{\infty}\sum_{m=-\infty}^{\infty}\frac{A_l^{\text{even}}}{2}\frac{P_{m-l}^{\text{odd}}}{2}\frac{1}{2}\sin([m]\omega t)$$

$$=\sum_{l=-\infty}^{\infty}\sum_{k=-\infty}^{\infty}\frac{A_l^{\text{odd}}}{2}\frac{P_{k-l}^{\text{even}}}{2}\frac{1}{2}\sin(k\omega t)\ +\sum_{l=-\infty}^{\infty}\sum_{k=-\infty}^{\infty}\frac{A_l^{\text{odd}}}{2}\frac{P_{k-l}^{\text{even}}}{2}\frac{1}{2}\sin(k\omega t)$$

$$+\sum_{l=-\infty}^{\infty}\sum_{k=-\infty}^{\infty}\frac{A_l^{\text{even}}}{2}\frac{P_{k-l}^{\text{odd}}}{2}\frac{1}{2}\sin(k\omega t)\ +\sum_{l=-\infty}^{\infty}\sum_{k=-\infty}^{\infty}\frac{A_l^{\text{even}}}{2}\frac{P_{k-l}^{\text{odd}}}{2}\frac{1}{2}\sin(k\omega t)$$

$$=\sum_{k=-\infty}^{\infty}\sum_{l=-\infty}^{\infty}\left(\frac{A_l^{\text{odd}}}{2}\frac{P_{k-l}^{\text{even}}}{2}+\frac{A_l^{\text{even}}}{2}\frac{P_{k-l}^{\text{odd}}}{2}\right)\sin(k\omega t)$$

■

The other terms on the RHS of Eqs. (2.12) $[\frac{d}{dt}\{P^{\text{odd}}(t)\}+P^{\text{even}}(t)R]$ are already linear combinations of $\cos(k\omega t)$, and those on the RHS of Eq. (2.13) are linear combinations of $\sin(k\omega t)$. By comparing the coefficients of $\cos(k\omega t)$ and $\sin(k\omega t)$, we obtain:

$$\cos(k\omega t):\frac{P_k^{\text{even}}}{2}R+\frac{P_k^{\text{odd}}}{2}k\omega=\sum_{l=-\infty}^{\infty}\left(\frac{A_l^{\text{even}}}{2}\frac{P_{k-l}^{\text{even}}}{2}-\frac{A_l^{\text{odd}}}{2}\frac{P_{k-l}^{\text{odd}}}{2}\right),\qquad(2.16)$$

$$\sin(k\omega t):\frac{P_k^{\text{odd}}}{2}R-\frac{P_k^{\text{even}}}{2}k\omega=\sum_{l=-\infty}^{\infty}\left(\frac{A_l^{\text{odd}}}{2}\frac{P_{k-l}^{\text{even}}}{2}+\frac{A_l^{\text{even}}}{2}\frac{P_{k-l}^{\text{odd}}}{2}\right).\qquad(2.17)$$

To eliminate negative indices of $P_l^{\text{even}}$ and $P_l^{\text{odd}}$, we re-index, using the even-odd properties from Eq. (A.5). Upon collecting the coefficients of $P_l^{\text{even}}$ and $P_l^{\text{odd}}$ and multiplying by two, we obtain, for $k = 0,1,2,3, \ldots$,

$$P_k^{\text{even}} R + P_k^{\text{odd}} k\omega = \tag{2.18}$$
$$\frac{A_k^{\text{even}}}{2} P_0^{\text{even}} + \sum_{l=1}^{\infty} \left[ \left( \frac{A_{k+l}^{\text{even}}}{2} + \frac{A_{k-l}^{\text{even}}}{2} \right) P_l^{\text{even}} + \left( \frac{A_{k+l}^{\text{odd}}}{2} - \frac{A_{k-l}^{\text{odd}}}{2} \right) P_l^{\text{odd}} \right]$$

for the even (cosine) terms, and:

$$P_k^{\text{odd}} R - P_k^{\text{even}} k\omega = \tag{2.19}$$
$$\frac{A_k^{\text{odd}}}{2} P_0^{\text{even}} + \sum_{l=1}^{\infty} \left[ \left( \frac{A_{k-l}^{\text{odd}}}{2} + \frac{A_{k+l}^{\text{odd}}}{2} \right) P_l^{\text{even}} + \left( \frac{A_{k-l}^{\text{even}}}{2} - \frac{A_{k+l}^{\text{even}}}{2} \right) P_l^{\text{odd}} \right]$$

for the odd (sine) terms. For $k = 0$ the even-odd properties [see Eq. (A.5)] lead to:

$$P_0^{\text{even}} R = \frac{A_0^{\text{even}}}{2} P_0^{\text{even}} + \sum_{l=1}^{\infty} \left[ A_l^{\text{even}} P_l^{\text{even}} + A_l^{\text{odd}} P_l^{\text{odd}} \right] \tag{2.20}$$

for the cosine (even) terms. The sine (odd) terms give zero:

$$P_0^{\text{odd}} R = \tag{2.21}$$
$$\frac{A_0^{\text{odd}}}{2} P_0^{\text{even}} + \sum_{l=1}^{\infty} \left[ \left( \frac{A_{-l}^{\text{odd}}}{2} + \frac{A_l^{\text{odd}}}{2} \right) P_l^{\text{even}} + \left( \frac{A_{-l}^{\text{even}}}{2} - \frac{A_l^{\text{even}}}{2} \right) P_l^{\text{odd}} \right] = 0.$$

Eqs. (2.18)–(2.20) can be represented as a semi-infinite block-system of equations:

$$\widetilde{\mathbf{A}}\widetilde{\mathbf{P}} = \widetilde{\mathbf{P}} R, \tag{2.22}$$

where:

$$\widetilde{\mathbf{P}} = \begin{bmatrix} P_0^{\text{even}} \\ P_1^{\text{even}} \\ P_1^{\text{odd}} \\ P_2^{\text{even}} \\ P_2^{\text{odd}} \\ \vdots \end{bmatrix}, \tag{2.23}$$

$$\widetilde{\mathbf{A}} = \begin{bmatrix} \frac{A_0^{\text{even}}}{2} & \frac{A_{0+1}^{\text{even}}}{2} + \frac{A_{0-1}^{\text{even}}}{2} & \frac{A_{0+1}^{\text{odd}}}{2} - \frac{A_{0-1}^{\text{odd}}}{2} & \frac{A_{0+2}^{\text{even}}}{2} + \frac{A_{0-2}^{\text{even}}}{2} & \frac{A_{0+2}^{\text{odd}}}{2} - \frac{A_{0-2}^{\text{odd}}}{2} & \cdots \\ \frac{A_1^{\text{even}}}{2} & \frac{A_{1+1}^{\text{even}}}{2} + \frac{A_{1-1}^{\text{even}}}{2} & \frac{A_{1+1}^{\text{odd}}}{2} - \frac{A_{1-1}^{\text{odd}}}{2} - 1\omega I_n & \frac{A_{1+2}^{\text{even}}}{2} + \frac{A_{1-2}^{\text{even}}}{2} & \frac{A_{1+2}^{\text{odd}}}{2} - \frac{A_{1-2}^{\text{odd}}}{2} & \cdots \\ \frac{A_1^{\text{odd}}}{2} & \frac{A_{1-1}^{\text{odd}}}{2} + \frac{A_{1+1}^{\text{odd}}}{2} + 1\omega I_n & \frac{A_{1-1}^{\text{even}}}{2} - \frac{A_{1+1}^{\text{even}}}{2} & \frac{A_{1-2}^{\text{odd}}}{2} + \frac{A_{1+2}^{\text{odd}}}{2} & \frac{A_{1-2}^{\text{even}}}{2} - \frac{A_{1+2}^{\text{even}}}{2} & \cdots \\ \frac{A_2^{\text{even}}}{2} & \frac{A_{2+1}^{\text{even}}}{2} + \frac{A_{2-1}^{\text{even}}}{2} & \frac{A_{2+1}^{\text{odd}}}{2} - \frac{A_{2-1}^{\text{odd}}}{2} & \frac{A_{2+2}^{\text{even}}}{2} + \frac{A_{2-2}^{\text{even}}}{2} & \frac{A_{2+2}^{\text{odd}}}{2} - \frac{A_{2-2}^{\text{odd}}}{2} - 2\omega I_n & \cdots \\ \frac{A_2^{\text{odd}}}{2} & \frac{A_{2-1}^{\text{odd}}}{2} + \frac{A_{2+1}^{\text{odd}}}{2} & \frac{A_{2-1}^{\text{even}}}{2} - \frac{A_{2+1}^{\text{even}}}{2} & \frac{A_{2-2}^{\text{odd}}}{2} + \frac{A_{2+2}^{\text{odd}}}{2} + 2\omega I_n & \frac{A_{2-2}^{\text{even}}}{2} - \frac{A_{2+2}^{\text{even}}}{2} & \cdots \\ \vdots & \vdots & \vdots & \vdots & \vdots & \ddots [\widetilde{\mathbf{A}}]_{kl} \ddots \end{bmatrix}, \tag{2.24}$$

In Eq. (3.15),

$$[\widetilde{\mathbf{A}}]_{kl} = \begin{bmatrix} \frac{A_{k+l}^{\text{even}}}{2} + \frac{A_{k-l}^{\text{even}}}{2} & \frac{A_{k+l}^{\text{odd}}}{2} - \frac{A_{k-l}^{\text{odd}}}{2} - k\omega I_n \delta_{kl} \\ \frac{A_{k-l}^{\text{odd}}}{2} + \frac{A_{k+l}^{\text{odd}}}{2} + k\omega I_n \delta_{kl} & \frac{A_{k-l}^{\text{even}}}{2} - \frac{A_{k+l}^{\text{even}}}{2} \end{bmatrix} \tag{2.25}$$

is the block matrix in row $k$, column $l$ that multiplies the element $\begin{bmatrix} P_l^{\text{even}} \\ P_l^{\text{odd}} \end{bmatrix}$.

**The key idea in this work is to address cases where in the Fourier series expansion of each $A(t)$ and $P(t)$, there is a finite number of harmonics. Then to exploit this property to, in order to construct the matrices $\widetilde{\mathrm{A}}$ and $\widetilde{\mathrm{P}}$ in Eq. (2.22) to be with finite dimensions. Solving Eq. (2.22) with infinite will be extremely difficult.**

## 2.3 Summary

This chapter presents a Fourier analysis of an LPTV system in cosine-sine form and derives algebraic equations containing the matrix $R$ and the Fourier coefficients of $P(t)$ as unknown variables. Referring to APPENDIX B, we note a similarity between the structures of the algebraic equations in the real-imaginary decomposition and in the even-odd decomposition. In the next chapter we show that a Floquet factorization (i.e., the pair $P(t)$ and $R$) is not unique. Based on this non-uniqueness, we derive some Floquet factorizations via a matrix similarity of the constant part $R$, via a matrix similarity of the LPTV system matrix $A(t)$ and via shifting the trace average of $A(t)$ to zero.

# CHAPTER 3. Nonunique Decomposition of LPTV Transition Matrix

## 3.1 Overview

On the one hand, a unique solution to an LTV ODE that depends on the transition matrix exists for every initial condition (this also holds for any LPTV ODE). On the other hand, the decomposition of the transition matrix into the product of the periodic-function matrix $P(t)$ and the constant matrix $R$ is not unique. We show this non-uniqueness on section 3.2 via a similarity transformation of the constant part $R$. In addition, on section 3.3 we examine the Floquet factorization under a transformation of the LPTV system matrix $A(t)$ [Eqs. (1.11)-(1.13)]. Section 3.4 shows some additional properties of an LPTV system matrix when its trace average is shifted to zero, and presents an alternative version of *Floquet theory* based on shifting the trace average of an LPTV system matrix to zero. Finally, this chapter is summarized in Section 3.5.

## 3.2 Transformation of *R*

Suppose that a constant matrix $R$ is decomposed into another similar constant matrix $\tilde{R}$ (in a sense of Footnote [1]) so that:

$$R = V\tilde{R}V^{-1}, \tag{3.1}$$

where $V$ is some arbitrary non-singular transformation matrix (a change of basis matrix), then the transition matrix $\Phi_A(t, t_0)$ can be calculated from Eq. (1.15) as follows:

$$\begin{aligned} \Phi_A(t, t_0) &= P(t)e^{(t-t_0)V\tilde{R}V^{-1}}P^{-1}(t_0) \\ &= [P(t)V]e^{(t-t_0)\tilde{R}}[P(t_0)V]^{-1} \\ &= [\tilde{P}(t)]e^{(t-t_0)\tilde{R}}[\tilde{P}(t_0)]^{-1}. \end{aligned} \tag{3.2}$$

Therefore, the periodic function matrix $\tilde{P}(t) = P(t)V$ and the constant matrix $\tilde{R} = V^{-1}RV$ can also be used to construct the transition matrix $\Phi_A(t, t_0)$ for any non-singular constant matrix (i.e., the decomposition of the transition matrix is not unique).

## 3.3 Transformation of *A*(*t*)

Let $A(t)$ ,$\tilde{A}(t) \in \mathbb{R}^{n\times n}$ be LPTV system matrices with transition matrices

$$\begin{aligned} \Phi_A(t, t_0) &= P(t)e^{(t-t_0)R}P^{-1}(t_0), \\ \Phi_{\tilde{A}}(t, t_0) &= \tilde{P}(t)e^{(t-t_0)\tilde{R}}\tilde{P}^{-1}(t_0). \end{aligned} \tag{3.3}$$

Consider for example that our goal is to obtain $\Phi_A(t, t_0)$, i.e., to find its floquet factorization [$P(t)$ and $R$]. Assume also that there exists a nonsingular periodic matrix $U(t) \in \mathbb{R}^{n\times n}$ that transforms an LPTV matrix $A(t)$ to a new LPTV matrix $\tilde{A}(t)$ in a sense of Eq. (1.13) [see e.g., (Colaneri, 2005) Remark 1] such that

$$\tilde{A}(t) = U^{-1}(t)A(t)U(t) - U^{-1}(t)\dot{U}(t). \tag{3.4}$$

If we know how to obtain $\Phi_{\tilde{A}}(t, t_0)$, then, by Eq. (1.13) we can write the floquet factorization of $\Phi_A(t, t_0)$ [i.e., $P(t)$ and $R$] in terms of the floquet factorization of $\Phi_{\tilde{A}}(t, t_0)$ [i.e., $\tilde{P}(t)$ and $\tilde{R}$] and $U(t)$ as follows

$$\begin{gathered} P(t)e^{(t-t_0)R}P^{-1}(t_0) = U(t)\tilde{P}(t)e^{(t-t_0)\tilde{R}}\tilde{P}^{-1}(t_0)U^{-1}(t_0), \\ \Rightarrow P(t) = U(t)\tilde{P}(t), R = \tilde{R}. \end{gathered} \tag{3.5}$$

.

In other words, $\tilde{P}(t) = U^{-1}(t)P(t)$ and $\tilde{R} = R$ are the floquert factorization of the LPTV system matrix $\tilde{A}(t) = U^{-1}(t)A(t)U(t) - U^{-1}(t)\dot{U}(t)$. Equivalently, $P(t) = U(t)\tilde{P}(t)$ and $R = \tilde{R}$ are the floquert factorization of the LPTV system matrix $A(t) = U(t)\tilde{A}(t)U^{-1}(t) + \dot{U}(t)U^{-1}(t)$. Note that the crossing between the triplet $\{A(t), P(t), R\}$ and the triplet $\{\tilde{A}(t), \tilde{P}(t), \tilde{R}\}$ is symmetric [see footnote[2]], if we let

$$\widetilde{U}(t) = U^{-1}(t),$$

then by a matrix calaculus we can obtain

$$\dot{U}(t)U^{-1}(t) = -\widetilde{U}^{-1}(t)\dot{\widetilde{U}}(t)$$

and rewrite $A(t)$ in term of $\tilde{A}(t)$ and $\widetilde{U}(t)$, analogically to writing $\tilde{A}(t)$ in terms of $A(t)$ and $U(t)$ as per Eq. (3.4). Nevertheless, the constant part in both triplets remain the same i.e., $R = \tilde{R}$.

We want to consider two special cases of this result. If

$$U(t) = P(t),$$

then we have the basic result of *Lyapunov reducibility theorem* [see Eqs. (1.18)-(1.20)], so that

$$\tilde{P}(t) \equiv I_n \,,$$
$$\tilde{A}(t) \equiv \tilde{R} = R.$$

If

$$U(t) \equiv U$$

is some non-singular constatnt matrix, then we have that $A(t)$ and $\tilde{A}(t)$ are similar matrices [in a sense of Eq. (1.11)] so that,

$$\tilde{A}(t) = U^{-1}A(t)U. \tag{3.6}$$

In this work (specially in CHAPTER 5) we use this matrix similarity to transform a given LPTV system to a new one so that the new LPTV system will have a canonical form (e.g., a real Jordan block form) when $\omega = 0$, i.e.,

$$\tilde{A}(t|\omega = 0) = U^{-1}A(t|\omega = 0)U, \tag{3.7}$$

Where $U$ is the similarity transformation matrix, $A(t|\omega = 0)$ and $\tilde{A}(t|\omega = 0)$ are constant matrices (time dependency is nullified), so $\tilde{A}(t|\omega = 0)$ has some canonical form.

To summarize, we have a fair degree of freedom to achieve the matrices $P(t)$ and $R$, for example, by assuming that $R$ is of canonical form, or $P(0) = I_n$ or equals any nonsingular matrix $P(t)$ that can be defined up to, e.g., a scaling factor or a rotation. In addition, we may consider transforming the system matrix to simplify the calculation of the required transition matrix.

## 3.4 LPTV System Matrix with Trace Average Shifted to Zero

To analyze an LTV system (whenever it is periodic or not), we consider shifting the diagonal of the matrix of the LTV system by some scalar function (Lewkowicz, 1999). Suppose $\psi(t) \in \mathbb{R}$ is some known scalar function. By shifting the diagonal of the square matrix $A(t)$ to produce a new system matrix $A(t) - \psi(t)I_n$, then this new system matrix has the transition matrix:

$$\Phi_{A-\psi I_n}(t,t_0) = e^{-\int_{t_0}^{t} \psi(\tau)d\tau}\Phi_A(t,t_0). \tag{3.8}$$

This section explores properties stemming from the trace of the LPTV system matrix $A(t)$ that translate into properties of $P(t)$ and thereby help us to find $P(t)$. Taking the trace of Eq. (1.17), subtracting from it Eq. (1.30), and using the trace property $\text{trace}\{P(t)RP^{-1}(t)\} = \text{trace}\{R\}$ [refer to (Lewkowicz, 2020)], we obtain the following identity:

$$\text{trace}\{A(t) - A_o\} = \text{trace}\{\dot{P}(t)P^{-1}(t)\}. \tag{3.9}$$

From the trace identity in Eq. (1.30) and *Jacobi–Liouville formula* in Eq. (1.27) we conclude that

$$e^{\int_{t_0}^{t} \text{trace}\{A(\tau)-A_o\}d\tau} = \frac{\det\{P(t)\}}{\det\{P(t_0)\}}. \tag{3.10}$$

Therefore, we conclude that

$$\text{trace}\{A(t) - A_o\} \equiv 0 \Leftrightarrow \det\{P(t)\} \equiv \text{const. } \forall\, t. \tag{3.11}$$

If the $\text{trace}\{A(t) - A_o\}$ is identically zero, then the integral over the trace within each interval $[t_0, t]$ is zero, so exponentiating this integral gives unity, which implies that $\det\{P(t)\} = \det\{P(t_0)\}$ for all $t_0$ and $t$. Since $t_0$ and $t$ are arbitrary, we must have $\det\{P(t)\} = \det\{P(t_0)\} \equiv c$ for some constant $c$ that does not depend on either $t_0$ or $t$. If $\text{trace}\{A(t) - A_o\}$ is a function of $t$ that is not identically zero, then the term $\det\{P(t_0)\}e^{\int_{t_0}^{t} \text{trace}\{A(\tau)-A_o\}d\tau}$ is a non-constant function of $t$, so $\det\{P(t)\}$ is a non-constant function of $t$.

The above property motivates us to find a family of LPTV system matrices $A(t)$ where the periodic part $P(t)$ of the transition matrix has a constant determinant, which should give us significant information for computing $P(t)$. We will see that the value of this determinant is arbitrary because, among other things, $P(t)$ can be defined up to a scaling factor. In addition, a mathematical manipulation can always produce a constant determinant for $P(t)$.

In the following theorem, we show how to obtain a transition matrix by using the principles of this section, and Eq. (3.8), refer to (Lewkowicz, 2020).

**Theorem 3.1:** If $A(t) = A(t+T)$ is an LPTV system matrix in the sense of **Theorem 1.3** [Eq. (1.15)], then the transition matrix $\Phi_A(t,t_0)$ can be decomposed as follows:

$$\Phi_A(t,t_0) = e^{\frac{1}{n}\int_{t_0}^{t} \mathrm{trace}\{A(\tau)\}d\tau} P(t) e^{(t-t_0)R} P^{-1}(t_0), \tag{3.12}$$

where $P(t) = P(t+T)$ is a periodic matrix with $\det\{P(t)\} = \mathrm{const.}$, and $R$ is a constant matrix with $\mathrm{trace}\{R\} = 0$.

**Proof:** We use the notion of shifting the LTV system matrix diagonal [see, e.g., (Lewkowicz, 1999) and Eq. (3.8)] and denote $\psi(t) = \frac{1}{n}\mathrm{trace}\{A(t)\}$, $\tilde{A}(t) = A(t) - \psi(t)I_n$. We know from *Floquet theory* that, in general, a transition matrix of $\tilde{A}(t)$ has the form $\Phi_{\tilde{A}}(t,t_0) = P(t)e^{(t-t_0)R}P^{-1}(t_0)$. Given that $\mathrm{trace}\{\tilde{A}(t)\} = \mathrm{trace}\{A(t) - \psi(t)I_n\} \equiv 0$, the following holds:

1. from Eq. (1.29): $\mathrm{trace}\{\tilde{A}_o\} = \mathrm{trace}\left\{\frac{1}{T}\int_{t_0}^{t_0+T} \tilde{A}(t)dt\right\} = 0$;
2. from Eq. (1.30) and item 1: $\mathrm{trace}\{R\} = 0$;
3. from Eq. (3.11) and item 1:
   $\mathrm{trace}\{\tilde{A}(t) - \tilde{A}_o\} \equiv 0 \Leftrightarrow \det\{P(t)\} \equiv \mathrm{const.}$;
4. from Eq (3.8):
   $\Phi_A(t,t_0) = e^{\int_{t_0}^{t}\psi(\tau)d\tau}\Phi_{A-\psi I_n}(t,t_0) = e^{\int_{t_0}^{t}\frac{1}{n}\mathrm{trace}\{A(\tau)\}d\tau}\Phi_{\tilde{A}}(t,t_0)$.

Finally, Eq. (3.12) is obtained given the conditions $\det\{P(t)\} = \text{const.}$ and $\text{trace}\{R\} = 0$

■.

Consider decomposing $\frac{1}{n}\text{trace}\{A(t)\}$ as follows:

$$\frac{1}{n}\text{trace}\{A(t)\} = \psi(t) = \psi_0 + \psi_1(t), \tag{3.13}$$

where $\psi_0$ is the constant part of $\psi(t)$ [i.e., the average of $\psi(t)$] and $\psi_1(t)$ is the periodic part of $\psi(t)$ with $\psi_1(t) = \psi_1(t+T)$. The anti-derivative[8] of $\psi_1(t)$ is denoted $\Psi_1(t)$ such that

$$\int_{t_0}^{t} \psi_1(\tau)d\tau = \Psi_1(t) - \Psi_1(t_0). \tag{3.14}$$

We can write Eq. (3.14) as follows:

$$\Phi_A(t,t_0) = \left[e^{\Psi_1(t)}P(t)\right]e^{(t-t_0)[R+\psi_0 I]}\left[e^{\Psi_1(t_0)}P(t_0)\right]^{-1}, \tag{3.15}$$

so that:

$$\tilde{P}(t) = \left[e^{\Psi_1(t)}P(t)\right] \text{ and } \tilde{R} = R + \psi_0 I_n \tag{3.16}$$

are the matrices to construct the transition matrix $\Phi_A(t,t_0)$.

[8] Assuming that $\psi_1(0)$ is defined, the antiderivative of $\psi_1(t)$ may be defined by $\Psi_1(t) = \int_0^t \psi_1(\tau)d\tau$; the antiderivative can thus be defined up to a constant offset.

## 3.5 Summary

In this chapter we show that Floquet factorization is not unique by, e.g., choosing a matrix similar to $R$. We may consider undertaking a coordinated change by, e.g., choosing a matrix similar to $A(t)$ and shifting the trace of $A(t)$ to zero to obtain $P(t)$ and $R$ with attractive properties that facilitate the search for $P(t)$ and $R$ .

The next chapter presents some examples that contain a finite or infinite number of harmonics, and the frequency $\omega$ as a free parameter in LPTV systems is used to examine how $\omega$ affects the solution for an LPTV system (e.g., stability).

# CHAPTER 4. Examples of Analyses of LPTV Systems

## 4.1 Overview

A variety of examples of matrices $A(t)$ of LPTV systems and their corresponding matrices $P(t)$ and $R$ obtained using *Floquet theory* are now shown. In all examples, the frequency of the LPTV system matrix $A(t)$ is set to be some arbitrary constant $\omega$. Without loss of generality, it is assumed that $\omega > 0$ (otherwise the parity of sine and cosine could be used to represent an LPTV system with a positive frequency). Our examples are also limited to those whose periodic matrices $A(t)$ and $P(t)$ are constructed by applying elementary functions[9], such as trigonometric polynomials formed by the set of the terms $\sin(k\omega t)$ and $\cos(k\omega t)$ such that $k \in \mathbb{Z}$ .

This chapter is organized as follows: Section 4.2 presents several examples of LPTV system matrices $A(t)$ and their corresponding matrices $P(t)$ and $R$ categorized into four cases according to the number of harmonics (infinite or finite) in $A(t)$ and $P(t)$. The motivation for this approach is to investigate situations in which both $A(t)$ and $P(t)$ have a finite number of harmonics. Section 4.3 outlines the notion of a frequency $\omega$ as a free parameter in LPTV systems. The examples below focus on the simple case in which the Fourier coefficients of $A(t)$ and $R$ are linear functions of $\omega$, whereas the Fourier coefficients of $P(t)$ are constant. Finally, the chapter is summarized in Section 4.4.

[9] Elementary functions may refer, e.g., to sums, differences, products, quotients, powers, exponentiations, or logarithms of hyperbolic or trigonometric function. However, piecewise functions, such as rectangular or sawtooth functions are beyond the scope of this chapter.

## 4.2 Number of Harmonics in LPTV Systems and in the Periodic Parts of their Transition Matrices

### 4.2.1 General

Let $L$ be the number of harmonics of the $T$-periodic matrix $A(t)$ for an LPTV system and $p$ be the number of harmonics of the periodic part $P(t)$ of its transition matrix. The following cases are considered:

- **Case 1**: $L$ is infinite and $p$ is infinite;
- **Case 2**: $L$ is infinite and $p$ is finite;
- **Case 3**: $L$ is finite and $p$ is infinite;
- **Case 4:** $L$ is finite and $p$ is finite.

The number of the harmonics may be affected by assuming $P(t)$ to be a $T$-periodic or a $2T$-periodic or (see section 1.4.4 above). We search for a family of matrices of LPTV systems that correspond to Case 4 (finite $L$) and for a procedure that uses *Floquet theory* to calculate the periodic matrix $P(t)$ (with finite $p$) and constant matrix $R$ to construct the transition matrix of this family. Each case itself may have some subcases, such as generating an infinite trigonometric polynomial by some elementary operation (e.g., dividing a finite trigonometric polynomial by another finite trigonometric polynomial, or exponentiating a finite trigonometric polynomial), or by some non-elementary operation (e.g., performing a periodic continuation of a non-periodic function). From Eq. (1.17), the condition $\det\{P(t)\} \equiv \text{const.}$ [Eq. (3.11)], and APPENDIX A (specifically **Lemma A.1**, **Lemma A.2**, and **Lemma A.3**), we assert that, if $P(t) \in \mathbb{R}^{n\times n}$ has a finite number $p$ of harmonics, then $A(t)$ has at most $np$ harmonics because the sum $\dot{P}(t) + P(t)R$ has $p$ harmonics, $P^{-1}(t)$ has at most $(n-1)p$ harmonics, and their product can produce at most $np$ harmonics. In particular, if $n = 2$, then $A(t)$ has at most $2p$ harmonics.

Consider solving an LPTV system by using Eq. (1.16). Since the RHS of this equation [i.e., $\dot{P}(t) + P(t)R$] has $p$ harmonics, then the LHS [i.e., $A(t)P(t)$ ] must have exactly $p$ harmonics. Otherwise, the Fourier coefficients would not match for all $t$. Normally, the number of harmonics of $A(t) \in \mathbb{R}^{n\times n}$ is given, whereas that of $P(t)$ is unknown. We search for a family of LPTV system matrices $A(t)$ with a finite number $L$ of harmonics that satisfies the conditions above [i.e., $P(t) \in \mathbb{R}^{n\times n}$ is periodic with a finite number $p$ of harmonics to be determined]. We hypothesize that $p = \left\lceil \frac{L}{n} \right\rceil$ with $L \geq 2$. If $L = 1$, i.e., $A(t)$'s periodic part has the form $A_c \cos(\omega_o t) + A_s \sin(\omega_o t)$, then, we can define $\omega$ such that $\omega_o = 2\omega$, denote $A_2^{\text{even}} = A_c$ and $A_2^{\text{odd}} = A_s$ the 2nd harmonics terms, and zero-pad the 1st i.e., $A_1^{\text{even}} = 0$ and $A_1^{\text{odd}} = 0$.

### 4.2.2 Examples

We explore here some examples of cases 1–4.

***Table 4-1*** *Examples of 2×2 LPTV Systems. Case 1: L is infinite, p is infinite.*

| Row | $A(t)$ | $P(t)$ | $R$ |
|---|---|---|---|
| A. | $\begin{bmatrix} \cos(\omega t) & e^{\frac{2}{\omega}\sin(\omega t)} \\ 0 & -\cos(\omega t) \end{bmatrix}$ | $\begin{bmatrix} e^{\frac{1}{\omega}\sin(\omega t)} & 0 \\ 0 & e^{-\frac{1}{\omega}\sin(\omega t)} \end{bmatrix}$ | $\begin{bmatrix} 0 & 1 \\ 0 & 0 \end{bmatrix}$ |
| B. | $\begin{bmatrix} \frac{a}{2} - 1 + \cos(2\omega t) + \frac{\omega \sin(\omega t)}{2 + \cos(\omega t)} & 1 - \frac{a}{2}\sin(2\omega t) \\ -1 - \frac{a}{2}\sin(2\omega t) & \frac{a}{2} - 1 - \cos(2\omega t) + \frac{\omega \sin(\omega t)}{2 + \cos(\omega t)} \end{bmatrix}$ | $\frac{1}{2 + \cos(\omega t)} \begin{bmatrix} \cos(\omega t) & -\sin(\omega t) \\ -\sin(\omega t) & -\cos(\omega t) \end{bmatrix}$ | $\begin{bmatrix} a-1 & \omega - 1 \\ 1 - \omega & -1 \end{bmatrix}$ |
| C. | $\begin{bmatrix} \sin(\omega t) + \frac{\omega \sin(\omega t)}{2 + \cos(\omega t)} & 1 + \cos(\omega t) \\ 0 & \sin(\omega t) + \frac{\omega \sin(\omega t)}{2 + \cos(\omega t)} \end{bmatrix}$ | $\frac{e^{-\frac{1}{\omega}\cos(\omega t)}}{2 + \cos(\omega t)} \begin{bmatrix} 1 & \frac{1}{\omega}\sin(\omega t) \\ 0 & 1 \end{bmatrix}$ | $\begin{bmatrix} 0 & 1 \\ 0 & 0 \end{bmatrix}$ |

***Table 4-2*** *Examples of 2×2 LPTV Systems. Case 2: L is infinite, p is finite.*

| Row | $A(t)$ | $P(t)$ | $R$ |
|---|---|---|---|
| A. | $\begin{bmatrix} \frac{\omega\cos(\omega t)}{2+\sin(\omega t)} & \frac{\sin(\omega t)+\omega\cos(\omega t)+2}{1-\frac{1}{4}\sin^2(\omega t)}-1 \\ 0 & \frac{-\omega\cos(\omega t)}{2-\sin(\omega t)} \end{bmatrix}$ | $\begin{bmatrix} 1+\frac{1}{2}\sin(\omega t) & \sin(\omega t) \\ 0 & 1-\frac{1}{2}\sin(\omega t) \end{bmatrix}$ | $\begin{bmatrix} 0 & 1 \\ 0 & 0 \end{bmatrix}$ |
| B. | $\begin{bmatrix} 3+\frac{(2-\sin(\omega t))\omega\cos(\omega t)}{3+\cos^2(\omega t)} & 1+\frac{2\omega\cos(\omega t)}{3+\cos^2(\omega t)} \\ 0 & 1-\frac{(2-\sin(\omega t))\omega\cos(\omega t)}{3+\cos^2(\omega t)} \end{bmatrix}$ | $\begin{bmatrix} 2+\sin(\omega t) & \sin(\omega t) \\ 0 & 2-\sin(\omega t) \end{bmatrix}$ | $\begin{bmatrix} 3 & 1 \\ 0 & 1 \end{bmatrix}$ |
| C. | $\begin{bmatrix} \frac{a}{2}-1+\frac{\omega\cos(\omega t)}{2+\sin(\omega t)}+\cos(2\omega t) & 1-\sin(2\omega t) \\ -1-\sin(2\omega t) & \frac{a}{2}-1+\frac{\omega\cos(\omega t)}{2+\sin(\omega t)}-\cos(2\omega t) \end{bmatrix}$ | $\begin{bmatrix} 2\cos(\omega t)+\frac{1}{2}\sin(\omega t) & -\frac{1}{2}+\frac{1}{2}\cos(\omega t)-2\sin(\omega t) \\ -\frac{1}{2}+\frac{1}{2}\cos(\omega t)-2\sin(\omega t) & -2\cos(\omega t)+\frac{1}{2}\sin(\omega t) \end{bmatrix}$ | $\begin{bmatrix} a-1 & \omega-1 \\ 1-\omega & -1 \end{bmatrix}$ |

***Table 4-3*** *Examples of 2×2 LPTV Systems. Case 3: L is finite, p is infinite.*

| Row | $A(t)$ | $P(t)$ | $R$ |
|---|---|---|---|
| A. | $\begin{bmatrix} \sin(\omega t) & 0 \\ 0 & -\sin(\omega t) \end{bmatrix}$ | $\begin{bmatrix} e^{-\frac{1}{\omega}(\cos(\omega t)-1)} & 0 \\ 0 & e^{\frac{1}{\omega}(\cos(\omega t)-1)} \end{bmatrix}$ | $\begin{bmatrix} 0 & 0 \\ 0 & 0 \end{bmatrix}$ |
| B. | $\begin{bmatrix} 1+\omega\cos(\omega t) & 2+2\omega\cos(\omega t) \\ 0 & 3-\omega\cos(\omega t) \end{bmatrix}$ | $\begin{bmatrix} e^{\sin(\omega t)} & e^{-\sin(\omega t)} \\ 0 & e^{-\sin(\omega t)} \end{bmatrix}$ | $\begin{bmatrix} 1 & 0 \\ 0 & 3 \end{bmatrix}$ |
| C. | $\begin{bmatrix} \psi(t)+\frac{a}{2}\cos(2\omega t) & 1-\frac{a}{2}\sin(2\omega t) \\ -1-\frac{a}{2}\sin(2\omega t) & \psi(t)-\frac{a}{2}\cos(2\omega t) \end{bmatrix}$ <br> $\psi(t)=-1+\frac{a}{2}+\omega\cos(\omega t)-\omega\sin(\omega t)$ | $e^{\sin(\omega t)+\cos(\omega t)-1}\begin{bmatrix} \cos(\omega t) & -\sin(\omega t) \\ -\sin(\omega t) & -\cos(\omega t) \end{bmatrix}$ | $\begin{bmatrix} a-1 & \omega-1 \\ 1-\omega & -1 \end{bmatrix}$ |
| D. | $\begin{bmatrix} \sin(\omega t) & -1-\cos(\omega t) \\ 1+\cos(\omega t) & \sin(\omega t) \end{bmatrix}$ | $e^{-\frac{1}{\omega}(\cos(\omega t)-1)}\begin{bmatrix} \cos\left(\frac{1}{\omega}\sin(\omega t)\right) & -\sin\left(\frac{1}{\omega}\sin(\omega t)\right) \\ \sin\left(\frac{1}{\omega}\sin(\omega t)\right) & \cos\left(\frac{1}{\omega}\sin(\omega t)\right) \end{bmatrix}$ | $\begin{bmatrix} 0 & -1 \\ 1 & 0 \end{bmatrix}$ |
| E. | $\begin{bmatrix} \sin(\omega t) & 1+\cos(\omega t) \\ 1+\cos(\omega t) & \sin(\omega t) \end{bmatrix}$ | $e^{-\frac{1}{\omega}(\cos(\omega t)-1)}\begin{bmatrix} \cosh\left(\frac{1}{\omega}\sin(\omega t)\right) & \sinh\left(\frac{1}{\omega}\sin(\omega t)\right) \\ \sinh\left(\frac{1}{\omega}\sin(\omega t)\right) & \cosh\left(\frac{1}{\omega}\sin(\omega t)\right) \end{bmatrix}$ | $\begin{bmatrix} 0 & 1 \\ 1 & 0 \end{bmatrix}$ |
| F. | $\begin{bmatrix} \sin(\omega t) & 1+\cos(\omega t) \\ 0 & \sin(\omega t) \end{bmatrix}$ | $e^{-\frac{1}{\omega}(\cos(\omega t)-1)}\begin{bmatrix} 1 & \frac{1}{\omega}\sin(\omega t) \\ 0 & 1 \end{bmatrix}$ | $\begin{bmatrix} 0 & 1 \\ 0 & 0 \end{bmatrix}$ |

***Table 4-4*** *Examples of 2×2 LPTV Systems. Case 4: L is finite, p is finite.*

| Row | $A(t)$ | $P(t)$ | $R$ |
|---|---|---|---|
| A. | $\begin{bmatrix} -1+2\sin(2\omega t) & -2\cos(\omega t)+\sin(\omega t)+\sin(3\omega t) \\ -4\sin(\omega t) & 1-2\sin(2\omega t) \end{bmatrix}$ <br> $+\omega\begin{bmatrix} -1-\cos(2\omega t) & -\frac{3}{2}\cos(\omega t)-\frac{1}{2}\cos(3\omega t)+\sin(\omega t) \\ 2\cos(\omega t) & 1+\cos(2\omega t) \end{bmatrix}$ | $\begin{bmatrix} \cos(\omega t) & 1-\sin(2\omega t) \\ -1 & 2\sin(\omega t) \end{bmatrix}$ | $\begin{bmatrix} 1 & 0 \\ 0 & -1 \end{bmatrix}$ |
| B. | $\begin{bmatrix} a+\cos(\omega t) & \frac{1}{2}+\frac{1}{2}\cos(2\omega t) \\ -1 & a-\cos(\omega t) \end{bmatrix}+$ <br> $\omega\begin{bmatrix} -\frac{1}{2}-\frac{1}{2}\cos(2\omega t) & -\frac{3}{4}\cos(\omega t)-\frac{1}{4}\cos(3\omega t)+\sin(\omega t) \\ -\cos(\omega t) & \frac{1}{2}+\frac{1}{2}\cos(2\omega t) \end{bmatrix}$ | $\begin{bmatrix} \cos(\omega t) & 1-\frac{1}{2}\sin(2\omega t) \\ -1 & \sin(\omega t) \end{bmatrix}$ | $\begin{bmatrix} a & 1 \\ 0 & a \end{bmatrix}$ |
| C. | $\frac{a}{2}\begin{bmatrix} -5\cos(\omega t)-\cos(3\omega t) & 5+4\cos(2\omega t)+\cos(4\omega t) \\ -\cos(2\omega t)-3 & 5\cos(\omega t)+\cos(3\omega t) \end{bmatrix}+$ <br> $\omega\begin{bmatrix} \sin(2\omega t) & -3\sin(\omega t)-\sin(3\omega t) \\ \sin(\omega t) & -\sin(2\omega t) \end{bmatrix}$ | $\begin{bmatrix} 2\cos(\omega t) & \cos(2\omega t) \\ 1 & \cos(\omega t) \end{bmatrix}$ | $\begin{bmatrix} 0 & a \\ -a & 0 \end{bmatrix}$ |
| D. | $\begin{bmatrix} 1+\frac{2\omega}{3}\sin(3\omega t) & \frac{7\omega}{3}+\frac{2\omega}{3}\cos(3\omega t) \\ -\frac{7\omega}{3}+\frac{2\omega}{3}\cos(3\omega t) & 1-\frac{2\omega}{3}\sin(3\omega t) \end{bmatrix}$ | $\begin{bmatrix} 2\cos(2\omega t)+\cos(\omega t) & 2\sin(2\omega t)-\sin(\omega t) \\ -2\sin(2\omega t)-\sin(\omega t) & 2\cos(2\omega t)-\cos(\omega t) \end{bmatrix}$ | $\begin{bmatrix} 1 & 0 \\ 0 & 1 \end{bmatrix}$ |
| E. | $\frac{1}{3}\begin{bmatrix} F(t) & -5-4\sin(\omega t)+G(t) \\ 5+4\sin(\omega t)+G(t) & -F(t) \end{bmatrix}+\omega\begin{bmatrix} 2\sin(3\omega t) & 4+2\cos(3\omega t) \\ -4+2\cos(3\omega t) & -2\sin(3\omega t) \end{bmatrix}$ <br> $F(t)=-\cos(2\omega t)+4\cos(4\omega t)-4\sin(3\omega t)$ <br> $G(t)=-4\cos(3\omega t)+\sin(2\omega t)-4\sin(4\omega t)$ | $\begin{bmatrix} 2\cos(2\omega t)+\cos(\omega t) & 2\sin(2\omega t)-\sin(\omega t) \\ -2\sin(2\omega t)-\sin(\omega t) & 2\cos(2\omega t)-\cos(\omega t) \end{bmatrix}$ | $\begin{bmatrix} 1 & \omega-1 \\ 1-\omega & -1 \end{bmatrix}$ |
| F. | $\begin{bmatrix} \frac{5a}{4}\cos(\omega t)-\frac{a}{4}\cos(3\omega t)-a\sin(\omega t) & \frac{13a}{8}+\frac{a}{2}\cos(2\omega t)-\frac{a}{8}\cos(4\omega t)-a\sin(2\omega t) \\ -\frac{3a}{2}-\frac{a}{2}\cos(2\omega t) & -\frac{5a}{4}\cos(\omega t)+\frac{a}{4}\cos(3\omega t)+a\sin(\omega t) \end{bmatrix}$ <br> $+\omega\begin{bmatrix} -\frac{1}{2}-\frac{1}{2}\cos(2\omega t) & -\frac{3}{4}\cos(\omega t)-\frac{1}{4}\cos(3\omega t)+\sin(\omega t) \\ \cos(\omega t) & \frac{1}{2}+\frac{1}{2}\cos(2\omega t) \end{bmatrix}$ | $\begin{bmatrix} \cos(\omega t) & 1-\frac{1}{2}\sin(2\omega t) \\ -1 & \sin(\omega t) \end{bmatrix}$ | $\begin{bmatrix} 0 & a \\ -a & 0 \end{bmatrix}$ |
| G. | $\begin{bmatrix} F(t) & \frac{3a}{2}-\frac{a}{2}\cos(2\omega t) \\ -\frac{25a}{8}+\frac{5a}{2}\cos(\omega t)-\frac{a}{2}\cos(3\omega t)+\frac{a}{8}\cos(4\omega t)-2a\sin(\omega t)+a\sin(2\omega t) & -F(t) \end{bmatrix}$ <br> $+\omega\begin{bmatrix} \frac{1}{2}-\cos(\omega t)+\frac{1}{2}\cos(2\omega t) & -\cos(\omega t) \\ -1+\frac{7}{4}\cos(\omega t)-2\cos(2\omega t)+\frac{1}{4}\cos(3\omega t)-\sin(\omega t) & -\frac{1}{2}+\cos(\omega t)-\frac{1}{2}\cos(2\omega t) \end{bmatrix}$ <br> $F(t)=\frac{3a}{2}-\frac{5a}{4}\cos(\omega t)-\frac{a}{2}\cos(2\omega t)+\frac{a}{4}\cos(3\omega t)+a\sin(\omega t)$ | $\begin{bmatrix} 1 & -\sin(\omega t) \\ -1+\cos(\omega t) & 1+\sin(\omega t)-\frac{1}{2}\sin(2\omega t) \end{bmatrix}$ | $\begin{bmatrix} 0 & a \\ -a & 0 \end{bmatrix}$ |
| H. | $aI_2+b\begin{bmatrix} 0 & 1 \\ -1 & 0 \end{bmatrix}+c\begin{bmatrix} \sin(2\omega t) & \cos(2\omega t) \\ \cos(2\omega t) & -\sin(2\omega t) \end{bmatrix}+d\begin{bmatrix} -\cos(2\omega t) & \sin(2\omega t) \\ \sin(2\omega t) & \cos(2\omega t) \end{bmatrix}$ <br> Generalized cases: $a,b,c,d \in \{C_0+\omega C_1 \mid C_0, C_1 \in \mathbb{R}\}$ | $\begin{bmatrix} \cos(\omega t) & \sin(\omega t) \\ -\sin(\omega t) & \cos(\omega t) \end{bmatrix}$ | $\begin{bmatrix} a-d & c+b-\omega \\ c-b+\omega & a+d \end{bmatrix}$ |

In the above examples, let $\omega$, $a$, $b$, $c$, and $d$ be real parameters that define the matrices $A(t)$ of LPTV systems and their corresponding matrices $P(t)$ and $R$. We search for a family of $n \times n$ real LPTV systems similar to those of **Table 4-4**, so that this family can be used with every real frequency $\omega$.

## 4.3 Notions of Frequency ω as a Free Parameter in LPTV Systems

### 4.3.1 General

We suggest using the frequency $\omega$ of an LPTV system as a free parameter (instead of plugging in a fixed value). This approach will allow for an examination of a family of solutions for LPTV systems and explore the system properties based on a variety of values of $\omega$. Such properties include stability criteria, low-frequency behavior (approaching an LTI system), and high-frequency behavior.

We limit ourselves to the case in which each Fourier coefficient of $A(t)$ is an ordinary polynomial in $\omega$, so that $R$ is a polynomial in $\omega$, but the Fourier coefficients of $P(t)$ [and, thus, also of $\det\{P(t)\}$] are not polynomials in $\omega$. This is done to simplify the problem and to assure the continuity for all $\omega$ of $A(t)$, $P(t)$, $R$, and $\Phi(t, t_0)$. For this demonstration, the polynomial in $\omega$ in the examples is linear affine in $\omega$.

### 4.3.2 Examples

**<u>Example 4.1</u>: Row H of Table 4-4**

Consider the following real LPTV system matrix [see (Wu M.-Y. , 1975) Example 2]:

$$A(t) = aI_2 + b\begin{bmatrix} 0 & 1 \\ -1 & 0 \end{bmatrix} + c\begin{bmatrix} \sin(2\omega t) & \cos(2\omega t) \\ \cos(2\omega t) & -\sin(2\omega t) \end{bmatrix} + d\begin{bmatrix} -\cos(2\omega t) & \sin(2\omega t) \\ \sin(2\omega t) & \cos(2\omega t) \end{bmatrix},$$
$$a = a_0 + \omega a_1;\ b = b_0 + \omega b_1; c = c_0 + \omega c_1; d = d_0 + \omega d_1. \quad (4.1)$$

The transition matrix of this LPTV system is constructed by the following pair of matrices $P(t)$ and $R$:

$$P(t) = \begin{bmatrix} \cos(\omega t) & \sin(\omega t) \\ -\sin(\omega t) & \cos(\omega t) \end{bmatrix}, \quad (4.2)$$

$$R = \begin{bmatrix} a-d & c+b-\omega \\ c-b+\omega & a+d \end{bmatrix}$$
$$= \begin{bmatrix} a_0 - d_0 + (a_1 - d_1)\omega & c_0 + b_0 + (c_1 + b_1 - 1)\omega \\ c_0 - b_0 + (c_1 - b_1 + 1)\omega & a_0 + d_0 + (a_1 + d_1)\omega \end{bmatrix}. \quad (4.3)$$

From Eq. (4.3), it follows that stability imposes the following two conditions:

i. $\mathrm{trace}\{R\} = 2a < 0$;
ii. $\det\{R\} = a^2 - d^2 + (b-\omega)^2 - c^2 > 0$.

Note that $P(t|\omega = 0) = P(0) = I_2$, so the transition matrix $\Phi_A(t, t_0)$ at $t_0 = 0$ is $\Phi_A(t, 0) = P(t)e^{tR}$, thus

$$\Phi_A(t,0) = \begin{bmatrix} \cos(\omega t) & \sin(\omega t) \\ -\sin(\omega t) & \cos(\omega t) \end{bmatrix} e^{\begin{bmatrix} a_0 - d_0 + (a_1 - d_1)\omega & c_0 + b_0 + (c_1 + b_1 - 1)\omega \\ c_0 - b_0 + (c_1 - b_1 + 1)\omega & a_0 + d_0 + (a_1 + d_1)\omega \end{bmatrix} t}. \quad (4.4)$$

This LPTV system is stable if and only if the real parts of the eigenvalues of $R$ , which depend on $\omega$, $a$, $b$, $c$, and $d$, are negative. Expressing $R$ in a closed form in $\omega$ and solving an LPTV system for some general frequency $\omega$ is equivalent to the finding a diagonal version of $R$ (which might be complex) for a specified frequency. Consider the same LPTV system, but at low frequency ($\omega = 0$). On the one hand,

$$A(t|\omega = 0) = \begin{bmatrix} a_0 - d_0 & c_0 + b_0 \\ c_0 - b_0 & a_0 + d_0 \end{bmatrix} = R|_{\omega=0}, \tag{4.5}$$

which is an LTI system with transition matrix:

$$\Phi_{A(\cdot|\omega=0)}(t,0) = e^{A(t|\omega=0)t} = e^{R|_{\omega=0}t} = e^{\begin{bmatrix} a_0-d_0 & c_0+b_0 \\ c_0-b_0 & a_0+d_0 \end{bmatrix}t}. \tag{4.6}$$

On the other hand, if we evaluate the transition matrix $\Phi_A(t,0) = P(t)e^{tR}$ of the original LPTV system at low frequency (i.e., $\omega$ approaches zero), we obtain:

$$\begin{aligned} \Phi_A(t,0|\omega = 0) &= P(t|\omega = 0)e^{tR|_{\omega=0}} \\ &= e^{\begin{bmatrix} a_0-d_0 & c_0+b_0 \\ c_0-b_0 & a_0+d_0 \end{bmatrix}t} = \Phi_A(t,0|\omega = 0). \end{aligned} \tag{4.7}$$

This LPTV system matrix can be generated according to Eq. (6.2.6) of (Colonius & Kliemann, 2014) with the following setup:

$$\begin{aligned} A(t) &= e^{-\omega tG}Be^{\omega tG}, \\ B &= \begin{bmatrix} a-d & c+b \\ c-b & a+d \end{bmatrix}, \\ G &= \begin{bmatrix} 0 & -1 \\ 1 & 0 \end{bmatrix}. \end{aligned} \tag{4.8}$$

$P(t)$ and $R$ can now be written in terms of $B$ and $G$ as follows:

$$P(t) = e^{-\omega t G},$$
$$R = B + \omega G. \tag{4.9}$$

Moreover, this example of an LPTV system matrix is a generalization of the examples of (Aggarwal & Infante, 1968), (Markus & Yamabe, 1960), and (Rosenbrook, 1963), which are cited in a number of articles and books, as previously stated in Section 1.4.1.

In the example of (Rosenbrook, 1963)

$$A(t) = \begin{bmatrix} -1 - 9\cos^2(6t) + 12\sin(6t)\cos(6t) & 12\cos^2(6t) + 9\sin(6t)\cos(6t) \\ -12\sin^2(6t) + 9\sin(6t)\cos(6t) & -1 - 9\sin^2(6t) - 12\sin(6t)\cos(6t) \end{bmatrix}$$
$$= \begin{bmatrix} -5.5 & 6 \\ -6 & -5.5 \end{bmatrix} + (6)\begin{bmatrix} \sin(12t) & \cos(12t) \\ \cos(12t) & -\sin(12t) \end{bmatrix} + (4.5)\begin{bmatrix} -\cos(12t) & \sin(12t) \\ \sin(12t) & \cos(12t) \end{bmatrix}.$$

we have $\omega = 6$, $a = -5.5$, $b = 6$, $c = 6$, and $d = 4.5$, thus, the flowing Floquet factorization is obtained

$$P(t) = \begin{bmatrix} \cos(6t) & \sin(6t) \\ -\sin(t) & \cos(6t) \end{bmatrix},$$

$$R = \begin{bmatrix} -5.5 - 4.5 & 6 + 6 - 6 \\ 6 - 6 + 6 & -5.5 + 4.5 \end{bmatrix} = \begin{bmatrix} -10 & 6 \\ 6 & -1 \end{bmatrix}.$$

In the example of (Aggarwal & Infante, 1968) (with some notation change to avoid confusion regarding to the parameter $a$):

$$A(t) = \begin{bmatrix} -1 + \beta\cos^2(t) & 1 - \beta\sin(t)\cos(t) \\ -1 - \beta\sin(t)\cos(t) & -1 + \beta\sin^2(t) \end{bmatrix}$$
$$= \left(-1 + \frac{\beta}{2}\right) I_2 + (1)\begin{bmatrix} 0 & 1 \\ -1 & 0 \end{bmatrix} + (0)\begin{bmatrix} \sin(2t) & \cos(2t) \\ \cos(2t) & -\sin(2t) \end{bmatrix} + \left(-\frac{\beta}{2}\right)\begin{bmatrix} -\cos(2t) & \sin(2t) \\ \sin(2t) & \cos(2t) \end{bmatrix}$$

we have $\omega = 1$, $a = -1 + \frac{\beta}{2}$, $b = 1$, $c = 0$, and $d = -\frac{\beta}{2}$, thus, the following Floquet factorization is obtained:

$$P(t) = \begin{bmatrix} \cos(t) & \sin(t) \\ -\sin(t) & \cos(t) \end{bmatrix},$$

$$R = \begin{bmatrix} -1 + \frac{\beta}{2} - \left(-\frac{\beta}{2}\right) & 0 + 1 - 1 \\ 0 + 1 - 1 & -1 + \frac{\beta}{2} + \left(-\frac{\beta}{2}\right) \end{bmatrix} = \begin{bmatrix} \beta - 1 & 0 \\ 0 & -1 \end{bmatrix},$$

Notice that if we define a new floquet factorization pair, such that we set $\omega = 1$ at $R$ but we remain $\omega$ to be a free parameter in $P(t)$, i.e.,

$$\hat{P}(t) = \begin{bmatrix} \cos(\omega t) & \sin(\omega t) \\ -\sin(\omega t) & \cos(\omega t) \end{bmatrix},$$

$$\hat{R} = \begin{bmatrix} \beta - 1 & 0 \\ 0 & -1 \end{bmatrix},$$

then, we have the following LPTV system matrix by applying Eq. (1.17)

$$\hat{A}(t) = \begin{bmatrix} -1 + \beta\cos^2(\omega t) & \omega - \beta\sin(\omega t)\cos(\omega t) \\ -\omega - \beta\sin(\omega t)\cos(\omega t) & -1 + \beta\sin^2(\omega t) \end{bmatrix}$$

Which is not identically to the LPTV system matrix $A(t)$ in. (Aggarwal & Infante, 1968) for any arbitrary $\omega$ [i.e., $A(t) \equiv \hat{A}(t)$ iif $\omega = 1$]. Notice that the example in (Markus & Yamabe, 1960) shown for $\beta = 1.5$.

On the next example we keep investigating the example (Aggarwal & Infante, 1968), but with a fixed constant $\beta$ [will be re-denoted as $a$ and assigned with a fixed value 1.5 such as in (Markus & Yamabe, 1960)] and considering the structure of the matrix expoential $e^{tR}$ corresponding to different values of the frequency $\omega$.

□.

**<u>Example 4.2</u>: *(Aggarwal & Infante, 1968), (Markus & Yamabe, 1960)***

Consider the following real LPTV system matrix:

$$A(t) = \begin{bmatrix} -1+\frac{a}{2}+\frac{a}{2}\cos(2\omega t) & 1-\frac{a}{2}\sin(2\omega t) \\ -1-\frac{a}{2}\sin(2\omega t) & -1+\frac{a}{2}-\frac{a}{2}\cos(2\omega t) \end{bmatrix} = \begin{bmatrix} -1+a\cos^2(\omega t) & 1-a\sin(\omega t)\cos(\omega t) \\ -1-a\sin(\omega t)\cos(\omega t) & -1+a\sin^2(\omega t) \end{bmatrix}, \quad (4.10)$$

with the following matrices $P(t)$ and $R$ as its solution:

$$P(t) = \begin{bmatrix} \cos(\omega t) & \sin(\omega t) \\ -\sin(\omega t) & \cos(\omega t) \end{bmatrix}, \quad (4.11)$$

$$R = \begin{bmatrix} a-1 & 1-\omega \\ \omega-1 & -1 \end{bmatrix}. \quad (4.12)$$

From Eq. (4.12), it follows that stability imposes the following two conditions:

i. $\text{trace}\{R\} = a-2 < 0$;
ii. $\det\{R\} = 1-a+(1-\omega)^2 > 0$.

Note that this example is a particular case of **<u>Example 4.1</u>** (row H of **Table 4-4**) with some parameters adjusted. Moreover, this example is a general case of (Aggarwal & Infante, 1968) with $\omega = 1$ and (Markus & Yamabe, 1960) with $\omega = 1$ and $a = 1.5$). The eigenvalues of $R$ depend on $\omega$ and $a$ and are given by

$$\lambda_{1,2} = \frac{a}{2} - 1 \pm \frac{1}{2}\sqrt{-(2\omega-a-2)(2\omega+a-2)}. \quad (4.13)$$

Previous works discuss the stability of this LPTV system based on a fixed value of $\omega$ (=1) with $a$ varying. For $\omega = 1$, we have a diagonal matrix:

$$R = \begin{bmatrix} a-1 & 0 \\ 0 & -1 \end{bmatrix},$$

which is a trivial case of matrix exponentiation, i.e.,

$$e^{tR} = \begin{bmatrix} e^{(a-1)t} & 0 \\ 0 & e^{-t} \end{bmatrix}.$$

Instead of varying $a$ and fixing $\omega$, we examine the opposite situation by fixing $a = 1.5$ and analyzing the stability of the LPTV system upon varying $\omega$. In this case, the eigenvalues of $R$, which depend on $\omega$, are:

$$\lambda_1 = -0.25 + 0.5\sqrt{\Delta}, \\ \lambda_2 = -0.25 - 0.5\sqrt{\Delta}, \tag{4.14}$$

Where:

$$\Delta = -(2\omega - 0.5)(2\omega - 3.5). \tag{4.15}$$

**Figure 4-1** and **Table 4-5** summarize the characteristics of $\lambda_{1,2}$ and $e^{tR}$ as functions of $\omega$ for frequencies determined by:

- the roots of $\Delta(\omega) = 0$ ($\omega \in \{0.25, 1.75\}$), which gives conditions on $\omega$ for complex eigenvalues (i.e., $\Delta < 0$);
- the roots of $\lambda_1(\omega) = 0$ [$4\Delta(\omega) = 1 \Longrightarrow \omega \in \left\{1 \pm \frac{\sqrt{2}}{2}\right\} \approx \{0.293, 1.707\}$], which gives conditions on $\omega$ for a stable $\lambda_1$ (i.e., $\lambda_1 < 0$ for $\Delta \geq 0$).

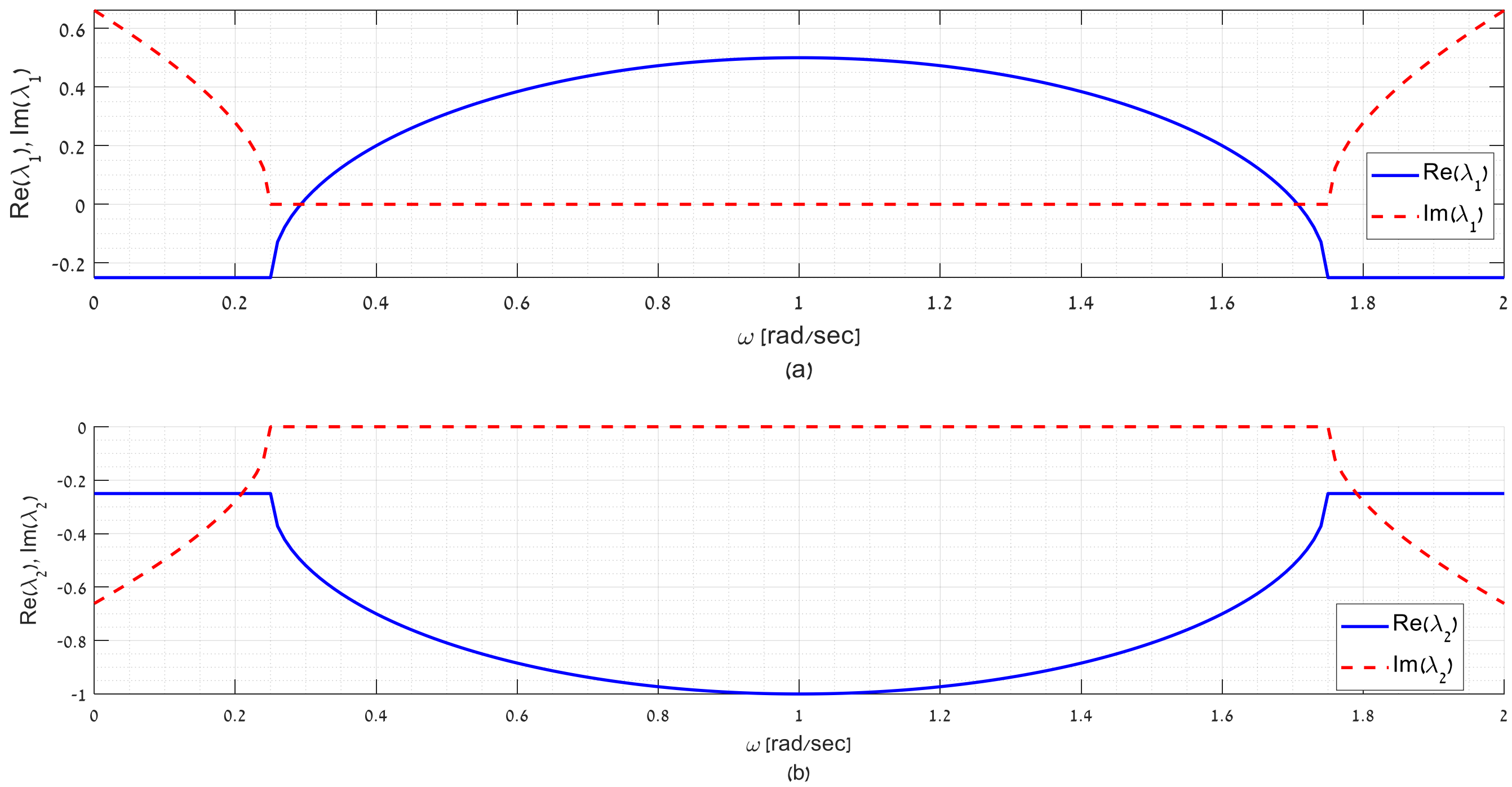

***Figure 4-1*** *Eigenvalue of R as functions of frequency (Markus & Yamabe, 1960). (a) $\lambda_1$; (b) $\lambda_2$.*

***Table 4-5*** *Stability status of $e^{tR}$ for different values of ω (Markus & Yamabe, 1960).*

| $\omega$ | $e^{tR}$ | Stability Status | Frequency Set |
|---|---|---|---|
| 0 | $\frac{e^{-\frac{1}{4}t}}{7}\begin{bmatrix} 7\cos\left(\frac{\sqrt{7}}{4}t\right)+3\sqrt{7}\sin\left(\frac{\sqrt{7}}{4}t\right) & 4\sqrt{7}\sin\left(\frac{\sqrt{7}}{4}t\right) \\ -4\sqrt{7}\sin\left(\frac{\sqrt{7}}{4}t\right) & 7\cos\left(\frac{\sqrt{7}}{4}t\right)-3\sqrt{7}\sin\left(\frac{\sqrt{7}}{4}t\right) \end{bmatrix}$ | Stable | $(-\infty, 0.25) \cup (1.75, \infty)$ |
| 0.25 | $\frac{e^{-\frac{1}{4}t}}{4}\begin{bmatrix} 4+3t & 3t \\ -3t & 4-3t \end{bmatrix}$ | Stable | $\{0.25, 1.75\}$ |
| 0.28 | $\frac{1}{7}\begin{bmatrix} 16e^{-\frac{1}{25}t}-9e^{-\frac{23}{50}t} & 12e^{-\frac{1}{25}t}-12e^{-\frac{23}{50}t} \\ 12e^{-\frac{23}{50}t}-12e^{-\frac{1}{25}t} & 16e^{-\frac{23}{50}t}-9e^{-\frac{1}{25}t} \end{bmatrix}$ | Stable | $\left(0.25, 1-\frac{\sqrt{2}}{2}\right) \cup \left(1+\frac{\sqrt{2}}{2}, 1.75\right)$ |
| $1-\frac{\sqrt{2}}{2}$ | $\begin{bmatrix} -e^{-\frac{1}{2}t}+2 & -\sqrt{2}e^{-\frac{1}{2}t}+\sqrt{2} \\ \sqrt{2}e^{-\frac{1}{2}t}-\sqrt{2} & 2e^{-\frac{1}{2}t}-1 \end{bmatrix}$ | Marginally Stable | $\left\{1 \pm \frac{\sqrt{2}}{2}\right\}$ |
| 1 | $\begin{bmatrix} e^{\frac{1}{2}t} & 0 \\ 0 & e^{-t} \end{bmatrix}$ | Unstable | $\left(1-\frac{\sqrt{2}}{2}, 1+\frac{\sqrt{2}}{2}\right)$ |
| $1+\frac{\sqrt{2}}{2}$ | $\begin{bmatrix} -e^{-0.5t}+2 & \sqrt{2}e^{-0.5t}-\sqrt{2} \\ -\sqrt{2}e^{-0.5t}+\sqrt{2} & 2e^{-0.5t}-1 \end{bmatrix}$ | Marginally Stable | $\left\{1 \pm \frac{\sqrt{2}}{2}\right\}$ |
| 1.72 | $\frac{1}{7}\begin{bmatrix} 16e^{-\frac{1}{25}t}-9e^{-\frac{23}{50}t} & 12e^{-\frac{23}{50}t}-12e^{-\frac{1}{25}t} \\ 12e^{-\frac{1}{25}t}-12e^{-\frac{23}{50}t} & 16e^{-\frac{23}{50}t}-9e^{-\frac{1}{25}t} \end{bmatrix}$ | Stable | $\left(0.25, 1-\frac{\sqrt{2}}{2}\right) \cup \left(1+\frac{\sqrt{2}}{2}, 1.75\right)$ |
| 1.75 | $\frac{e^{-0.25t}}{4}\begin{bmatrix} 4+3t & 3t \\ -3t & 4-3t \end{bmatrix}$ | Stable | $\{0.25, 1.75\}$ |
| 2 | $\frac{e^{-\frac{1}{4}t}}{7}\begin{bmatrix} 7\cos\left(\frac{\sqrt{7}}{4}t\right)+3\sqrt{7}\sin\left(\frac{\sqrt{7}}{4}t\right) & -4\sqrt{7}\sin\left(\frac{\sqrt{7}}{4}t\right) \\ 4\sqrt{7}\sin\left(\frac{\sqrt{7}}{4}t\right) & 7\cos\left(\frac{\sqrt{7}}{4}t\right)-3\sqrt{7}\sin\left(\frac{\sqrt{7}}{4}t\right) \end{bmatrix}$ | Stable | $(-\infty, 0.25) \cup (1.75, \infty)$ |

The eigenvalue frequencies are symmetric about some frequencies ($\omega = 1$ in this example), so that $e^{tR}$ is reflected about $\omega = 1$ (i.e., $e^{tR}|_{\omega=1+\omega_0} = \left[e^{tR}|_{\omega=1-\omega_0}\right]^T \ \forall \ \omega_0$). The point $\omega = 1$ is thus a critical point in the sense that the real part of $\lambda_1$ is maximal (=0.5) and the real part of $\lambda_2$ is minimal (=−1). Given the complex conjugate pair property (if $a + ib$ is an eigenvalue of a real matrix, then also $a - ib$), the imaginary part varies when the real part is constant (−0.25 in this example), and the imaginary part is zero when the real part varies

□.

## 4.4 Summary

This chapter outlines examples of LPTV systems $A(t)$ and their solutions obtained by $P(t)$ and $R$. We distinguish between four cases of infinite and finite harmonics in $A(t)$ and $P(t)$. We focus on the last case [finite harmonics in both $A(t)$ and $P(t)$] and analyze how the frequency $\omega$ affects the structure of $P(t)$ and $R$. We also addressed the case where both $A(t)$ and $P(t)$ have a finite number of harmonics and observed that the Fourier coefficients of $A(t)$ are linear functions of $\omega$, those of $P(t)$ are constant (independent of $\omega$), and those of $R$ are linear functions of $\omega$. This fact motivates the next chapter, in which we compare powers of $\omega$. This means that we treat the frequency $\omega$ as a free parameter in LPTV systems, and not as a fixed number.

# <u>CHAPTER 5.</u> LPTV Systems with Coefficients as Polynomials in ω

## 5.1 Overview

This chapter introduces a family of periodic matrices for $A(t)$ with a finite number of harmonics, so that each Fourier coefficient of $A(t)$ is a polynomial in $\omega$. As per *Floquet theory*, we search for some conditions so that the corresponding matrices $P(t)$ $and$ $R$ produce a transition matrix $A(t)$ such that the number of harmonics of $P(t)$ is also finite.

**<u>Remark</u>**: To simplify the problem, we assume, without loss of generality, that $\text{trace}\{A(t)\} \equiv 0$. If not, then we can shift the trace of the system matrix to zero and use the trace version of *Floquet theory* **<u>(Theorem 3.1</u>**) so that $\det\{P(t)\} =$ const., $\text{trace}\{R\} = 0$, and recast $A(t)$ as $\tilde{A}(t) = A(t) - \frac{1}{n}\text{trace}\{A(t)\}I_n$ so that $\text{trace}\{\tilde{A}(t)\} \equiv 0$.

This chapter is organized as follows: Section 5.2 describes how to express Fourier coefficients as polynomials in $\omega$ for a cosine-sine Fourier series of an LPTV system matrix $A(t)$ and the periodic part $P(t)$ of its solution. Section 5.3 relates an LPTV system with Fourier coefficients expressed as polynomials in $\omega$ and its LTI version for the LPTV system evaluated at $\omega = 0$. Section 5.4 transforms an LPTV system so that, at $\omega = 0$, we obtain a canonical form of an LTI system (e.g., Jordan block diagonal). Section 5.5 compares powers of $\omega$ to solve an LPTV system. Finally, Section 5.6 summarizes the chapter.

## 5.2 Cosine and Sine Fourier Coefficients as Polynomials in ω

Suppose that an LPTV system matrix $A(t) = A\left(t + \frac{2\pi}{\omega}\right) \in \mathbb{R}^{n\times n}$ is defined as follows:

$$A(t) = \sum_{r=0}^{N} \omega^r A^{\{r\}}(t), \tag{5.1}$$

$$A^{\{r\}}(t) = \sum_{l=-L}^{L} \left( \frac{A^{\{r\}}{}_l^{\text{even}}}{2} \cos(l\omega t) + \frac{A^{\{r\}}{}_l^{\text{odd}}}{2} \sin(l\omega t) \right), \tag{5.2}$$

where $n$ is the size of the matrix $A(t)$, $N$ is the degree of the polynomial in $\omega^r$, $r = 0,1,\dots,N$, $A^{\{r\}}(t)$ is the $r$th Fourier series such that $A^{\{r\}}{}_l^{\text{even}}$ and $A^{\{r\}}{}_l^{\text{odd}}$ are Fourier coefficients that do not depend on $\omega$, and $L$ is the number of the harmonics of $A(t)$. Inserting Eq. (5.2) into Eq. (5.1) and changing the order of summation gives:

$$A(t) = \sum_{l=-L}^{L} \sum_{r=0}^{N} \left( \omega^r \frac{A^{\{r\}}{}_l^{\text{even}}}{2} \right) \cos(l\omega t) + \sum_{l=-L}^{L} \sum_{r=0}^{N} \left( \omega^r \frac{A^{\{r\}}{}_l^{\text{odd}}}{2} \right) \sin(l\omega t). \tag{5.3}$$

For convenience, we denote the Fourier coefficients of $A(t)$ as follows:

$$\frac{A_l^{\text{even}}}{2} = \sum_{r=0}^{N} \left( \omega^r \frac{A^{\{r\}}{}_l^{\text{even}}}{2} \right), \frac{A_l^{\text{odd}}}{2} = \sum_{r=0}^{N} \left( \omega^r \frac{A^{\{r\}}{}_l^{\text{odd}}}{2} \right). \tag{5.4}$$

Inserting in the new definition above for $A(t)$ into Eq. (1.16) gives:

$$\sum_{r=0}^{N} \omega^r \left[ \sum_{l=-L}^{L} \left( \frac{A^{\{r\}}{}_l^{\text{even}}}{2} \right) \cos(l\omega t) \right] P(t) + \sum_{r=0}^{N} \omega^r \left[ \sum_{l=-L}^{L} \left( \frac{A^{\{r\}}{}_l^{\text{odd}}}{2} \right) \sin(l\omega t) \right] P(t) = \dot{P}(t) + P(t)R. \tag{5.5}$$

It might be convenient to have a degree-$N$ polynomial in powers of $\omega^r$, so we assume that $R$ is a degree-$N$ polynomial in $\omega^r$, but not $P(t)$. Thus,

$$P(t) = \sum_{k=-p}^{p} \left( \frac{P_k^{\text{even}}}{2} \cos(k\omega t) + \frac{P_k^{\text{odd}}}{2} \sin(k\omega t) \right),$$
$$R = \sum_{r=0}^{N} \omega^r R^{\{r\}}\,. \tag{5.6}$$

As done at the beginning of this chapter, we use *Floquet Theory* (**Theorem 3.1**) so that $\det\{P(t)\} = \text{const.}$ and $\text{trace}\{R\} = 0$, which gives $\text{trace}\{R^{\{r\}}\} = 0$ for $r = 0,1,2,3, \dots, N$. By using the chain rule, we obtain $\dot{P}(t) = \omega \frac{\partial P}{\partial(\omega t)}$, which provides a linear term (proportional to $\omega^1$) on the LHS of Eq. (5.5) so that both sides of Eq. (5.5) can be presented as a degree-$N$ polynomial in $\omega^r$. Combining the above relations with Eqs. (2.18) and (2.19) gives the new cosine and sine coefficients for comparison. For $k = 0,1,2,3, \dots, p$, the cosine (even) terms are:

$$P_k^{\text{even}} \left[ \sum_{r=0}^{N} \omega^r R^{\{r\}} \right] + P_k^{\text{odd}} k\omega = \sum_{r=0}^{N} \left( \omega^r \frac{A^{\{r\}}{}_k^{\text{even}}}{2} \right) P_0^{\text{even}}$$
$$+ \sum_{l=1}^{L} \left( \sum_{r=0}^{N} \left( \omega^r \left[ \frac{A^{\{r\}}{}_{k+l}^{\text{even}}}{2} + \frac{A^{\{r\}}{}_{k-l}^{\text{even}}}{2} \right] \right) \right) P_l^{\text{even}} + \sum_{l=1}^{L} \left( \sum_{r=0}^{N} \left( \omega^r \left[ \frac{A^{\{r\}}{}_{k+l}^{\text{odd}}}{2} - \frac{A^{\{r\}}{}_{k-l}^{\text{odd}}}{2} \right] \right) \right) P_l^{\text{odd}}, \tag{5.7}$$

and the sine (odd) terms are:

$$P_k^{\text{odd}} \left[ \sum_{r=0}^{N} \omega^r R^{\{r\}} \right] - P_k^{\text{even}} k\omega = \sum_{r=0}^{N} \left( \omega^r \frac{A^{\{r\}}{}_k^{\text{odd}}}{2} \right) P_0^{\text{even}} +$$
$$+ \sum_{l=1}^{L} \left( \sum_{r=0}^{N} \left( \omega^r \left[ \frac{A^{\{r\}}{}_{k-l}^{\text{odd}}}{2} + \frac{A^{\{r\}}{}_{k+l}^{\text{odd}}}{2} \right] \right) \right) P_l^{\text{even}} + \sum_{l=1}^{L} \left( \sum_{r=0}^{N} \left( \omega^r \left[ \frac{A^{\{r\}}{}_{k-l}^{\text{even}}}{2} - \frac{A^{\{r\}}{}_{k+l}^{\text{even}}}{2} \right] \right) \right) P_l^{\text{odd}}. \tag{5.8}$$

Comparing powers of $\omega^r$ gives the following set of algebraic equations (for $k = 0,1,2,3, \dots, p$ and $r = 0,1,2,3, \dots N$):

$$P_k^{\text{even}} R^{\{r\}} + P_k^{\text{odd}} k\delta_{r1} = \frac{{A^{\{r\}}}_k^{\text{even}}}{2} P_0^{\text{even}} + \quad (5.9)$$
$$+ \sum_{l=1}^{L} \left\{ \left[ \frac{{A^{\{r\}}}_{k+l}^{\text{even}}}{2} + \frac{{A^{\{r\}}}_{k-l}^{\text{even}}}{2} \right] P_l^{\text{even}} + \left[ \frac{{A^{\{r\}}}_{k+l}^{\text{odd}}}{2} - \frac{{A^{\{r\}}}_{k-l}^{\text{odd}}}{2} \right] P_l^{\text{odd}} \right\}$$

for the cosine (even) and $\omega^r$ terms, and:

$$P_k^{\text{odd}} R^{\{r\}} - P_k^{\text{even}} k\delta_{r1} = \frac{{A^{\{r\}}}_k^{\text{odd}}}{2} P_0^{\text{even}} + \quad (5.10)$$
$$+ \sum_{l=1}^{L} \left\{ \left[ \frac{{A^{\{r\}}}_{k-l}^{\text{odd}}}{2} + \frac{{A^{\{r\}}}_{k+l}^{\text{odd}}}{2} \right] P_l^{\text{even}} + \left[ \frac{{A^{\{r\}}}_{k-l}^{\text{even}}}{2} - \frac{{A^{\{r\}}}_{k+l}^{\text{even}}}{2} \right] P_l^{\text{odd}} \right\}$$

for the sine (odd) and $\omega^r$ terms. In Eqs. (5.9) and (5.10), $\delta_{r1}$ is the Kronecker delta function and multiplies the term $\omega^1$. By using Eqs. (2.22)–(2.25) and comparing powers of $\omega$, Eqs. (5.9) and (5.10) can be presented as a matrix-vector form of algebraic equations i.e.,

$$\widetilde{\mathbf{A}}^{\{r\}} \widetilde{\mathbf{P}} = \widetilde{\mathbf{P}} R^{\{r\}}. \quad (5.11)$$

We thus need to solve Eqs. (5.9) and (5.10) for each term in $R^{\{r\}}$ (for each $r = 0,1,2,3,\dots,N$) and for each $P_k^{\text{even}}$ and $P_k^{\text{odd}}$ (for each $k = 0,1,2,3,\dots,p$). However, assuming that each $P_k^{\text{even}}$ and $P_k^{\text{odd}}$ is independent of $\omega$, then the solution $\{P_k^{\text{even}}, P_k^{\text{odd}}, k = 0,1,2,3,\dots,p\}$ must be compatible with each $R^{\{r\}}$ term (for each $r = 0,1,2,3,\dots,N$) if we find a set of solutions $\{P_k^{\text{even}}, P_k^{\text{odd}}, k = 0,1,2,3,\dots,p\}$ that is compatible with some solution $R^{\{r_0\}}$ for an arbitrary $r_0 \in \{0,1,\dots,N\}$.

Moreover, suppose that, for an arbitrary $r_0 \in \{0,1,\dots,N\}$, we find a solution $R^{\{r_0\}}$ and a concurrent set $\{P_k^{\text{even}}, P_k^{\text{odd}}, k = 0,1,2,3,\dots,p\}$. Since we know all the terms in $P_k^{\text{even}}$ and $P_k^{\text{odd}}$, we can construct the periodic part $P(t)$ of the solution, and we can use Eq. (1.18) to find the constant part $R$. **Theorem 3.1** of *Floquet theory* indicates that our solution is correct:

1. $\det\{P(t)\} = \text{const.}$;
2. the RHS of Eq. (1.18) $[P^{-1}(t)\{A(t)P(t) - \dot{P}(t)\}]$ is a constant matrix (equal to $R$);
3. the result for $R$ in item 2 above is the sum $\sum_{r=0}^{N} \omega^r R^{\{r\}}$ [see Eq. (5.6)], so
    a. $\text{trace}\{R^{\{r\}}\} = 0$ for all $r = 0,1,2,3,\dots,N$;
    b. the term $R^{\{r_0\}}$ (that was calculated to find the set $\{P_k^{\text{even}}, P_k^{\text{odd}}, k = 0,1,2,3,\dots,p\}$) is the same term calculated from the sum $\sum_{r=0}^{N} \omega^r R^{\{r\}}$ for $r = r_0$.

## 5.3 Relation to LTI Systems

We now consider $\omega \to 0$, in which case the LPTV system is not periodic. In this case, we find that the system becomes an LTI system. We want to find the relationship between the original LPTV system with its transition matrix (for some $\omega > 0$) and the related LTI system with its transition matrix (when $\omega \to 0$). We first need to examine the scalar term $e^{\frac{1}{n}\int_{t_0}^{t} \text{trace}\{A(\tau)\}d\tau}$ in **Theorem 3.1** of *Floquet theory* to verify that it does not diverge when $\omega \to 0$ for fixed times $t$ and $t_0$.

**Lemma 5.1:** For all constants $a, b$:

$$\lim_{\omega\to 0} \frac{\cos(\omega b) - \cos(\omega a)}{\omega} = 0 \text{ and } \lim_{\omega\to 0} \frac{\sin(\omega b) - \sin(\omega a)}{\omega} = b - a.$$

This can be trivially proved by using elementary trigonometric identities and limits:

$$\cos(\omega b) - \cos(\omega a) = -2\sin\left(\frac{b-a}{2}\omega\right)\sin\left(\frac{b+a}{2}\omega\right),$$

$$\sin(\omega b) - \sin(\omega a) = 2\sin\left(\frac{b-a}{2}\omega\right)\cos\left(\frac{b+a}{2}\omega\right),$$

$$\lim_{\theta\to 0}\cos(\theta) = 1, \quad \lim_{\theta\to 0}\sin(\theta) = 0, \qquad \lim_{\theta\to 0}\frac{\sin(\theta)}{\theta} = 1\,.$$

This is not detailed in this work

□.

**Corollary 5.2**: If $\psi(t|\omega) = \frac{1}{n}\operatorname{trace}\{A(t|\omega)\}$ has the form:

$$\psi(t|\omega) = \frac{\psi_0^{\text{even}}}{2} + \sum_{l=1}^{L}\left(\psi_l^{\text{even}}\cos(l\omega t) + \psi_l^{\text{odd}}\sin(l\omega t)\right),$$

where none of the terms $\psi_l^{\text{even}}$ or $\psi_l^{\text{odd}}$depend on $\omega$, then:

$$e^{\int_{t_0}^{t}\psi(\tau|\omega=0)d\tau} = \lim_{\omega\to 0} e^{\int_{t_0}^{t}\psi(\tau|\omega)d\tau}.$$

**Proof:** by exponentiaon of $\psi(t|\omega)$ , we have

$$e^{\int_{t_0}^{t}\psi(\tau|\omega)d\tau} = e^{\frac{\psi_0^{\text{even}}}{2}(t-t_0)+\sum_{l=1}^{L}\left(\psi_l^{\text{even}}\frac{\sin(l\omega t)-\sin(l\omega t_0)}{l\omega} - \psi_l^{\text{odd}}\frac{\cos(l\omega t)-\cos(l\omega t_0)}{l\omega}\right)}.$$

Therefore, on the one hand:

$$\lim_{\omega\to 0} e^{\int_{t_0}^{t}\psi(\tau|\omega)d\tau}$$

$$= e^{(t-t_0)\frac{\psi_0^{\text{even}}}{2}+\sum_{l=1}^{L}\left(\psi_l^{\text{even}}\lim_{\omega\to 0}\frac{\sin(l\omega t)-\sin(l\omega t_0)}{l\omega} - \psi_l^{\text{odd}}\lim_{\omega\to 0}\frac{\cos(l\omega t)-\cos(l\omega t_0)}{l\omega}\right)},$$

$$\lim_{\omega\to 0} e^{\int_{t_0}^{t}\psi(\tau|\omega)d\tau} = e^{(t-t_0)\frac{\psi_0^{\text{even}}}{2}+\sum_{l=1}^{L}\left(\psi_l^{\text{even}}\frac{(lt-lt_0)}{l}\right)} = e^{(t-t_0)\left[\frac{\psi_0^{\text{even}}}{2}+\sum_{l=1}^{L}\psi_l^{\text{even}}\right]},$$

but, on the other hand,

$$\psi(t|\omega=0) = \frac{\psi_0^{\text{even}}}{2} + \sum_{l=1}^{L}\left(\psi_l^{\text{even}}\cos(l(0)t) + \psi_l^{\text{odd}}\sin(l(0)t)\right),$$

$$\psi(t|\omega=0) = \frac{\psi_0^{\text{even}}}{2} + \sum_{l=1}^{L}\psi_l^{\text{even}} = \text{const.},$$

$$e^{\int_{t_0}^{t} \psi(\tau|\omega=0)d\tau} = e^{(t-t_0)\left[\frac{\psi_0^{\text{even}}}{2}+\sum_{l=1}^{L} \psi_l^{\text{even}}\right]} = \lim_{\omega\to 0} e^{\int_{t_0}^{t} \psi(\tau|\omega)d\tau}.$$

■

We conclude that, for a scalar LPTV system $\psi(t|\omega)$, the transition function $\Phi_{\psi(\cdot|\omega)}(t, t_0) = e^{\int_{t_0}^{t} \psi(\tau|\omega)d\tau}$ is a continuous function of $\omega$ and, in particular, at the point $\omega = 0$. In addition, $\psi(t|\omega = 0) = \text{const.}$ produces a scalar LTI system with a transition function $\Phi_{\psi(\cdot|\omega=0)}(t, t_0)$.

We now extend the notion of this corollary to the matrix case of LPTV systems and their solution (see **Theorem 3.1** of *Floquet theory*). For a frequency analysis, we change the notation "$(t)$" used to indicate time dependence to a notation using time and frequency, "$(t|\omega)$" [for example, $A(t) \Rightarrow A(t|\omega)$]. Suppose that:

$$A(t|\omega) = \sum_{l=-L}^{L}\sum_{r=0}^{N}\left(\omega^r \frac{A^{\{r\}}{}_l^{\text{even}}}{2}\right)\cos(l\omega t) + \sum_{l=-L}^{L}\sum_{r=0}^{N}\left(\omega^r \frac{A^{\{r\}}{}_l^{\text{odd}}}{2}\right)\sin(l\omega t), \quad (5.12)$$

and, without loss of generality (see the beginning of the chapter),

$$\text{trace}\{A(t|\omega)\} \equiv 0.$$

We denote the *Floquet theory* solution by using the periodic matrix $P(t|\omega)$ and the constant matrix $R(\omega)$ as follows:

$$A(t|\omega)P(t|\omega) = \frac{\partial}{\partial t}\{P(t|\omega)\} + P(t|\omega)R(\omega), \quad (5.13)$$

Where:

$$P(t|\omega) = \sum_{k=-p}^{p}\left(\frac{P_k^{\text{even}}}{2}\cos(k\omega t) + \frac{P_k^{\text{odd}}}{2}\sin(k\omega t)\right),$$
$$R(\omega) = \sum_{r=0}^{N}\omega^r R^{\{r\}}, \quad (5.14)$$

$$\det\{P(t|\omega)\} \equiv \text{const.} \neq 0, \quad \text{trace}\{R(\omega)\} \equiv 0,$$

so that the transition matrix is:

$$\Phi_A(t, t_0|\omega) = P(t|\omega)e^{(t-t_0)R(\omega)}P^{-1}(t_0|\omega). \tag{5.15}$$

On the one hand, if we insert $\omega = 0$ into Eq. (5.12), we obtain:

$$A(t|\omega = 0) = \sum_{l=-L}^{L} \frac{A^{\{0\}}{}_l^{\text{even}}}{2} = \frac{A^{\{0\}}{}_0^{\text{even}}}{2} + \sum_{l=1}^{L} A^{\{0\}}{}_l^{\text{even}} \equiv \text{const.}, \\ \text{trace}\{A(t|\omega = 0)\} \equiv 0, \tag{5.16}$$

which is a constant matrix that can represent an LTI system with a transition matrix:

$$\Phi_{A(\cdot|\omega=0)}(t, t_0) = e^{(t-t_0)A(t|\omega=0)}. \tag{5.17}$$

On the other hand, if we insert $\omega = 0$ into Eqs. (5.13)–(5.15), we have the following equation involving $P(t|\omega = 0)$ and $R(0)$:

$$A(t|\omega = 0)P(t|\omega = 0) = P(t|\omega = 0)R(0), \tag{5.18}$$

Where:

$$P(t|\omega = 0) = \sum_{k=-p}^{p} \frac{P_k^{\text{even}}}{2} = \frac{P_0^{\text{even}}}{2} + \sum_{k=1}^{p} P_k^{\text{even}} \equiv \text{const.}, \\ R(0) = R^{\{0\}}, \\ \frac{\partial}{\partial t}\{P(t|\omega = 0)\} \equiv 0, \quad \det\{P(t|\omega = 0)\} \equiv \text{const.} \neq 0, \\ \text{trace}\{R(0)\} = 0, \tag{5.19}$$

so that the transition matrix is:

$$\Phi_A(t,t_0|\omega=0)=P(t|\omega=0)e^{(t-t_0)R(0)}P^{-1}(t_0|\omega=0). \tag{5.20}$$

Note that, in solving LPTV systems, we have some degree of freedom to determine $P(t|\omega)$ (up to a scaling factor, an invertible rotation or reflection transformation, etc. see CHAPTER 3), so we may determine $P(t|\omega)$ by using one of the following options:

1. Choose $R(0)$ to have some canonical form (e.g., a real Jordan matrix block) so that $A(t|\omega=0)$ is constructed such that $A(t|\omega=0)V=VR(0)$ for an invertible matrix $V$, and then find $P(t|\omega)$ such that $P(t|\omega=0)=V$.

2. Set $R(0)=A(t|\omega=0)$ and find $P(t|\omega)$ such that $P(t|\omega=0)=I_n$.

Using either option 1 or option 2, we obtain:

$$\Phi_A(t,t_0|\omega=0)=\Phi_{A(\cdot|\omega=0)}(t,t_0), \tag{5.21}$$

which is the transition matrix generated by the LPTV system matrix $A(t|\omega)$ and then evaluated at $\omega=0$ [denoted $\Phi_A(t,t_0|\omega=0)$; see Eq. (5.20)]. $A(t|\omega)$ is the same as the transition matrix generated by the related LTI system matrix $A(t|\omega=0)$ [denoted $\Phi_{A(\cdot|\omega=0)}(t,t_0)$; see Eq. (5.17)].

## 5.4 Transformation of LPTV Matrix to Canonical Form at ω = 0

Suppose that $A(t|\omega)$ is an LPTV system matrix that is continuous in frequency $\omega$ (e.g., with Fourier coefficients that are polynomial in $\omega$). Consider the eigen decomposition of $A(t|\omega=0)$:

$$A(t|\omega = 0)U = UJ, \tag{5.22}$$

where $J$ has a real Jordan canonical block form [i.e., we can represent eigenvalues of $A(t|\omega = 0)$], and $U$ is the corresponding eigenvector matrix. By using the invertible matrix $U$, we transform $A(t|\omega)$ into a new similar matrix [see section 3.3]:

$$\hat{A}(t|\omega) = U^{-1}A(t|\omega)U, \tag{5.23}$$

which has the canonical form at $\omega = 0$ [i.e., $\hat{A}(t|\omega = 0) = J$]. Given that $\hat{A}(t|\omega)$ and $A(t|\omega)$ are similar matrices, they have the same trace. We denote $\psi(t|\omega) = \frac{1}{n}\mathrm{trace}\{A(t|\omega)\} \equiv \frac{1}{n}\mathrm{trace}\{\hat{A}(t|\omega)\}$. If $\psi(t|\omega) \not\equiv 0$, we can zero the trace by using section 3.4:

$$\tilde{A}(t|\omega) = U^{-1}[A(t|\omega) - \psi(t|\omega)I_n]U. \tag{5.24}$$

Note that we can interchange the eigen decomposition and the trace-zeroing; that is, we can first do the eigen decomposition as per Eq. (5.23) and then zero the trace as per Eq. (5.24), or vice versa. The eigenvector matrix $U$ remains the same, as required, and we have a new canonical form for $J = \tilde{A}(t|\omega = 0)$ such that $\mathrm{trace}\{\tilde{A}(t|\omega = 0)\} \equiv 0$ as required. Applying *Floquet theory* to the LPTV system matrix $\tilde{A}(t|\omega)$, we have the matrices $\tilde{P}(t|\omega)$ and $\tilde{R}(\omega)$ such that

$$\begin{gathered}\Phi_{\tilde{A}}(t, t_0|\omega) = \tilde{P}(t|\omega)e^{(t-t_0)\tilde{R}(\omega)}\tilde{P}^{-1}(t_0|\omega),\\ \tilde{A}(t|\omega)\tilde{P}(t|\omega) = \dot{\tilde{P}}(t|\omega) + \tilde{P}(t|\omega)\tilde{R}(\omega),\end{gathered} \tag{5.25}$$

where $\tilde{P}(t|\omega)$ is the periodic matrix of the solution with $\det\{\tilde{P}(t|\omega)\} \equiv \text{const.} \neq 0$, and $\tilde{R}(\omega)$ is the constant part of the solution with $\mathrm{trace}\{\tilde{R}(\omega)\} = 0$. $\tilde{P}(t|\omega)$. $\tilde{R}(\omega)$ can be chosen with the following properties:

$$\tilde{P}(t|\omega=0)=I_n,\\ \tilde{R}(0)=\tilde{A}(t|\omega=0). \tag{5.26}$$

When using $\omega=0$ in Eq. (5.25), we obtain an LTI system represented by the constant matrix $\tilde{R}(0)=\tilde{A}(t|\omega=0)$. In addition, $\tilde{P}(t|\omega=0)$ is a constant matrix at $\omega=0$, so choosing $\tilde{P}(t|\omega=0)=I_n$ is compatible with the equations above. The motivation to perform these transformations is to simplify the structure of the matrices $\tilde{P}(t|\omega)$ and $\tilde{R}(\omega)$ and to simplify the procedure to find them. In particular, if the Fourier coefficients of $\tilde{A}(t|\omega)$ are polynomial in $\omega$ so that $\tilde{R}(\omega)$ is polynomial in $\omega$, but the Fourier coefficients of $\tilde{P}(t|\omega)$ are independent of $\omega$, it would be convenient to find the Fourier coefficients of $\tilde{P}(t|\omega)$ by comparing powers of $\omega$ between Eqs. (5.9) and (5.10) and then solving for $\tilde{R}(\omega)$ by using Eq. (5.25).

From the properties of Eqs. (3.5), (3.6) and (3.12), the transition matrix $\Phi_A(t,t_0|\omega)$ can be decomposed as follows based on the solution for $\tilde{P}(t|\omega)$ and $\tilde{R}(\omega)$:

$$\Phi_A(t,t_0|\omega)=e^{\int_{t_0}^{t}\psi(\tau|\omega)d\tau}\left\{U\left[\tilde{P}(t|\omega)e^{(t-t_0)\tilde{R}(\omega)}\tilde{P}^{-1}(t_0|\omega)\right]U^{-1}\right\}. \tag{5.27}$$

The factor $e^{\int_{t_0}^{t}\psi(\tau|\omega)d\tau}\{\cdot\}$ is responsible for zeroing the trace for Eq. (5.24), and $U[\cdot]U^{-1}$ inverts the canonical form decomposition [see Eq. (5.23)]. Note that:

$$\hat{P}(t|\omega)=U\tilde{P}(t|\omega) \text{ and } \hat{R}(\omega)=\tilde{R}(\omega) \tag{5.28}$$

are the matrices used to obtain the transition matrix $\Phi_{A-\psi I}(t,t_0|\omega)$. Based on Eqs. (3.14) and (3.15), we can decompose the scalar function $\psi(t|\omega)$ into a periodic part [denoted by $\psi_1(t|\omega)$ with anti-derivative $\Psi_1(t|\omega)$] and a constant part (denoted by $\psi_0$). We then construct the matrices

$$P(t|\omega)=e^{\Psi_1(t)}U\tilde{P}(t|\omega) \text{ and } R(\omega)=\tilde{R}(\omega)+\psi_0 I_N. \tag{5.29}$$

For the transition matrix $\Phi_A(t,t_0|\omega)$, see **Theorem 1.3** of *Floquet theory*.

The next section shows several examples of solving an LPTV system by comparing powers of $\omega$.

## 5.5 Examples

**Example 5.3: Row H of Table 4-4**

Consider the following real LPTV system matrix:

$$A(t) = \begin{bmatrix} a & b \\ -b & a \end{bmatrix} + c\begin{bmatrix} \sin(2\omega t) & \cos(2\omega t) \\ \cos(2\omega t) & -\sin(2\omega t) \end{bmatrix} + d\begin{bmatrix} -\cos(2\omega t) & \sin(2\omega t) \\ \sin(2\omega t) & \cos(2\omega t) \end{bmatrix},$$
$$a = a_0 + \omega a_1;\ b = b_0 + \omega b_1; c = c_0 + \omega c_1; d = d_0 + \omega d_1. \tag{5.30}$$

In general, $\text{trace}\{A(t)\} \not\equiv 0$ . However, based on Eq (3.8) and Section 3.4, we can solve for the case where $a = 0$ and apply the following equation to obtain:

$$\hat{A}(t) = A(t) - aI_2 \leftrightarrow \Phi_A(t,t_0) = e^{(t-t_0)a}\Phi_{\hat{A}}(t,t_0) \tag{5.31}$$

The factor $a$ plays the role of the function $\psi(\text{t}) = \frac{1}{n}\text{trace}\{A(t)\} \equiv a$ [see Eq. (3.13)] with $n = 2$. We now solve for $a = 0$ and omit the hat sign,

$$A(t) = b\begin{bmatrix} 0 & 1 \\ -1 & 0 \end{bmatrix} + c\begin{bmatrix} \sin(2\omega t) & \cos(2\omega t) \\ \cos(2\omega t) & -\sin(2\omega t) \end{bmatrix} + d\begin{bmatrix} -\cos(2\omega t) & \sin(2\omega t) \\ \sin(2\omega t) & \cos(2\omega t) \end{bmatrix},$$
$$a = a_0 + \omega a_1;\ b = b_0 + \omega b_1; c = c_0 + \omega c_1; d = d_0 + \omega d_1. \tag{5.32}$$

We assume that the transition matrix is decomposed according to *Floquet theory* into a periodic part $P(t) = P\left(t + \frac{2\pi}{\omega}\right)$ and a constant part $R = R^{\{0\}} + \omega R^{\{1\}}$ such that $\det\{P(t)\} \equiv \text{const.} \neq 0$ and $\text{trace}\{R\} \equiv 0$ for all $\omega$ (i.e., $\text{trace}\{R^{\{0\}}\} = \text{trace}\{R^{\{1\}}\} = 0$). The Fourier coefficients of $A(t)$, ${A^{\{r\}}}_k^{\text{even}}$ and ${A^{\{r\}}}_k^{\text{odd}}$, are defined for $k = 0, 2$ but not for $k = 1$, so we assume that the Fourier coefficients of $P(t)$, $P_k^{\text{even}}$ and $P_k^{\text{odd}}$, are defined for $k = 0, 1, 2$. Recall that the superscript $\{r\}$ denotes the power of $\omega$ (i.e., $A^{\{r\}}{}_k^{\text{even}}$ and $A^{\{r\}}{}_k^{\text{odd}}$ are the Fourier coefficients of $\omega^r$ for $r = 0, 1$). By evaluating $A(t)$ at $\omega = 0$ we obtain:

$$A(t|\omega = 0) = \begin{bmatrix} -d_0 & c_0 + b_0 \\ c_0 - b_0 & d_0 \end{bmatrix}. \tag{5.33}$$

We can use this information to obtain $R^{\{0\}}$ because $A(t|\omega = 0)P(t|\omega = 0) = P(t|\omega = 0)R^{\{0\}}$ [Eq. (5.18)]. Note that the eigenvalues of $A(t|\omega = 0)$ are:

$$\lambda_{1,2} = \pm\sqrt{d_0^2 + c_0^2 - b_0^2}, \tag{5.34}$$

which might be complex. Conversely, $A^2(t|\omega = 0)$ is a square matrix,

$$A^2(t|\omega = 0) = (d_0^2 + c_0^2 - b_0^2)I_2, \tag{5.35}$$

which has two identical eigenvalues $(d_0^2 + c_0^2 - b_0^2)$. Therefore, we use $P(t|\omega = 0) = I_2$ and $R^{\{0\}} = A(t|\omega = 0)$. By using Eqs. (2.22)–(2.25), we have the following algebraic matrix-vector form:

$$\widetilde{\mathbf{A}}\widetilde{\mathbf{P}} = \widetilde{\mathbf{P}}R, \tag{5.36}$$

With:

$$\widetilde{\mathbf{A}} = \begin{bmatrix} \begin{bmatrix} 0 & b \\ -b & 0 \end{bmatrix} & 0 & 0 & \begin{bmatrix} -d & c \\ c & d \end{bmatrix} & \begin{bmatrix} c & d \\ d & -c \end{bmatrix} \\ 0 & \frac{1}{2}\begin{bmatrix} -d & c \\ c & d \end{bmatrix} + \begin{bmatrix} 0 & b \\ -b & 0 \end{bmatrix} & \frac{1}{2}\begin{bmatrix} c & d \\ d & -c \end{bmatrix} - \omega I_2 & 0 & 0 \\ 0 & \frac{1}{2}\begin{bmatrix} c & d \\ d & -c \end{bmatrix} + \omega I_2 & \begin{bmatrix} 0 & b \\ -b & 0 \end{bmatrix} - \frac{1}{2}\begin{bmatrix} -d & c \\ c & d \end{bmatrix} & 0 & 0 \\ \frac{1}{2}\begin{bmatrix} -d & c \\ c & d \end{bmatrix} & 0 & 0 & \begin{bmatrix} 0 & b \\ -b & 0 \end{bmatrix} & -2\omega I_2 \\ \frac{1}{2}\begin{bmatrix} c & d \\ d & -c \end{bmatrix} & 0 & 0 & +2\omega I_2 & \begin{bmatrix} 0 & b \\ -b & 0 \end{bmatrix} \end{bmatrix}, \tag{5.37}$$

$$\widetilde{\mathbf{P}} = \begin{bmatrix} P_0^{\text{even}} \\ P_1^{\text{even}} \\ P_1^{\text{odd}} \\ P_2^{\text{even}} \\ P_2^{\text{odd}} \end{bmatrix}, \tag{5.38}$$

where $R$ and the terms $P_k^{\text{even}}$ and $P_k^{\text{odd}}$ are unknown and must be found. To simplify the structure, we solve for the coefficient of $\omega^0$ in Eqs. (5.36)–(5.38) [or, equivalently, consider $\omega = 0$ and retain all coefficients of $\omega^0$; e.g., $R \mapsto R^{\{0\}}, b \mapsto b_0$; see Eq. (5.11)], and obtain:

$$\widetilde{\mathbf{A}}^{\{0\}}\widetilde{\mathbf{P}} = \widetilde{\mathbf{P}}R^{\{0\}}. \tag{5.39}$$

By multiplying Eq. (5.39) by $\widetilde{\mathbf{A}}^{\{0\}}$ from the left and using the associative property of matrix multiplication, we have a new LHS multiplied by $\widetilde{\mathbf{A}}^{\{0\}}$ from the left and a new RHS multiplied by $R^{\{0\}}$, such that:

$$\left(\widetilde{\mathbf{A}}^{\{0\}}\right)^2\widetilde{\mathbf{P}} = \widetilde{\mathbf{P}}\left(R^{\{0\}}\right)^2, \tag{5.40}$$

where $\left(R^{\{0\}}\right)^2 = (d_0^2 + c_0^2 - b_0^2)I_2$, and $\left(\widetilde{\mathbf{A}}^{\{0\}}\right)^2$ is:

$$\left(\widetilde{\mathbf{A}}^{\{0\}}\right)^2 = \begin{bmatrix} \sigma & 0 & 0 & 0 & 0 & 0 & 0 & 0 & 0 & 0 \\ 0 & \sigma & 0 & 0 & 0 & 0 & 0 & 0 & 0 & 0 \\ 0 & 0 & \beta & 0 & 0 & -\alpha & 0 & 0 & 0 & 0 \\ 0 & 0 & 0 & \beta & \alpha & 0 & 0 & 0 & 0 & 0 \\ 0 & 0 & 0 & \alpha & \beta & 0 & 0 & 0 & 0 & 0 \\ 0 & 0 & -\alpha & 0 & 0 & \beta & 0 & 0 & 0 & 0 \\ 0 & 0 & 0 & 0 & 0 & 0 & \beta & 0 & 0 & -\alpha \\ 0 & 0 & 0 & 0 & 0 & 0 & 0 & \beta & \alpha & 0 \\ 0 & 0 & 0 & 0 & 0 & 0 & 0 & \alpha & \beta & 0 \\ 0 & 0 & 0 & 0 & 0 & 0 & -\alpha & 0 & 0 & \beta \end{bmatrix}, \tag{5.41}$$

$$\sigma = d_0^2 + c_0^2 - b_0^2, \quad \alpha = 0.5, \quad \beta = \alpha - b^2.$$

The eigen decomposition of $\left(\widetilde{\mathbf{A}}^{\{0\}}\right)^2$is given by:

$$\left(\widetilde{\mathbf{A}}^{\{0\}}\right)^2 \mathbf{V} = \mathbf{V}\mathbf{J}, \tag{5.42}$$

where:

$$\mathbf{J} = \text{diag}\{[-b^2 \quad -b^2 \quad -b^2 \quad -b^2 \quad \sigma \quad \sigma \quad \sigma \quad \sigma \quad \sigma \quad \sigma]\},$$

$$\mathbf{V} = \begin{bmatrix} 0 & 0 & 0 & 0 & 1 & 0 & 0 & 0 & 0 & 0 \\ 0 & 0 & 0 & 0 & 0 & 1 & 0 & 0 & 0 & 0 \\ 0 & 1 & 0 & 0 & 0 & 0 & \mathbf{1} & 0 & 0 & 0 \\ -1 & 0 & 0 & 0 & 0 & 0 & 0 & \mathbf{1} & 0 & 0 \\ 1 & 0 & 0 & 0 & 0 & 0 & 0 & \mathbf{1} & 0 & 0 \\ 0 & 1 & 0 & 0 & 0 & 0 & -\mathbf{1} & 0 & 0 & 0 \\ 0 & 0 & 0 & 1 & 0 & 0 & 0 & 0 & 0 & -1 \\ 0 & 0 & -1 & 0 & 0 & 0 & 0 & 0 & 1 & 0 \\ 0 & 0 & 1 & 0 & 0 & 0 & 0 & 0 & 1 & 0 \\ 0 & 0 & 0 & 1 & 0 & 0 & 0 & 0 & 0 & 1 \end{bmatrix}. \tag{5.43}$$

Given that $\left(R^{\{0\}}\right)^2 = \sigma I_2$, we construct $\widetilde{\mathbf{P}}$ by a linear combination of the columns of $\mathbf{V}$, which are eigenvectors corresponding to the eigenvalue $\sigma$. If we set

$$\widetilde{\mathbf{P}} = \begin{bmatrix} 0 \\ P_1^{\text{even}} \\ P_1^{\text{odd}} \\ 0 \\ 0 \end{bmatrix}, \tag{5.44}$$

with

$$P_1^{\text{even}} = \begin{bmatrix} 1 & 0 \\ 0 & 1 \end{bmatrix}, \quad P_1^{\text{odd}} = \begin{bmatrix} 0 & 1 \\ -1 & 0 \end{bmatrix}, \tag{5.45}$$

then we use the following guess for $P(t)$:

$$P(t) = \begin{bmatrix} \cos(\omega t) & \sin(\omega t) \\ -\sin(\omega t) & \cos(\omega t) \end{bmatrix}. \tag{5.46}$$

Note that the above guess for $P(t)$ satisfies $\det\{P(t)\} \equiv 1 = \text{const.} \neq 0$ and $P(t|\omega = 0) = I_2$, which is consistent with our assumptions.

We now reintroduce the hat superscript [e.g., $\hat{P}(t) \coloneqq P(t)$]. Inserting our guess for $\hat{P}(t)$ into the RHS of Eq. (1.18), we can check if the value of the constant part $\hat{R}$ of the solution is valid:

$$\begin{aligned} \hat{R} &= \hat{P}^{-1}(t)\left[\hat{A}(t)\hat{P}(t) - \dot{\hat{P}}(t)\right] = \\ &= \begin{bmatrix} -d_0 - d_1\omega & c_0 + b_0 + (c_1 + b_1 - 1)\omega \\ c_0 - b_0 + (c_1 - b_1 + 1)\omega & d_0 + d_1\omega \end{bmatrix} \\ &= \begin{bmatrix} -d_0 & c_0 + b_0 \\ c_0 - b_0 & d_0 \end{bmatrix} + \omega \begin{bmatrix} -d_1 & c_1 + b_1 - 1 \\ c_1 - b_1 + 1 & d_1 \end{bmatrix}. \end{aligned} \tag{5.47}$$

We find that $\hat{R}$ is not a function $t$. In addition, the coefficient of $\omega^0$ in the expression for $\hat{R}$ is identical to our assumption $\hat{R}^{\{0\}} = \hat{A}(t|\omega = 0)$. Therefore, the matrices $\hat{P}(t)$ and $\hat{R}$ are a solution for constructing the transition matrix of the shifted version of $A(t)$ [i.e., $\Phi_{\hat{A}}(t, t_0)$, see Eq. (5.32)]. To obtain the solution of the original LPTV system $A(t)$, we use Eq. (3.16); given $\psi(\mathrm{t}) = \frac{1}{2}\mathrm{trace}\{A(t)\} \equiv a$, the following matrices $P(t)$ and $R$ are a solution to construct $\Phi_A(t, t_0)$:

$$P(t) = \begin{bmatrix} \cos(\omega t) & \sin(\omega t) \\ -\sin(\omega t) & \cos(\omega t) \end{bmatrix},$$
$$R = \begin{bmatrix} a_0 - d_0 + (a_1 - d_1)\omega & c_0 + b_0 + (c_1 + b_1 - 1)\omega \\ c_0 - b_0 + (c_1 - b_1 + 1)\omega & a_0 + d_0 + (a_1 + d_1)\omega \end{bmatrix}, \tag{5.48}$$

which is consistent with the data given in row H of **Table 4-4**.

■

The procedure used here to solve this parametric example can be used to solve a family of LPTV systems from the literature (Aggarwal & Infante, 1968), (Markus & Yamabe, 1960), (Rosenbrook, 1963). Note that a numerical example does not require the eigen decomposition in Eq. (5.40); the eigen decomposition in Eq. (5.39) suffices. In the following examples, we solve LPTV systems with a larger number of harmonics in $A(t)$ and $P(t)$ and finish with a 3×3 example.

**Example 5.4: Row E of Table 4-4**

Consider the following real 2×2 matrix for an LPTV system ($n = 2$):

$$A(t) = \frac{1}{3}\begin{bmatrix} -\cos(2\omega t) + 4\cos(4\omega t) - 4\sin(3\omega t) & -5 - 4\cos(3\omega t) - 4\sin(\omega t) + \sin(2\omega t) - 4\sin(4\omega t) \\ 5 - 4\cos(3\omega t) + 4\sin(\omega t) + \sin(2\omega t) - 4\sin(4\omega t) & \cos(2\omega t) - 4\cos(4\omega t) + 4\sin(3\omega t) \end{bmatrix}$$
$$+\omega \begin{bmatrix} 2\sin(3\omega t) & 4 + 2\cos(3\omega t) \\ -4 + 2\cos(3\omega t) & -2\sin(3\omega t) \end{bmatrix}, \tag{5.49}$$

which has $L = 4$ harmonics. By writing $A(t) = A^{\{0\}}(t) + \omega A^{\{1\}}(t)$, we obtain:

$$A^{\{0\}}(t) = \frac{1}{3}\begin{bmatrix} -\cos(2\omega t) + 4\cos(4\omega t) - 4\sin(3\omega t) & -5 - 4\cos(3\omega t) - 4\sin(\omega t) + \sin(2\omega t) - 4\sin(4\omega t) \\ 5 - 4\cos(3\omega t) + 4\sin(\omega t) + \sin(2\omega t) - 4\sin(4\omega t) & \cos(2\omega t) - 4\cos(4\omega t) + 4\sin(3\omega t) \end{bmatrix},$$
$$A^{\{1\}}(t) = \begin{bmatrix} 2\sin(3\omega t) & 4 + 2\cos(3\omega t) \\ -4 + 2\cos(3\omega t) & -2\sin(3\omega t) \end{bmatrix}. \tag{5.50}$$

Note that $\mathrm{trace}\{A(t)\} \equiv 0$ and $\mathrm{trace}\{A^{\{0\}}(t)\} \equiv \mathrm{trace}\{A^{\{1\}}(t)\} \equiv 0$; thus, the trace need not be zeroed. In addition, the matrix $A(t)$ evaluated at $\omega = 0$,

$$A(t|\omega = 0) = \begin{bmatrix} 1 & -3 \\ \frac{1}{3} & -1 \end{bmatrix}, \tag{5.51}$$

is decomposed as follows:

$$A(t|\omega = 0) = UJU^{-1}, \tag{5.52}$$

where

$$\begin{aligned} J &= \begin{bmatrix} 0 & 1 \\ 0 & 0 \end{bmatrix}, \\ U &= \begin{bmatrix} 1 & 1 \\ \frac{1}{3} & 0 \end{bmatrix}. \end{aligned} \tag{5.53}$$

We assume that the transition matrix is decomposed according to *Floquet theory* into a periodic part $P(t) = P\left(t + \frac{2\pi}{\omega}\right)$ and a constant part $R = R^{\{0\}} + \omega R^{\{1\}}$. Since $A(t)$ has $L = 4$ harmonics and is of size $n = 2$, we hypothesize that $P(t)$ has $p = \left\lceil \frac{4}{2} \right\rceil = 2$ harmonics. After expressing $A(t)$ in terms of its Fourier coefficients ${A^{\{r\}}}_k^{\text{even}}$ and ${A^{\{r\}}}_k^{\text{odd}}$ and in powers of $\omega$, we use Eqs. (2.22)–(2.25) and (5.11) to compare powers of $\omega$, which gives the following algebraic matrix-vector equation for $\omega^1$:

$$\widetilde{\mathbf{A}}^{\{1\}}\widetilde{\mathbf{P}} = \widetilde{\mathbf{P}}R^{\{1\}}, \tag{5.54}$$

with

$$\widetilde{\mathbf{A}}^{\{1\}} = \begin{bmatrix} \begin{bmatrix} 0 & 4 \\ -4 & 0 \end{bmatrix} & \begin{bmatrix} 0 & 0 \\ 0 & 0 \end{bmatrix} & \begin{bmatrix} 0 & 0 \\ 0 & 0 \end{bmatrix} & \begin{bmatrix} 0 & 0 \\ 0 & 0 \end{bmatrix} & \begin{bmatrix} 0 & 0 \\ 0 & 0 \end{bmatrix} \\ \begin{bmatrix} 0 & 0 \\ 0 & 0 \end{bmatrix} & \begin{bmatrix} 0 & 4 \\ -4 & 0 \end{bmatrix} & \begin{bmatrix} -1 & 0 \\ 0 & -1 \end{bmatrix} & \begin{bmatrix} 0 & 1 \\ 1 & 0 \end{bmatrix} & \begin{bmatrix} 1 & 0 \\ 0 & -1 \end{bmatrix} \\ \begin{bmatrix} 0 & 0 \\ 0 & 0 \end{bmatrix} & \begin{bmatrix} 1 & 0 \\ 0 & 1 \end{bmatrix} & \begin{bmatrix} 0 & 4 \\ -4 & 0 \end{bmatrix} & \begin{bmatrix} 1 & 0 \\ 0 & -1 \end{bmatrix} & \begin{bmatrix} 0 & -1 \\ -1 & 0 \end{bmatrix} \\ \begin{bmatrix} 0 & 0 \\ 0 & 0 \end{bmatrix} & \begin{bmatrix} 0 & 1 \\ 1 & 0 \end{bmatrix} & \begin{bmatrix} 1 & 0 \\ 0 & -1 \end{bmatrix} & \begin{bmatrix} 0 & 4 \\ -4 & 0 \end{bmatrix} & \begin{bmatrix} -2 & 0 \\ 0 & -2 \end{bmatrix} \\ \begin{bmatrix} 0 & 0 \\ 0 & 0 \end{bmatrix} & \begin{bmatrix} 1 & 0 \\ 0 & -1 \end{bmatrix} & \begin{bmatrix} 0 & -1 \\ -1 & 0 \end{bmatrix} & \begin{bmatrix} 2 & 0 \\ 0 & 2 \end{bmatrix} & \begin{bmatrix} 0 & 4 \\ -4 & 0 \end{bmatrix} \end{bmatrix}, \tag{5.55}$$

$$\widetilde{\mathbf{P}} = \begin{bmatrix} P_0^{\text{even}} \\ P_1^{\text{even}} \\ P_1^{\text{odd}} \\ P_2^{\text{even}} \\ P_2^{\text{odd}} \end{bmatrix}, \tag{5.56}$$

where $R^{\{1\}}$ and the terms $P_k^{\text{even}}$ and $P_k^{\text{odd}}$ are unknown and must be found. Consider the following decomposition of the matrix $\widetilde{\mathbf{A}}^{\{1\}}$:

$$\widetilde{\mathbf{A}}^{\{1\}}\mathbf{V} = \mathbf{V}\mathbf{J}, \tag{5.57}$$

where

$$\mathbf{J} = \text{blkdiag}\left\{\begin{bmatrix} 0 & 5 \\ -5 & 0 \end{bmatrix}, \begin{bmatrix} 0 & 6 \\ -6 & 0 \end{bmatrix}, \begin{bmatrix} 0 & 2 \\ -2 & 0 \end{bmatrix}, \begin{bmatrix} 0 & 4 \\ -4 & 0 \end{bmatrix}, \begin{bmatrix} 0 & 1 \\ -1 & 0 \end{bmatrix}\right\},$$
$$\mathbf{V} = \begin{bmatrix} \begin{bmatrix} 0 & 0 \\ 0 & 0 \end{bmatrix} & \begin{bmatrix} 0 & 0 \\ 0 & 0 \end{bmatrix} & \begin{bmatrix} 0 & 0 \\ 0 & 0 \end{bmatrix} & \begin{bmatrix} 0 & 2 \\ -2 & 0 \end{bmatrix} & \begin{bmatrix} 0 & 0 \\ 0 & 0 \end{bmatrix} \\ \begin{bmatrix} -2 & 0 \\ 0 & -2 \end{bmatrix} & \begin{bmatrix} 0 & 0 \\ 0 & 0 \end{bmatrix} & \begin{bmatrix} 4 & 0 \\ 0 & 4 \end{bmatrix} & \begin{bmatrix} 0 & 0 \\ 0 & 0 \end{bmatrix} & \begin{bmatrix} 1 & 0 \\ 0 & -1 \end{bmatrix} \\ \begin{bmatrix} 0 & 2 \\ -2 & 0 \end{bmatrix} & \begin{bmatrix} 0 & 0 \\ 0 & 0 \end{bmatrix} & \begin{bmatrix} 0 & 4 \\ -4 & 0 \end{bmatrix} & \begin{bmatrix} 0 & 0 \\ 0 & 0 \end{bmatrix} & \begin{bmatrix} 0 & -1 \\ -1 & 0 \end{bmatrix} \\ \begin{bmatrix} 0 & 0 \\ 0 & 0 \end{bmatrix} & \begin{bmatrix} -2 & 0 \\ 0 & -2 \end{bmatrix} & \begin{bmatrix} 2 & 0 \\ 0 & -2 \end{bmatrix} & \begin{bmatrix} 0 & 0 \\ 0 & 0 \end{bmatrix} & \begin{bmatrix} 2 & 0 \\ 0 & 2 \end{bmatrix} \\ \begin{bmatrix} 0 & 0 \\ 0 & 0 \end{bmatrix} & \begin{bmatrix} 0 & 2 \\ -2 & 0 \end{bmatrix} & \begin{bmatrix} 0 & -2 \\ -2 & 0 \end{bmatrix} & \begin{bmatrix} 0 & 0 \\ 0 & 0 \end{bmatrix} & \begin{bmatrix} 0 & 2 \\ -2 & 0 \end{bmatrix} \end{bmatrix}. \tag{5.58}$$

The function blkdiag$\{M_1, \dots, M_n\}$ arranges the matrices $M_1, \dots, M_n$ into a block diagonal form (also known as a *direct sum*). Each 2×2 block on **J**'s diagonal is a generalized "eigenvalue" that can be assigned to $R^{\{1\}}$ and corresponds to each column block in **V** that is a generalized "eigenvector" that can be assigned to $\widetilde{\mathbf{P}}$. In this particular case, each 2×2 block on **J**'s diagonal is a matrix representation of the eigenvalues $\lambda_k = -a_k i$, (e.g., $J_5 = \begin{bmatrix} 0 & 1 \\ -1 & 0 \end{bmatrix} \leftrightarrow \lambda_5 = -i$). Since we have no prior information regarding $R^{\{1\}}$ [in contrast with $R^{\{0\}}$, which can be obtained from the eigenvalues of $A(t|\omega = 0)$], we need to examine each block generalized eigenvalue and the corresponding generalized eigenvector. To simplify the notation, we use:

$$\begin{gathered} \mathbf{J} = \text{blkdiag}\{J_1, J_2, \dots, J_5\}, \\ \mathbf{V} = [V_1, V_2, \dots V_5]\,, \end{gathered} \tag{5.59}$$

where $J_k$ is a generalized eigenvalue corresponding to a compatible generalized eigenvector. If we set $R^{\{1\}} \coloneqq J_5 = \begin{bmatrix} 0 & 1 \\ -1 & 0 \end{bmatrix}$ and $\widetilde{\mathbf{P}} = V_5$, then a potential solution for $P(t)$ is:

$$P(t) = \begin{bmatrix} 2\cos(2\omega t) + \cos(\omega t) & 2\sin(2\omega t) - \sin(\omega t) \\ -2\sin(2\omega t) - \sin(\omega t) & 2\cos(2\omega t) - \cos(\omega t) \end{bmatrix}. \tag{5.60}$$

It is easy to check that $\det\{P(t)\} \equiv 3 = \text{const.} \neq 0$. In addition, by inserting our guess for $P(t)$ into the RHS of Eq. (1.18), we can check whether the value for the constant part $R$ is valid:

$$R = P^{-1}(t)\big[A(t)P(t) - \dot{P}(t)\big] = \begin{bmatrix} 1 & \omega - 1 \\ 1 - \omega & -1 \end{bmatrix}. \tag{5.61}$$

We thus have a potential solution for the constant part $R$, which shows that $R$ is not a function of $t$. This solution for $R$ can be decomposed into the following sum:

$$R = R^{\{0\}} + \omega R^{\{1\}},$$

$$\begin{gathered} R = R^{\{0\}} + \omega R^{\{1\}}, \\ R^{\{1\}} = \begin{bmatrix} 0 & 1 \\ -1 & 0 \end{bmatrix}, \\ R^{\{0\}} = \begin{bmatrix} 1 & -1 \\ 1 & -1 \end{bmatrix} = \begin{bmatrix} 1 & 1 \\ 1 & 0 \end{bmatrix} \begin{bmatrix} 0 & 1 \\ 0 & 0 \end{bmatrix} \begin{bmatrix} 1 & 1 \\ 1 & 0 \end{bmatrix}^{-1}. \end{gathered} \tag{5.62}$$

Note that $R^{\{1\}}$ in this equation is identical to $R^{\{1\}}$ derived from Eqs. (5.58)–(5.60). In addition, it is easily verified that $R^{\{0\}}$ and $A(t|\omega = 0)$ are similar matrices because they share the same Jordan block matrix $\begin{bmatrix} 0 & 1 \\ 0 & 0 \end{bmatrix}$, which is similar to each matrix $R^{\{0\}}$ and $A(t|\omega = 0)$ individually.

To summarize, we conclude that a solution to construct the transition matrix $\Phi_A(t, t_0)$ of the matrix $A(t)$ representing the LPTV system is formed from the matrices $P(t)$ and $R$, where:

$$\begin{gathered} P(t) = \begin{bmatrix} 2\cos(2\omega t) + \cos(\omega t) & 2\sin(2\omega t) - \sin(\omega t) \\ -2\sin(2\omega t) - \sin(\omega t) & 2\cos(2\omega t) - \cos(\omega t) \end{bmatrix}, \\ R = \begin{bmatrix} 1 & \omega - 1 \\ 1 - \omega & -1 \end{bmatrix}. \end{gathered} \tag{5.63}$$

This solution is consistent with the data given in row E of **Table 4-4**.

■

**<u>Example 5.5</u>: Row F of Table 4-4**

Consider the following real 2×2 matrix for an LPTV system $(n = 2)$:

$$\begin{aligned} A(t) = &\begin{bmatrix} \frac{5a}{4}\cos(\omega t) - \frac{a}{4}\cos(3\omega t) - a\sin(\omega t) & \frac{13a}{8} + \frac{a}{2}\cos(2\omega t) - \frac{a}{8}\cos(4\omega t) - a\sin(2\omega t) \\ -\frac{3a}{2} - \frac{a}{2}\cos(2\omega t) & -\frac{5a}{4}\cos(\omega t) + \frac{a}{4}\cos(3\omega t) + a\sin(\omega t) \end{bmatrix}, \\ &+\omega \begin{bmatrix} -\frac{1}{2} - \frac{1}{2}\cos(2\omega t) & -\frac{3}{4}\cos(\omega t) - \frac{1}{4}\cos(3\omega t) + \sin(\omega t) \\ \cos(\omega t) & \frac{1}{2} + \frac{1}{2}\cos(2\omega t) \end{bmatrix}, \end{aligned} \tag{5.64}$$

which has $L = 4$ harmonics. By writing $A(t) = A^{\{0\}}(t) + \omega A^{\{1\}}(t)$, we obtain:

$$\begin{gathered}
A^{\{0\}}(t) = \begin{bmatrix} \frac{5a}{4}\cos(\omega t) - \frac{a}{4}\cos(3\omega t) - a\sin(\omega t) & \frac{13a}{8} + \frac{a}{2}\cos(2\omega t) - \frac{a}{8}\cos(4\omega t) - a\sin(2\omega t) \\ -\frac{3a}{2} - \frac{a}{2}\cos(2\omega t) & -\frac{5a}{4}\cos(\omega t) + \frac{a}{4}\cos(3\omega t) + a\sin(\omega t) \end{bmatrix}, \\
A^{\{1\}}(t) = \begin{bmatrix} -\frac{1}{2} - \frac{1}{2}\cos(2\omega t) & -\frac{3}{4}\cos(\omega t) - \frac{1}{4}\cos(3\omega t) + \sin(\omega t) \\ \cos(\omega t) & \frac{1}{2} + \frac{1}{2}\cos(2\omega t) \end{bmatrix}.
\end{gathered} \tag{5.65}$$

Note that $\text{trace}\{A(t)\} \equiv 0$ and $\text{trace}\{A^{\{0\}}(t)\} \equiv \text{trace}\{A^{\{1\}}(t)\} \equiv 0$, so it is not necessary to zero the trace. In addition, the matrix $A(t)$ evaluated at $\omega = 0$,

$$A(t|\omega = 0) = \begin{bmatrix} a & 2a \\ -a & -a \end{bmatrix}, \tag{5.66}$$

is decomposed as follows:

$$A(t|\omega = 0) = UJU^{-1}, \tag{5.67}$$

where

$$\begin{gathered}
J = \begin{bmatrix} 0 & a \\ -a & 0 \end{bmatrix}, \\
U = \begin{bmatrix} -1 & -1 \\ 1 & 0 \end{bmatrix}.
\end{gathered} \tag{5.68}$$

To recast the matrix, we use $A_F(t) \coloneqq A(t)$, $A_G(t) \coloneqq U^{-1}A_F(t)U$ to obtain

$$A_G(t) = A_G^{\{0\}}(t) + \omega A_G^{\{1\}}(t), \tag{5.69}$$

with

$$\begin{gathered}
A_G^{\{0\}}(t) = \begin{bmatrix} \frac{3a}{2} - \frac{5a}{4}\cos(\omega t) - \frac{a}{2}\cos(2\omega t) + \frac{a}{4}\cos(3\omega t) + a\sin(\omega t) & \frac{3a}{2} - \frac{a}{2}\cos(2\omega t) \\ -\frac{25a}{8} + \frac{5a}{2}\cos(\omega t) - \frac{a}{2}\cos(3\omega t) + \frac{a}{8}\cos(4\omega t) - 2a\sin(\omega t) + a\sin(2\omega t) & -\frac{3a}{2} + \frac{5a}{4}\cos(\omega t) + \frac{a}{2}\cos(2\omega t) - \frac{a}{4}\cos(3\omega t) - a\sin(\omega t) \end{bmatrix}, \\
A_G^{\{1\}}(t) = \begin{bmatrix} \frac{1}{2} - \cos(\omega t) + \frac{1}{2}\cos(2\omega t) & -\cos(\omega t) \\ -1 + \frac{7}{4}\cos(\omega t) - 2\cos(2\omega t) + \frac{1}{4}\cos(3\omega t) - \sin(\omega t) & -\frac{1}{2} + \cos(\omega t) - \frac{1}{2}\cos(2\omega t) \end{bmatrix}.
\end{gathered} \tag{5.70}$$

Note that $A_G(t)$ is exactly the LPTV system matrix defined in row G of **Table 4-4**, and that $\mathrm{trace}\{A_G(t)\} \equiv 0$ and $A_G(t|\omega = 0) = J$. We now solve for $A_G(t)$ and, to simplify the problem, we omit the subscript $G$ [e.g., $A(t) \coloneqq A_G(t)$].

We assume the transition matrix is decomposed according to *Floquet theory* into a periodic part $P(t) = P\left(t + \frac{2\pi}{\omega}\right)$ and a constant part $R = R^{\{0\}} + \omega R^{\{1\}}$. Since $A(t)$ has $L = 4$ harmonics and is of size $n = 2$, we hypothesize that $P(t)$ has $p = \left\lceil \frac{4}{2} \right\rceil = 2$ harmonics. After expressing $A(t)$ in terms of its Fourier coefficients ${A^{\{r\}}}_k^{\text{even}}$ and ${A^{\{r\}}}_k^{\text{odd}}$, we use Eqs. (2.22)–(2.25) and (5.11) to compare powers of $\omega$, which gives the following algebraic matrix-vector equation for $\omega^0$

$$\widetilde{\mathbf{A}}^{\{0\}}\widetilde{\mathbf{P}} = \widetilde{\mathbf{P}}R^{\{0\}}, \tag{5.71}$$

with

$$\widetilde{\mathbf{A}}^{\{0\}} = \frac{a}{16}\begin{bmatrix} \begin{bmatrix} 24 & 24 \\ -50 & -24 \end{bmatrix} & \begin{bmatrix} -20 & 0 \\ 40 & 20 \end{bmatrix} & \begin{bmatrix} 16 & 0 \\ -32 & -16 \end{bmatrix} & \begin{bmatrix} -8 & -8 \\ 0 & 8 \end{bmatrix} & \begin{bmatrix} 0 & 0 \\ 16 & 0 \end{bmatrix} \\ \begin{bmatrix} -10 & 0 \\ 20 & 10 \end{bmatrix} & \begin{bmatrix} 20 & 20 \\ -50 & -20 \end{bmatrix} & \begin{bmatrix} 0 & 0 \\ 8 & 0 \end{bmatrix} & \begin{bmatrix} -8 & 0 \\ 16 & 8 \end{bmatrix} & \begin{bmatrix} 8 & 0 \\ -16 & -8 \end{bmatrix} \\ \begin{bmatrix} 8 & 0 \\ -16 & -8 \end{bmatrix} & \begin{bmatrix} 0 & 0 \\ 8 & 0 \end{bmatrix} & \begin{bmatrix} 28 & 28 \\ -50 & -28 \end{bmatrix} & \begin{bmatrix} 10 & 0 \\ -20 & -10 \end{bmatrix} & \begin{bmatrix} -12 & 0 \\ 24 & 12 \end{bmatrix} \\ \begin{bmatrix} -4 & -4 \\ 0 & 4 \end{bmatrix} & \begin{bmatrix} -8 & 0 \\ 16 & 8 \end{bmatrix} & \begin{bmatrix} -8 & 0 \\ 16 & 8 \end{bmatrix} & \begin{bmatrix} 24 & 24 \\ -49 & -24 \end{bmatrix} & \begin{bmatrix} 0 & 0 \\ 0 & 0 \end{bmatrix} \\ \begin{bmatrix} 0 & 0 \\ 8 & 0 \end{bmatrix} & \begin{bmatrix} 8 & 0 \\ -16 & -8 \end{bmatrix} & \begin{bmatrix} -12 & 0 \\ 24 & 12 \end{bmatrix} & \begin{bmatrix} 0 & 0 \\ 0 & 0 \end{bmatrix} & \begin{bmatrix} 24 & 24 \\ -51 & -24 \end{bmatrix} \end{bmatrix}, \tag{5.72}$$

$$\widetilde{\mathbf{P}} = \begin{bmatrix} P_0^{\text{even}} \\ P_1^{\text{even}} \\ P_1^{\text{odd}} \\ P_2^{\text{even}} \\ P_2^{\text{odd}} \end{bmatrix}, \tag{5.73}$$

where $R^{\{0\}}$ and the terms $P_k^{\text{even}}$ and $P_k^{\text{odd}}$ are unknown and must be found. Note that $\tilde{A}$ has eigenvalues $\lambda = \pm ai$ that can be represented by the 2×2 matrix $J = \begin{bmatrix} 0 & a \\ -a & 0 \end{bmatrix}$.

By taking

$$R^{\{0\}} = \begin{bmatrix} 0 & a \\ -a & 0 \end{bmatrix}, \tag{5.74}$$

the terms $P_k^{\text{even}}$ and $P_k^{\text{odd}}$ must be defined by the corresponding real and imaginary parts of the eigenvector $\mathrm{v}_{-ai}$, so $P(t|\omega = 0) = P(0) = cI_2$ and $\det\{P(t)\} \equiv \text{const.} \neq 0$.

By setting

$$\widetilde{\mathbf{P}} = [\operatorname{Im}(\mathrm{v}_{-ai}), \operatorname{Re}(\mathrm{v}_{-ai})] = \begin{bmatrix} \begin{bmatrix} -4 & 0 \\ 4 & -4 \end{bmatrix} \\ \begin{bmatrix} 0 & 0 \\ -2 & 0 \end{bmatrix} \\ \begin{bmatrix} 0 & 2 \\ 0 & -2 \end{bmatrix} \\ \begin{bmatrix} 0 & 0 \\ 0 & 0 \end{bmatrix} \\ \begin{bmatrix} 0 & 0 \\ 0 & 1 \end{bmatrix} \end{bmatrix}, \tag{5.75}$$

Eq. (5.72) is satisfied. Let $P_k^{\text{even}}$ and $P_k^{\text{odd}}$ be the blocks for $\widetilde{\mathbf{P}}$ as follows:

$$\frac{P_0^{\text{even}}}{2} = \begin{bmatrix} -2 & 0 \\ 2 & -2 \end{bmatrix}, \quad P_1^{\text{even}} = \begin{bmatrix} 0 & 0 \\ -2 & 0 \end{bmatrix}, \quad P_1^{\text{odd}} = \begin{bmatrix} 0 & 2 \\ 0 & -2 \end{bmatrix},$$
$$P_2^{\text{even}} = \begin{bmatrix} 0 & 0 \\ 0 & 0 \end{bmatrix}, \quad P_2^{\text{odd}} = \begin{bmatrix} 0 & 0 \\ 0 & 1 \end{bmatrix}. \tag{5.76}$$

This implies that:

$$P(t) = \begin{bmatrix} -2 & 2\sin(\omega t) \\ 2 - 2\cos(\omega t) & -2 - 2\sin(1\omega t) + 1\sin(2\omega t) \end{bmatrix}. \tag{5.77}$$

Note that $P(t|\omega = 0) = -2I_2$ and $\det\{P(t)\} \equiv 4 \equiv \text{const.} \neq 0$ , which indicates that our choice for $P(t)$ is correct. Since $P(t)$ can be defined up to a scale factor, we can, without loss of generality, multiply our solution for $P(t)$ by −1/2 to normalize the result so that $P(t|\omega = 0) = I_2$, i.e.,

$$P(t) = \begin{bmatrix} 1 & -\sin(\omega t) \\ -1 + \cos(\omega t) & 1 + \sin(\omega t) - \frac{1}{2}\sin(2\omega t) \end{bmatrix}. \tag{5.78}$$

We now insert our guess for $P(t)$ into the RHS of Eq. (1.18) to check if the constant part $R$ is valid. The result is:

$$R = P^{-1}(t)(A(t)P(t) - \dot{P}(t)) = \begin{bmatrix} 0 & a \\ -a & 0 \end{bmatrix}. \tag{5.79}$$

We thus conclude that the matrices $P_g(t)\ and\ R_G$ form a solution to construct the transition matrix $\Phi_{A_G}(t, t_0)$ for the matrix $A_G(t)$ of the LPTV system, with:

$$P_g(t) = \begin{bmatrix} 1 & -\sin(\omega t) \\ -1 + \cos(\omega t) & 1 + \sin(\omega t) - \frac{1}{2}\sin(2\omega t) \end{bmatrix},$$
$$R_g = \begin{bmatrix} 0 & a \\ -a & 0 \end{bmatrix}. \tag{5.80}$$

This solution is consistent with the data given in row G of **Table 4-4**. To calculate the transition matrix $\Phi_{A_F}(t, t_0)$ of the original example, we use Eq. (5.28), which gives:

$$P_F(t) = UP_g(t) \text{ and } R_F = R_G. \tag{5.81}$$

Since $P_F(t)$ is defined up to a scale factor, we can multiply $P_F(t)$ by −1 so that our final result is the pair of matrices $P(t) \coloneqq -P_F(t)$ and $R \coloneqq R_F$ for the LPTV system $A(t) \coloneqq A_F(t)$ such that:

$$P(t) = \begin{bmatrix} \cos(\omega t) & 1 - \frac{1}{2}\sin(2\omega t) \\ -1 & \sin(\omega t) \end{bmatrix}, \quad R = \begin{bmatrix} 0 & a \\ -a & 0 \end{bmatrix}, \tag{5.82}$$

which is consistent with the data given in row F of **Table 4-4**.

■

**<u>Example 5.6</u>: A 3×3 example**

Consider the following real matrix for an LPTV system:

$$A(t) = A^{\{0\}}(t) + \omega A^{\{1\}}(t) = \tag{5.83}$$

$$= \frac{1}{8}\left(\begin{bmatrix} 72 & -81 & -3 \\ 396 & 169 & -147 \\ -572 & 249 & -241 \end{bmatrix} + \begin{bmatrix} 434 & -84 & -286 \\ 256 & 156 & -248 \\ 434 & -216 & -590 \end{bmatrix} \mathrm{c}_{1\omega t} + \begin{bmatrix} 256 & -564 & 236 \\ 742 & -360 & 1034 \\ 256 & -1780 & 104 \end{bmatrix} \mathrm{s}_{1\omega t} + \begin{bmatrix} 248 & -36 & -132 \\ 396 & 836 & -396 \\ 572 & 0 & 558 \end{bmatrix} \mathrm{c}_{2\omega t} + \begin{bmatrix} 192 & 278 & 138 \\ 572 & -294 & 426 \\ -396 & 998 & 102 \end{bmatrix} \mathrm{s}_{2\omega t}\right.$$
$$+ \begin{bmatrix} -154 & -108 & -223 \\ -256 & -79 & -232 \\ -154 & 24 & 233 \end{bmatrix} \mathrm{c}_{3\omega t} + \begin{bmatrix} 256 & -223 & 108 \\ -154 & -232 & 79 \\ 256 & 233 & -24 \end{bmatrix} \mathrm{s}_{3\omega t} + \begin{bmatrix} 0 & 45 & -329 \\ 0 & 491 & -249 \\ 0 & -249 & -491 \end{bmatrix} \mathrm{c}_{4\omega t} + \begin{bmatrix} 0 & -329 & -45 \\ 0 & -249 & -491 \\ 0 & -491 & 249 \end{bmatrix} \mathrm{s}_{4\omega t} + \begin{bmatrix} 0 & 128 & 77 \\ 0 & -77 & 128 \\ 0 & 128 & 77 \end{bmatrix} \mathrm{c}_{5\omega t} + \left.\begin{bmatrix} 0 & 77 & -128 \\ 0 & 128 & 77 \\ 0 & 77 & 128 \end{bmatrix} \mathrm{s}_{5\omega t}\right)$$

$$+ \frac{\omega}{8}\left(\begin{bmatrix} -12 & 4 & 28 \\ 4 & 0 & 8 \\ 12 & -4 & 12 \end{bmatrix} + \begin{bmatrix} 8 & 8 & -24 \\ -4 & 0 & 20 \\ 8 & 8 & -8 \end{bmatrix} \mathrm{c}_{1\omega t} + \begin{bmatrix} -4 & 0 & -12 \\ -8 & 16 & -24 \\ -4 & 16 & -12 \end{bmatrix} \mathrm{s}_{1\omega t} + \begin{bmatrix} -12 & 8 & 8 \\ 4 & 12 & 4 \\ -12 & 0 & 0 \end{bmatrix} \mathrm{c}_{2\omega t} + \begin{bmatrix} -4 & -4 & -4 \\ -12 & 0 & 0 \\ -4 & -12 & 4 \end{bmatrix} \mathrm{s}_{2\omega t}\right.$$
$$+ \begin{bmatrix} 8 & 10 & -4 \\ 4 & -4 & 14 \\ 18 & 10 & -4 \end{bmatrix} \mathrm{c}_{3\omega t} + \begin{bmatrix} -4 & -4 & -10 \\ 8 & 14 & 4 \\ -4 & -4 & -10 \end{bmatrix} \mathrm{s}_{3\omega t} + \begin{bmatrix} 0 & -4 & 12 \\ 0 & -12 & -4 \\ 0 & -4 & 12 \end{bmatrix} \mathrm{c}_{4\omega t} + \begin{bmatrix} 0 & 12 & 4 \\ 0 & -4 & 12 \\ 0 & 12 & 4 \end{bmatrix} \mathrm{s}_{4\omega t} + \begin{bmatrix} 0 & -2 & -4 \\ 0 & 4 & -2 \\ 0 & -2 & -4 \end{bmatrix} \mathrm{c}_{5\omega t} + \left.\begin{bmatrix} 0 & -4 & 2 \\ 0 & -2 & 4 \\ 0 & -4 & 2 \end{bmatrix} \mathrm{s}_{5\omega t}\right),$$

where $\mathrm{c}_\theta = \cos(\theta)$, $\mathrm{s}_\theta = \sin(\theta)$, and $A^{\{0\}}(t)$ and $A^{\{1\}}(t)$ are $\frac{2\pi}{\omega}$-periodic 3×3 matrices ($n = 3$) with five harmonics ($L = 5$) given by the following Fourier series: For $A^{\{0\}}(t)$,

$$A^{\{0\}}(t) = \frac{A^{\{0\}}{}_0^{\text{even}}}{2} + \sum_{l=1}^{5}\left(A^{\{0\}}{}_l^{\text{even}}\cos(l\omega t) + A^{\{0\}}{}_l^{\text{odd}}\sin(l\omega t)\right)$$

$$A^{\{0\}}{}_1^{\text{even}} = \frac{1}{8}\begin{bmatrix} 434 & -84 & -286 \\ 256 & 156 & -248 \\ 434 & -216 & -590 \end{bmatrix}; A^{\{0\}}{}_1^{\text{odd}} = \frac{1}{8}\begin{bmatrix} 256 & -564 & 236 \\ 742 & -360 & 1034 \\ 256 & -1780 & 104 \end{bmatrix}; A^{\{0\}}{}_2^{\text{even}} = \frac{1}{8}\begin{bmatrix} 248 & -36 & -132 \\ 396 & 836 & -396 \\ 572 & 0 & 558 \end{bmatrix}; A^{\{0\}}{}_2^{\text{odd}} = \frac{1}{8}\begin{bmatrix} 192 & 278 & 138 \\ 572 & -294 & 426 \\ -396 & 998 & 102 \end{bmatrix};$$
$$A^{\{0\}}{}_3^{\text{even}} = \frac{1}{8}\begin{bmatrix} -154 & -108 & -223 \\ -256 & -79 & -232 \\ -154 & 24 & 233 \end{bmatrix}; A^{\{0\}}{}_3^{\text{odd}} = \frac{1}{8}\begin{bmatrix} 256 & -223 & 108 \\ -154 & -232 & 79 \\ 256 & 233 & -24 \end{bmatrix}; A^{\{0\}}{}_4^{\text{even}} = \frac{1}{8}\begin{bmatrix} 0 & 45 & -329 \\ 0 & 491 & -249 \\ 0 & -249 & -491 \end{bmatrix}; A^{\{0\}}{}_4^{\text{odd}} = \frac{1}{8}\begin{bmatrix} 0 & -329 & -45 \\ 0 & -249 & -491 \\ 0 & -491 & 249 \end{bmatrix}; \tag{5.84}$$
$$A^{\{0\}}{}_5^{\text{even}} = \frac{1}{8}\begin{bmatrix} 0 & 128 & 77 \\ 0 & -77 & 128 \\ 0 & 128 & 77 \end{bmatrix}; A^{\{0\}}{}_5^{\text{odd}} = \frac{1}{8}\begin{bmatrix} 0 & 77 & -128 \\ 0 & 128 & 77 \\ 0 & 77 & 128 \end{bmatrix}; \frac{A^{\{0\}}{}_0^{\text{even}}}{2} = \frac{1}{8}\begin{bmatrix} 72 & -81 & -3 \\ 396 & 169 & -147 \\ -572 & 249 & -241 \end{bmatrix}.$$

For $A^{\{1\}}(t)$,

$$A^{\{1\}}(t) = \frac{A^{\{1\}}{}_0^{\text{even}}}{2} + \sum_{l=1}^{5}\left(A^{\{1\}}{}_l^{\text{even}}\cos(l\omega t) + A^{\{1\}}{}_l^{\text{odd}}\sin(l\omega t)\right)$$
$$A^{\{1\}}{}_1^{\text{even}} = \frac{1}{8}\begin{bmatrix} 8 & 8 & -24 \\ -4 & 0 & 20 \\ 8 & 8 & -8 \end{bmatrix}; A^{\{1\}}{}_1^{\text{odd}} = \frac{1}{8}\begin{bmatrix} -4 & 0 & -12 \\ -8 & 16 & -24 \\ -4 & 16 & -12 \end{bmatrix}; A^{\{1\}}{}_2^{\text{even}} = \frac{1}{8}\begin{bmatrix} -12 & 8 & 8 \\ 4 & 12 & 4 \\ -12 & 0 & 0 \end{bmatrix}; A^{\{1\}}{}_2^{\text{odd}} = \frac{1}{8}\begin{bmatrix} -4 & -4 & -4 \\ -12 & 0 & 0 \\ -4 & -12 & 4 \end{bmatrix};$$
$$A^{\{1\}}{}_3^{\text{even}} = \frac{1}{8}\begin{bmatrix} 8 & 10 & -4 \\ 4 & -4 & 14 \\ 18 & 10 & -4 \end{bmatrix}; A^{\{1\}}{}_3^{\text{odd}} = \frac{1}{8}\begin{bmatrix} -4 & -4 & -10 \\ 8 & 14 & 4 \\ -4 & -4 & -10 \end{bmatrix}; A^{\{1\}}{}_4^{\text{even}} = \frac{1}{8}\begin{bmatrix} 0 & -4 & 12 \\ 0 & -12 & -4 \\ 0 & -4 & 12 \end{bmatrix}; A^{\{1\}}{}_4^{\text{odd}} = \frac{1}{8}\begin{bmatrix} 0 & 12 & 4 \\ 0 & -4 & 12 \\ 0 & 12 & 4 \end{bmatrix}; \tag{5.85}$$
$$A^{\{1\}}{}_5^{\text{even}} = \frac{1}{8}\begin{bmatrix} 0 & -2 & -4 \\ 0 & 4 & -2 \\ 0 & -2 & -4 \end{bmatrix}; A^{\{1\}}{}_5^{\text{odd}} = \frac{1}{8}\begin{bmatrix} 0 & -4 & 2 \\ 0 & -2 & 4 \\ 0 & -4 & 2 \end{bmatrix}; \frac{A^{\{1\}}{}_0^{\text{even}}}{2} = \frac{1}{8}\begin{bmatrix} -12 & 4 & 28 \\ 4 & 0 & 8 \\ 12 & -4 & 12 \end{bmatrix}.$$

For simplicity, we present all $A^{\{r\}}{}_l^{\text{even}}$ and $A^{\{r\}}{}_l^{\text{odd}}$ with the common denominator of eight, so we did not simplify the fractions. Observe that $\text{trace}\{A(t)\} \equiv 0$ and, in addition, $A(t)$ evaluates at $\omega = 0$ to:

$$A(t|\omega = 0) = \begin{bmatrix} 75 & -17 & -112 \\ 99 & -22 & -143 \\ 35 & -8 & -53 \end{bmatrix}. \tag{5.86}$$

This constant matrix $A(t|\omega = 0)$ has the following eigen decomposition:

$$V^{-1}A(t|\omega = 0)V = J\ ,$$
$$V = \begin{bmatrix} 22 & 75 & 1 \\ 242 & 99 & 0 \\ -22 & 35 & 0 \end{bmatrix},\ J = \begin{bmatrix} 0 & 1 & 0 \\ 0 & 0 & 1 \\ 0 & 0 & 0 \end{bmatrix}, \tag{5.87}$$

where $J$ is the Jordan block corresponding to the eigenvalues $\{0,0,0\}$, and $V$ is the corresponding matrix of eigenvectors of $J$. We can find matrices $P(t) = P\left(t + \frac{2\pi}{\omega}\right)$ and $R = R^{\{0\}} + \omega R^{\{1\}}$ for $A(t)$ [see Eq. (1.16)], where:

$$A(t)P(t) = \dot{P}(t) + P(t)R. \tag{5.88}$$

Note that multiplying from the right by $V$ and using the associative property of matrix multiplication gives:

$$A(t)[P(t)V] = \frac{d}{dt}[P(t)V] + [P(t)V](V^{-1}RV). \tag{5.89}$$

We can find an equivalent pair for the solution, $\tilde{P}(t) = P(t)V$ and $\tilde{R} = V^{-1}RV$. For $\omega = 0$, we take $R^{\{0\}} = A(t|\omega = 0)$, $\tilde{R}^{\{0\}} = J$, and $P(t|\omega = 0) \equiv I_3$. In addition, we must have $\det\{P(t)\} \equiv \text{const.} \neq 0$ and $\text{trace}\{R\} \equiv 0$ [and $\det\{\tilde{P}(t)\} \equiv \text{const.} \neq 0$ and $\text{trace}\{\tilde{R}\} \equiv 0$]. Since $A(t)$ has $L = 5$ harmonics and is of size $n = 3$, we hypothesize that the number of the harmonics of $P(t)$ is $p = \left\lceil \frac{5}{3} \right\rceil = 2$. After expressing $A(t)$ in terms of its Fourier coefficients ${A^{\{r\}}}_k^{\text{even}}$ and ${A^{\{r\}}}_k^{\text{odd}}$ and in powers of ω, we use Eqs. (2.22)–(2.25) and (5.11) to compare powers of $\omega$, obtaining the following algebraic matrix-vector equation for $\omega^0$:

$$\widetilde{\mathbf{A}}^{\{0\}}\widetilde{\mathbf{P}} = \widetilde{\mathbf{P}}\tilde{R}^{\{0\}}, \tag{5.90}$$

with

$$\widetilde{\mathbf{A}}^{\{0\}} = \frac{1}{16}\begin{bmatrix}
\begin{bmatrix} 144 & -162 & -6 \\ 792 & 338 & -294 \\ -1144 & 498 & 482 \end{bmatrix} & \begin{bmatrix} 868 & -168 & -572 \\ 512 & 312 & -496 \\ 868 & -432 & -1180 \end{bmatrix} & \begin{bmatrix} 512 & -1128 & 472 \\ 1484 & -720 & 2068 \\ 512 & -3560 & 208 \end{bmatrix} & \begin{bmatrix} 496 & -72 & -264 \\ 792 & -1672 & -792 \\ 1144 & 0 & 1176 \end{bmatrix} & \begin{bmatrix} 384 & 556 & 276 \\ 1144 & -588 & 852 \\ -792 & 1996 & 204 \end{bmatrix} \\
\begin{bmatrix} 434 & -84 & -286 \\ 256 & 156 & -248 \\ 434 & -216 & -590 \end{bmatrix} & \begin{bmatrix} 392 & 198 & -138 \\ 1188 & -498 & -690 \\ -572 & 498 & 106 \end{bmatrix} & \begin{bmatrix} 192 & 278 & 138 \\ 572 & -294 & 426 \\ -396 & 998 & 102 \end{bmatrix} & \begin{bmatrix} 280 & -192 & -509 \\ 0 & 77 & -480 \\ 280 & -192 & -357 \end{bmatrix} & \begin{bmatrix} 512 & -787 & 344 \\ 588 & -592 & 1113 \\ 512 & -1547 & 80 \end{bmatrix} \\
\begin{bmatrix} 256 & -564 & 236 \\ 742 & -360 & 1034 \\ 256 & -1780 & 104 \end{bmatrix} & \begin{bmatrix} 192 & 278 & 138 \\ 572 & -294 & 426 \\ -396 & 998 & 102 \end{bmatrix} & \begin{bmatrix} -104 & -126 & 126 \\ 396 & 1174 & 102 \\ -1716 & 498 & -1070 \end{bmatrix} & \begin{bmatrix} 0 & 341 & -128 \\ -896 & 128 & -955 \\ 0 & 2013 & -128 \end{bmatrix} & \begin{bmatrix} 588 & 24 & -63 \\ 512 & 235 & -16 \\ 588 & -240 & -823 \end{bmatrix} \\
\begin{bmatrix} 248 & -36 & -132 \\ 396 & -836 & -396 \\ 572 & 0 & 558 \end{bmatrix} & \begin{bmatrix} 280 & -192 & -509 \\ 0 & 77 & -480 \\ 280 & -192 & -357 \end{bmatrix} & \begin{bmatrix} 0 & 341 & -128 \\ -896 & 128 & -955 \\ 0 & 2013 & -128 \end{bmatrix} & \begin{bmatrix} 144 & -117 & -335 \\ 792 & 829 & -543 \\ -1144 & 249 & -973 \end{bmatrix} & \begin{bmatrix} 0 & -329 & -45 \\ 0 & -249 & -491 \\ 0 & -491 & 249 \end{bmatrix} \\
\begin{bmatrix} 192 & 278 & 138 \\ 572 & -294 & 426 \\ -396 & 998 & 102 \end{bmatrix} & \begin{bmatrix} 512 & -787 & 344 \\ 588 & -592 & 1113 \\ 512 & -1547 & 80 \end{bmatrix} & \begin{bmatrix} 588 & 24 & -63 \\ 512 & 235 & -16 \\ 588 & -240 & -823 \end{bmatrix} & \begin{bmatrix} 0 & -329 & -45 \\ 0 & -249 & -491 \\ 0 & -491 & 249 \end{bmatrix} & \begin{bmatrix} 144 & -207 & 323 \\ 792 & -153 & -45 \\ -1144 & 747 & 9 \end{bmatrix}
\end{bmatrix}, \tag{5.91}$$

$$\widetilde{\mathbf{P}} = \begin{bmatrix} P_0^{\text{even}} V \\ P_1^{\text{even}} V \\ P_1^{\text{odd}} V \\ P_2^{\text{even}} V \\ P_2^{\text{odd}} V \end{bmatrix}. \tag{5.92}$$

The matrix $\widetilde{\mathbf{A}}^{\{0\}}$ can be decomposed by using the following eigen decomposition:

$$\widetilde{\mathbf{A}}^{\{0\}} \underline{\underline{\mathbf{V}}} = \underline{\underline{\mathbf{V}}}\,\underline{\underline{\mathbf{J}}}\,, \tag{5.93}$$

where

$$\underline{\underline{\mathbf{J}}} = \begin{bmatrix} J & * \\ 0 & * \end{bmatrix}, \tag{5.94}$$

$$\underline{\underline{\mathbf{V}}} = \begin{bmatrix} \begin{bmatrix} 0 & 0 & 1 \\ 1 & 0 & 0 \\ 0 & 0 & 1 \end{bmatrix} V & \\ \begin{bmatrix} 1 & 0 & -1 \\ 0 & 1 & 0 \\ 0 & 0 & 0 \end{bmatrix} V & \\ \begin{bmatrix} -1 & 1 & -1 \\ 0 & 0 & 0 \\ 0 & -1 & 0 \end{bmatrix} V & * \ldots \\ \begin{bmatrix} 0 & 0 & 0.5 \\ -0.5 & 0 & 0 \\ 0 & 0 & 0.5 \end{bmatrix} V & \\ \begin{bmatrix} 0.5 & 0 & 0 \\ 0 & 0 & 0.5 \\ 0.5 & 0 & 0 \end{bmatrix} V & \end{bmatrix}. \tag{5.95}$$

We only consider eigenvectors corresponding to eigenvalues $\lambda = \{0,0,0\}$ that are compatible with the Jordan block $J$. If we set $\widetilde{\mathbf{P}}$ such that $\underline{\underline{\mathbf{V}}} = [\widetilde{\mathbf{P}} \ldots *]$, then we have the following solution for $P(t)$:

$$P(t)=\begin{bmatrix}\cos(\omega t)-\sin(\omega t)+\frac{1}{2}\sin(2\omega t) & \sin(\omega t) & \frac{1}{2}-\cos(\omega t)-\sin(\omega t)+\frac{1}{2}\cos(2\omega t)\\ \frac{1}{2}-\frac{1}{2}\cos(2\omega t) & \cos(\omega t) & \frac{1}{2}\sin(2\omega t)\\ \frac{1}{2}\sin(2\omega t) & -\sin(\omega t) & \frac{1}{2}+\frac{1}{2}\cos(2\omega t)\end{bmatrix}, \qquad (5.96)$$

which has the properties $\det\{P(t)\}\equiv 1\neq 0$ and $P(t|\omega=0)\equiv I_3$ so that we can take $R^{\{0\}}=A(t|\omega=0)$. We insert $P(t)$ into the RHS of Eq. (1.18)] to check if the constant part $R$ is valid; the result is:

$$R=\begin{bmatrix}75-\omega & 2\omega-17 & 3\omega-112\\ \omega+99 & -22 & 3\omega-143\\ \omega+35 & 2\omega-8 & \omega-53\end{bmatrix}. \qquad (5.97)$$

The solution of the constant part $R$ is valid because it is constant (i.e., independent of $t$). We obtain $R^{\{0\}}=A(t|\omega=0)$ [see Eq.(5.86)] and $\text{trace}\{R\}\equiv 0$.

We conclude that the periodic part $P(t)$ in Eq. (5.96) and the constant part $R$ in Eq. (5.97) are the solution matrices needed to construct the transition matrix of the matrix $A(t)$ [Eq. (5.83)] of the LPTV system. Based on Eq. (5.89) (up to a scale factor for the transformation matrix $V$) we can use other matrices $\tilde{P}(t)$ and $\tilde{R}$ as solutions for the LPTV system of Eq. (5.83) e.g.:

$$\tilde{P}(t)=P(t)(2V)= \qquad (5.98)$$

$$\begin{bmatrix}88\cos(\omega t)-22\cos(2\omega t)+484\sin(\omega t)+22\sin(2\omega t)-22 & 80\cos(\omega t)+35\cos(2\omega t)-22\sin(\omega t)+75\sin(2\omega t)+35 & 2\cos(\omega t)-2\sin(\omega t)+\sin(2\omega t)\\ 488\cos(\omega t)-22\cos(2\omega t)-22\sin(2\omega t)+22 & 198\cos(\omega t)-75\cos(2\omega t)+35\sin(2\omega t)+75 & 1-\cos(\omega t)\\ -22\cos(2\omega t)-484\sin(\omega t)+22\sin(2\omega t)-22 & 35\cos(2\omega t)-198\sin(\omega t)+75\sin(2\omega t)+35 & \sin(2\omega t)\end{bmatrix},$$

$$\tilde{R}=(2V)^{-1}R(2V)=\begin{bmatrix}-\frac{1025}{242}\omega & 1-\frac{4671}{5324}\omega & \frac{67}{5324}\omega\\ \frac{109}{11}\omega & \frac{959}{242}\omega & 1-\frac{5}{242}\omega\\ -254\omega & -\frac{549}{11}\omega & \frac{3}{11}\omega\end{bmatrix}. \qquad (5.99)$$

□

## 5.6 Summary

This chapter assumes that $A(t)$, $P(t)$, and $R$ may be set up from a linear function in $\omega$ of the Fourier coefficients of $A(t)$. The Fourier coefficients of $A(t)$ are constant (independent of $\omega$) and $R$ is a linear function of $\omega$. By treating the frequency $\omega$ as a free parameter, we expand the algebraic equations for the Fourier coefficients of the LPTV system into powers of $\omega$, which allows the coefficients to be matched to solve the system.

By setting $\omega = 0$ in the matrix $A(t)$ (under the assumption of continuity at $\omega = 0$), we obtain a constant (time invariant) matrix $A(t|\omega = 0)$ that gives us information regarding $R|_{\omega=0} = R^{\{0\}}$. This information significantly reduces the computation of the matrix $R$, so that we can choose $R^{\{0\}}$ to be any matrix that is similar to $A(t|\omega = 0)$ (in terms of matrix similarity). It is also possible to obtain $R$ from higher powers of $\omega$ (e.g., $R^{\{1\}}$), although greater computational power is required to select a compatible $R^{\{1\}}$ within the possible options generated by the eigen decomposition.

To find matrices $P(t)$ and $R$ such that $P(t|\omega = 0) = I_n$ and $A(t|\omega = 0) = R^{\{0\}}$, it is convenient to apply a similarity transform to the matrix $A(t)$ of the LPTV system such that $A(t|\omega = 0)$ will have a Jordan canonical form [in addition to zero trace of $A(t)$].

The bottleneck of the proposed approach is that we are limited to a finite number of harmonics for $P(t)$. In addition, the method requires more exploration to obtain the matrices $A(t)$ for LPTV systems and their periodic part $P(t)$ so that both have a finite number of harmonics. Once conditions are found that produce $A(t)$ with a finite number of harmonics, the next step would be to generalize the structure of LPTV systems to solve them when $A(t)$ and\or $P(t)$ have infinite harmonics.

# CHAPTER 6. Discussion

## 6.1 Contribution of the Research

This work makes the following contributions to the study of LPTV systems:

1. **Decomposition into similar real-imaginary and even-odd Fourier series:** This work reveals a similarity between algebraic equations in the real-imaginary Fourier decomposition and the even-odd Fourier decomposition. Depending on the application, we can choose between the two methods to solve LPTV systems, although even-odd decomposition might prove more convenient to use on a real LPTV system.

2. **Stability analysis based on a variable frequency $\boldsymbol{\omega}$**: This allows us to define a condition on the frequency $\omega$ so that the LPTV system is stable, in contrast with previous works that rely on a fixed frequency.

3. **$\boldsymbol{A(t)}, \boldsymbol{P(t)},$ and $\boldsymbol{R}$ are polynomials in $\boldsymbol{\omega}$ with a finite number of harmonics:** To address the case in which both $A(t)$ and $P(t)$ have a finite number of harmonics, the Fourier coefficients of $A(t)$ are polynomial in $\omega$, the Fourier coefficients of $P(t)$ are independent of $\omega$, and $R$ is polynomial in $\omega$. This setup allows us to solve the algebraic equations of the LPTV system by comparing the coefficients of powers of $\omega$ .

4. **Matrix similarity between $\boldsymbol{A(t)}$ and $\boldsymbol{R}$ at $\boldsymbol{\omega = 0}$**: At $\omega = 0$, $A(t|\omega = 0)$ is constant and similar to $R^{\{0\}}$ in the sense of matrix similarity. It might be convenient to transform $A(t)$ to a new similar matrix $\hat{A}(t)$ such that the matrix $\hat{A}(t|\omega = 0)$ has a real Jordan canonical form and to set $R^{\{0\}} = \hat{A}(t|\omega = 0)$. This transformation will reduce the required computational power since $R^{\{0\}}$ is already obtained. From $R^{\{0\}}$, we can obtain $P(t)$ with the condition $P(t|\omega = 0) = I_n$ and then obtain $R$ directly, as discussed in the previous paragraph. In addition, it is convenient to zero the trace of $A(t)$ so that, for all $\omega$, $\det\{P(t)\} \equiv \text{const.} \neq 0$ and $\text{trace}\{R\} = 0$.

**5. The frequency $\boldsymbol{\omega}$ may be varied for all $\boldsymbol{\omega}$**: In contrast with Yakubovich and Starzhinskii (Yakubovich & Starzhinskii, 1975), who use a perturbation analysis of an LPTV system with some small parameter $\varepsilon > 0$ in Chapter 4 of their work, all analysis and solution procedures under the assumptions applied in this work (including continuity in $\omega$) are valid for all $\omega \in \mathbb{R}$. Under these assumptions, $\omega = 0$ is valid since $\Phi_A(t, t_0 | \omega = 0) = \Phi_{A(\cdot|\omega=0)}(t, t_0)$ (i.e., inserting $\omega = 0$ before solving the ODE of the LPTV system or after solving it gives the same transition matrix). In addition, negative values of $\omega$ are valid because we can use the parity of cosine and sine functions to produce a new LPTV system reflected to positive frequencies.

## 6.2 Suggestions for Future Research

We suggest studying the following issues for LPTV systems:

**1. Determine the conditions for $\boldsymbol{A(t)}$ such that both $\boldsymbol{A(t)}$ and $\boldsymbol{P(t)}$ have a finite number of harmonics:** To determine these conditions, we may explore the following questions [$A(t)$ has $L$ harmonics, $P(t)$ has $p$ harmonics, and $n$ is the dimension of $A(t)$ and $P(t)$]:

a. If $L, p < \infty$, is $L \geq p$ ? Should we speculate $p = \left\lceil \frac{L}{n} \right\rceil$ ?

b. If $L < \infty$ and $\text{trace}\{A(t)\} \equiv \text{const.}$ + another condition, then is $p < \infty$ ?

c. Suppose that the Fourier coefficients of $A(t)$ are polynomial in $\omega$ [e.g., $A(t) = A^{\{0\}}(t) + \omega A^{\{1\}}(t) + \cdots + \omega^N A^{\{N\}}(t)$] such that we have $L, p < \infty$. If we swap components among $\{A^{\{r\}}(t)\}_{r=0}^{N}$ [e.g., $A(t) = A^{\{1\}}(t) + \omega A^{\{0\}}(t)$+…] or swap sub-components among $\left\{\{A^{\{r\}}{}_l^{\text{even}\backslash\text{odd}}\}_{r=0}^{N}\right\}_{l=0}^{L}$, with $L < \infty$, does it imply $p < \infty$? Under which conditions is this true?

**2. Generalize the structures of $A(t), P(t)$, and $R$:** We suggest exploring other possible structures of $A(t), P(t)$, and $R$ such that the procedure of solving the ODEs of LPTV systems by comparing powers of $\omega$ remains valid. For example,

a. The Fourier coefficients of $P(t)$ are polynomial in $\omega$.

b. The denominators of the components $A(t), P(t)$, and\or $R$ are parameterized with some arbitrary constant parameter or are polynomial in $\omega$. Multiplying by the common denominator may be useful to obtain the solution.

c. $A(t)$ and\or $P(t)$ can be presented as a quotient of a finite Fourier series in numerators of the matrix divided by a finite Fourier series in the denominators. Multiplying by the common denominator may be useful to obtain the solution.

**3. Studying LPTV systems under number systems other than ℝ and ℂ:** In this work, we focus on real LPTV systems [i.e., $A(t) \in \mathbb{R}$]. In addition, we cover complex LPTV systems [i.e., $A(t) \in \mathbb{C}$] by using a real 2×2 matrix representation $a + ib \leftrightarrow \begin{bmatrix} a & -b \\ b & a \end{bmatrix}$. We suggest studying LPTV systems in different number systems, such as, split complex numbers, dual numbers, or any variation thereof [see, e.g., (Akar, Yüce, & Şahin, 2018), (Dattoli, Licciardi, Pidatella, & Sabia, 2018)].

a. Split complex numbers (also known as hyperbolic complex numbers) $\mathbb{H} = \{a + jb: a, b \in \mathbb{R}, j^2 = +1, j \neq \pm 1\}$. In special cases, it is possible to use these in even-odd decomposition (see APPENDIX C). Note that $a + jb \leftrightarrow \begin{bmatrix} a & b \\ b & a \end{bmatrix}$.

b. Dual numbers $\mathbb{D} = \{a + \varepsilon b: a, b \in \mathbb{R}, \varepsilon^2 = 0, \varepsilon \neq 0\}$ or any isomorphic number system [see, e.g., (Klein & Maimon, 2019) Section 6.2]. This number system might be useful for a small-perturbation approach, as suggested by (Yakubovich & Starzhinskii, 1975) in Chapter 4. Note that $a + \varepsilon b \leftrightarrow \begin{bmatrix} a & b \\ 0 & a \end{bmatrix}$.

4. **Extend the notion of LTI tools in LPTV systems (Wereley, 1991):** In his Ph.D. thesis (Wereley, 1991), Wereley focuses mainly on defining the Toeplitz transform, harmonic transfer functions, zeros and poles, etc., which are related to the exponential representation of infinite Fourier series for LPTV systems and are especially used with the Hill Equation. I suggest extending these definitions for use in a cosine-sine representation of Fourier series and to explore cases with a finite number of harmonics.

# APPENDIX A. Fourier Series for Matrices

Suppose that matrix $A(t)$ is a periodic function with period $T$ or, equivalently, with a frequency $\omega = \frac{2\pi}{T}$. In this case, $A(t)$ can be decomposed into a Fourier series. In this work, we use the finite summation version of the Fourier series decomposition to focus on cases in which the LPTV system matrix and its transition matrix involve a finite Fourier series decomposition [suppose $A(t)$ has $L$ harmonics, then take $L \to \infty$ for the infinite version of the Fourier series].

Traditionally, there are two forms of a Fourier series decomposition:
The exponential form is given by:

$$A(t) = \sum_{l=-L}^{L} A_l e^{il\omega t}\,, \qquad \text{(A.1)}$$

where $i = \sqrt{-1}$ is the imaginary unit and $A_l$ is the Fourier series coefficient and is computed by:

$$A_l = \frac{1}{T}\int_T A(t) e^{-il\omega t} dt\,, \quad l \in \mathbb{Z}. \qquad \text{(A.2)}$$

If $A(t)$ is real, then $A_{-l} = \overline{A_l}$.

The cosine-sine form is given by:

$$A(t) = \frac{A_0^{\text{even}}}{2} + \sum_{l=1}^{L} \left[A_l^{\text{even}} \cos(l\omega t) + A_l^{\text{odd}} \sin(l\omega t)\right], \qquad \text{(A.3)}$$

where

$$A_0^{\text{even}} = \frac{2}{T}\int_T A(t) dt, \qquad \text{(A.4)}$$

$$A_l^{\text{even}} = \frac{2}{T}\int_T A(t)\cos(l\omega t)\,dt,$$
$$A_l^{\text{odd}} = \frac{2}{T}\int_T A(t)\sin(l\omega t)\,dt,$$

for $l \in \mathbb{N}$. We can expand the sum in Eq. (A.3) by exploiting the following properties of even-odd symmetry from Eq. (A.4):

$$\begin{gathered} A_{-l}^{\text{even}} = A_l^{\text{even}} \Rightarrow A_{-l}^{\text{even}}\cos(-l\omega t) = A_l^{\text{even}}\cos(l\omega t), \\ A_{-l}^{\text{odd}} = -A_l^{\text{odd}} \Rightarrow A_{-l}^{\text{odd}}\sin(-l\omega t) = A_l^{\text{odd}}\sin(l\omega t), \\ A_0^{\text{odd}} = 0. \end{gathered} \tag{A.5}$$

By using these properties, we rewrite Eq. (A.3) as follows:

$$A(t) = \sum_{l=-L}^{L}\left(\frac{A_l^{\text{even}}}{2}\cos(l\omega t) + \frac{A_l^{\text{odd}}}{2}\sin(l\omega t)\right). \tag{A.6}$$

We use the exponential form and the cosine-sine form for different purposes: The exponential form, due to its compactness, is used to define or prove general properties when neither a real analysis nor the even-odd properties are involved. The cosine-sine form is used to define or prove properties when either a real analysis or the even-odd properties are involved (e.g., to analyze the transition matrix when the system matrix is real).

**Lemma A.1**: If $P(t)$ and $Q(t)$ are square matrices (with the same size and time period), such that $P(t)$ has $p$ harmonics and $Q(t)$ has $q$ harmonics, then $P(t)Q(t)$ has at most $p+q$ harmonics.

**Proof:** Let $P(t) = \sum_{k=-p}^{p} P_k e^{ik\omega t}$ and $Q(t) = \sum_{r=-q}^{q} Q_r e^{ir\omega t}$, then

$$P(t)Q(t) = \sum_{r=-q}^{q}\sum_{k=-p}^{p} P_k Q_r e^{i\omega(k+r)t} = \sum_{l=-(q+p)}^{q+p}\left(\sum_{k=-p}^{p} P_k Q_{l-k}\right)e^{i\omega l t}.$$

From the equation above, we see that the Fourier coefficient of $e^{i\omega lt}$ is given by the sum $\sum_{k=-p}^{p} P_k Q_{l-k}$ for all $l \in \mathbb{Z}$ such that $-(q+p) \leq l \leq q+p$. Note that some Fourier coefficients may be zero, so the number of the harmonics of $P(t)Q(t)$ is at most $p+q$.

■

**Lemma A.2:** For any $n \in \mathbb{N}$ and for any matrix $P(t) \in \mathbb{R}^{n\times n}$, if $P(t)$ has a finite number $p$ of harmonics, then $\det\{P(t)\}$ has *at most np* harmonics.

**Proof:** We use proof by induction, for $n \in \mathbb{N}$ ($p < \infty$ is fixed). For $n = 1$ we have $\det\{P(t)\} = P(t)$ (for $p = 1$ harmonics, the result is trivially obtained), therefore we start our base case by $n = 2$.

Base Case: $n = 2$

If $P(t) = \begin{bmatrix} p_{11}(t) & p_{12}(t) \\ p_{21}(t) & p_{22}(t) \end{bmatrix}$ has $p$ harmonics, then $\det\{P(t)\} = p_{11}(t)p_{22}(t) - p_{12}(t)p_{21}(t)$ has *at most* $2p$ harmonics because each element $p_{ij}(t)$ has at most $p$ harmonics, so any product of two elements $p_{ij}(t)$ can produce a term with *at most* $2p$ harmonics (and, therefore, any linear combination of this product). We conclude that the lemma is true for the base case ($n = 2$).

Induction Hypothesis: Assume the lemma is true for $n = N$.
Note that, for $P(t) \in \mathbb{R}^{N\times N}$, we have

$\det\{P(t)\} = \sum_{j=1}^{N} (-1)^{i+j} p_{ij}(t) \det\{\mathrm{M}_{ij}\{P(t)\}\}$ for fixed $i$,

where $\mathrm{M}_{ij}\{P(t)\}$ is the minor matrix of $P(t)$ generated by removing row $i$ and column $j$ (therefore, $\mathrm{M}_{ij}\{P(t)\} \in \mathbb{R}^{N-1\times N-1}$).

Assume that any $P(t) \in \mathbb{R}^{N\times N}$ with $p$ harmonics, $\det\{P(t)\}$ has *at most* $Np$ harmonics.

<u>Induction Step</u>: Prove that the hypothesis holds for $n = N + 1$.
Note that, for $P(t) \in \mathbb{R}^{N+1\times N+1}$ we have

$$\det\{P(t)\} = \sum_{j=1}^{N+1}(-1)^{i+j}p_{ij}(t)\det\{\mathrm{M}_{ij}\{P(t)\} \quad \text{for fixed } i.$$

In this case, $\mathrm{M}_{ij}\{P(t)\} \in \mathbb{R}^{N\times N}$. By the induction hypothesis, we assume that all minors of $P(t)$, $\mathrm{M}_{ij}\{P(t)\}$ in $\mathbb{R}^{N\times N}$ for all $i, j = 1, \dots, N$, obtain the property $\det\{\mathrm{M}_{ij}\{P(t)\}$ of having *at most* $Np$ harmonics. Since each element $p_{ij}(t)$ has at most $p$ harmonics, then each product $p_{ij}(t)\det\{\mathrm{M}_{ij}\{P(t)\}$ (and any linear combination thereof) has *at most* $(N + 1)p$ harmonics.

<u>Conclusion</u>: We have proven the lemma for the base case ($n = 2$ or $n = 1$ for the trivial case). The validity of the assumption of the induction hypothesis ($n = N$) implies that the induction step ($n = N + 1$) is true for any arbitrary $N \in \mathbb{N}$. Therefore, by induction, the lemma is proven.

■

**<u>Lemma A.3</u>:** For any $n \in N$: $n \geq 2$ and for any matrix $P(t) \in \mathbb{R}^{n\times n}$, if $P(t)$ has a finite number $p$ of harmonics, then the adjoint $\{P(t)\}$ has *at most* $(n - 1)p$ harmonics.

**<u>Proof</u>:** The adjoint $\{P(t)\}$ is defined as the transpose of the cofactor matrix of $P(t)$, which is constructed from the minor elements $(-1)^{i+j}\det\{\mathrm{M}_{ij}\{P(t)\}$, i.e.,

$$[\mathrm{adjoint}\{P(t)\}]_{ji} = (-1)^{i+j}\det\{\mathrm{M}_{ij}\{P(t)\}.$$

Referring to **Lemma A.2**, each determinant above has *at most* $(n-1)p$ harmonics, so the matrix adjoint $\{P(t)\}$ has *at most* $(n-1)p$ harmonics.

■

**Observations**:

1. We have consistently stated above "*at most np* harmonics" because the exact number of the harmonics might be less than expected. For example, consider an $n \times n$ periodic matrix $P(t)$ with $p$ harmonics: it is possible that the number of harmonics given by $\det\{P(t)\}$ is $(n-1)p$ or even zero.

2. If $\det\{P(t)\} = \text{const.} \neq 0$, then $P^{-1}(t)$ has *at most* $(n-1)p$ harmonics since $P^{-1}(t) = \frac{\text{adjoint}\{P(t)\}}{\det\{P(t)\}} = \frac{\text{adjoint}\{P(t)\}}{const}$ and $\text{adjoint}\{P(t)\}$ has *at most* $(n-1)p$ harmonics.

3. If (in addition to item 2) $n = 2$, then $P^{-1}(t)$ has *exactly* $p$ harmonics because, for a $2 \times 2$ matrix, no multiplication is involved to compute the adjoint, only a change of position or sign; i.e., $\text{adjoint}\{P(t)\} = \begin{bmatrix} p_{22}(t) & -p_{12}(t) \\ -p_{21}(t) & p_{11}(t) \end{bmatrix}$.

# APPENDIX B. Exponential Fourier Analysis for LPTV Systems

## B.1 General

Suppose that the periodic matrix $P(t)$ and $R$ are the pair of matrices that solve Eq. (1.16) and construct the transition matrix $\Phi_A(t, t_0)$. Consider the following exponential Fourier series decomposition for $A(t)$ and $P(t)$:

$$A(t) = \sum_{l=-\infty}^{\infty} A_l e^{i\omega lt}\,, \tag{B.1}$$

$$P(t) = \sum_{k=-\infty}^{\infty} P_k e^{i\omega kt}\,, \tag{B.2}$$

where $A_l$ and $P_l$ are the complex Fourier coefficients of $e^{i\omega lt}$ for the matrices $A(t)$ and $P(t)$, respectively. Inserting the above definition into Eq. (1.16), re-indexing the LHS of that equation by $k \mapsto k - l$, and collecting factors of $e^{i\omega kt}$ gives:

$$\sum_{k=-\infty}^{\infty} \left( \sum_{l=-\infty}^{\infty} A_l P_{k-l} \right) e^{i\omega kt} = \sum_{k=-\infty}^{\infty} P_k (i\omega k I_n + R) e^{i\omega kt}, \tag{B.3}$$

Given that Eq. (B.3) holds for all $t$, by comparring coefficients of $e^{i\omega kt}$, we obtain the following set of algebraic equations in $P_k$ and $R$:

$$\sum_{l=-\infty}^{\infty} A_l P_{k-l} = P_k (i\omega k I_n + R), \qquad k \in \mathbb{Z}. \tag{B.4}$$

Equivalently, by re-indexing the LHS, we obtain

$$\sum_{l=-\infty}^{\infty} A_{k-l}P_l = P_k(i\omega k I_n + R), \qquad k \in \mathbb{Z}, \tag{B.5}$$

which can be represented as an infinite block-system of equations

$$\widetilde{\mathbf{A}}\widetilde{\mathbf{P}} = \widetilde{\mathbf{P}}R, \tag{B.6}$$

where

$$\widetilde{\mathbf{P}} = \begin{bmatrix} \vdots \\ P_{-2} \\ P_{-1} \\ P_0 \\ P_1 \\ P_2 \\ \vdots \end{bmatrix}, \tag{B.7}$$

and

$$\widetilde{\mathbf{A}} = \begin{bmatrix} \ddots & \vdots & \vdots & \vdots & \vdots & \vdots & ⋰ \\ \cdots & A_0 + i2\omega I_n & A_{-1} & A_{-2} & A_{-3} & A_{-4} & \cdots \\ \cdots & A_1 & A_0 + i1\omega I_n & A_{-1} & A_{-2} & A_{-3} & \cdots \\ \cdots & A_2 & A_1 & A_0 + i0\omega I_n & A_{-1} & A_{-2} & \cdots \\ \cdots & A_3 & A_2 & A_1 & A_0 - i1\omega I_n & A_{-1} & \cdots \\ \cdots & A_4 & A_3 & A_2 & A_1 & A_0 - i2\omega I_n & \cdots \\ ⋰ & \vdots & \vdots & \vdots & \vdots & \vdots & \ddots \end{bmatrix}. \tag{B.8}$$

Based on (Wereley, 1991), we decompose $\widetilde{\mathbf{A}}$ into the difference between a *Toeplitz transform* of $A(t)$ and the block diagonal form of the terms $\{il\omega I_n, l \in \mathbb{Z}\}$ so that:

$$\widetilde{\mathbf{A}} = \boldsymbol{\mathcal{A}} - \boldsymbol{\mathcal{N}},$$

$$\boldsymbol{\mathcal{A}} = \boldsymbol{\mathcal{T}}\left(A(t)\right) \equiv \begin{bmatrix} \ddots & \vdots & \vdots & \vdots & ⋰ \\ \cdots & A_0 & A_{-1} & A_{-2} & \cdots \\ \cdots & A_1 & A_0 & A_{-1} & \cdots \\ \cdots & A_2 & A_1 & A_0 & \cdots \\ ⋰ & \vdots & \vdots & \vdots & \ddots \end{bmatrix}, \tag{B.9}$$

$$\boldsymbol{\mathcal{N}} = \text{blkdiag}(\dots, i(-1)\omega I_n, i(0)\omega I_n, i(1)\omega I_n, \dots) \equiv \text{blkdiag}(\{il\omega I_n, l \in \mathbb{Z}\}),$$

where $\boldsymbol{\mathcal{T}}\,(\cdot)$ is the *Toeplitz transform* operator, which maps an infinite exponential Fourier series with matrix coefficients into an infinite block matrix.

In general, the LPTV system may be a complex system [i.e., $A(t)$ is a complex matrix], so $P(t)$ and $R$ might also be complex. It is also possible that $P(t)$ and $R$ are complex even though $A(t)$ is real. However, we limit ourselves in this work to finding procedures that obtain real constant matrices $R$ from real LPTV systems and to using properties of Fourier series applicable to real functions (e.g., $A_{-l} = \overline{A_l}$).

## B.2 Real-Imaginary Decomposition

In the following real-imaginary decomposition, we denote $A_l$ and $P_k$ as:

$$A_l = A_l^{\text{real}} + iA_l^{\text{im}}, \tag{B.10}$$

$$P_k = P_k^{\text{real}} + iP_k^{\text{im}}. \tag{B.11}$$

Using the assumption that $A(t)$ and $P(t)$ are real, we have:

$$A_{-l} = \overline{A_l} = A_l^{\text{real}} - iA_l^{\text{im}} \Leftrightarrow A_{-l}^{\text{real}} = A_l^{\text{real}}, A_{-l}^{\text{im}} = -A_l^{\text{im}}, \tag{B.12}$$

$$P_{-k} = \overline{P_k} = P_k^{\text{real}} - iP_k^{\text{im}} \Leftrightarrow P_{-k}^{\text{real}} = P_k^{\text{real}}, P_{-k}^{\text{im}} = -P_k^{\text{im}}. \tag{B.13}$$

From the property of the imaginary part for $l = 0$ in Eq. (B.12) and $k = 0$ in Eq. (B.13), we conclude that $A_0^{\text{im}} = P_0^{\text{im}} = 0$. Inserting Eqs. (B.12) and (B.13) into Eq. (B.4) and comparing the real and imaginary parts gives the following set of algebraic equations in $P_k^{\text{real}}$, $P_k^{\text{im}}$, and $R$:

$$\begin{aligned} 1&: \sum_{l=-\infty}^{\infty} \left(A_l^{\text{real}} P_{k-l}^{\text{real}} - A_l^{\text{im}} P_{k-l}^{\text{im}}\right) = P_k^{\text{real}} R - P_k^{\text{im}} \omega k I_n, \\ i&: \sum_{l=-\infty}^{\infty} \left(A_l^{\text{real}} P_{k-l}^{\text{im}} + A_l^{\text{im}} P_{k-l}^{\text{real}}\right) = P_k^{\text{im}} R + P_k^{\text{real}} \omega k I_n. \end{aligned} \tag{B.14}$$

Equivalently, by re-indexing the LHS of Eq. (B.14) [cf. Eq. (B.5)],

$$\begin{aligned} 1&: \sum_{l=-\infty}^{\infty} \left(A_{k-l}^{\text{real}} P_l^{\text{real}} - A_{k-l}^{\text{im}} P_l^{\text{im}}\right) = P_k^{\text{real}} R - P_k^{\text{im}} \omega k I_n, \\ i&: \sum_{l=-\infty}^{\infty} \left(A_{k-l}^{\text{real}} P_l^{\text{im}} + A_{k-l}^{\text{im}} P_l^{\text{real}}\right) = P_k^{\text{im}} R + P_k^{\text{real}} \omega k I_n. \end{aligned} \tag{B.15}$$

To reduce the negative indices of $P_k^{\text{real}}$ and $P_k^{\text{im}}$, we integrate the properties of Eqs. (B.12) and (B.13) into Eq. (B.15) and re-index the LHS of Eq. (B.15) to collect factors of $P_l^{\text{real}}$ and $P_k^{\text{im}}$ for $l = 0,1,2, ....$ The result is:

$$\begin{aligned} 1&: A_k^{\text{real}} P_0^{\text{real}} + \sum_{l=1}^{\infty} \left[\left(A_{k-l}^{\text{real}} + A_{k+l}^{\text{real}}\right) P_l^{\text{real}} + \left(A_{k+l}^{\text{im}} - A_{k-l}^{\text{im}}\right) P_l^{\text{im}}\right] = P_k^{\text{real}} R - P_k^{\text{im}} \omega k I_n, \\ i&: A_k^{\text{im}} P_0^{\text{real}} + \sum_{l=1}^{\infty} \left[\left(A_{k+l}^{\text{im}} + A_{k-l}^{\text{im}}\right) P_l^{\text{real}} + \left(A_{k-l}^{\text{real}} - A_{k+l}^{\text{real}}\right) P_l^{\text{im}}\right] += P_k^{\text{im}} R + P_k^{\text{real}} \omega k I_n, \end{aligned} \tag{B.16}$$

for $k = 0,1,2,\ldots.$

For $k = 0$, we have:

$$\begin{aligned} 1&: A_0^{\text{real}} P_0^{\text{real}} + \sum_{l=1}^{\infty} \left[\left(2A_l^{\text{real}}\right) P_l^{\text{real}} + \left(2A_l^{\text{im}}\right) P_l^{\text{im}}\right] = P_0^{\text{real}} R, \\ i&: A_k^{\text{im}} P_0^{\text{real}} + \sum_{l=1}^{\infty} \left[\left(A_l^{\text{im}} + A_{-l}^{\text{im}}\right) P_l^{\text{real}} + \left(A_{-l}^{\text{real}} - A_l^{\text{real}}\right) P_l^{\text{im}}\right] P_0^{\text{im}} R = 0, \end{aligned} \tag{B.17}$$

so, equation $i$ [of Eqs. (B.17)] is identically zero due to its real matrices $A(t)$ and $P(t)$:

- $A_0^{\text{im}} = P_0^{\text{im}} = 0;$
- $A_l^{\text{im}} + A_{-l}^{\text{im}} = 0;$
- $A_{-l}^{\text{real}} - A_l^{\text{real}} = 0;$
- $P_0^{\text{real}} \omega(0) I_n = 0.$

Eq. (B.16) can be represented as a semi-infinite block-system of the equation

$$\widetilde{\mathbf{A}}\widetilde{\mathbf{P}} = \widetilde{\mathbf{P}}R, \tag{B.18}$$

such that

$$\widetilde{\mathbf{P}} = \begin{bmatrix} P_0^{\text{real}} \\ P_1^{\text{real}} \\ P_1^{\text{im}} \\ P_2^{\text{real}} \\ P_2^{\text{im}} \\ \vdots \end{bmatrix}, \tag{B.19}$$

$$\widetilde{\mathbf{A}} = \begin{bmatrix} A_0 & A_{0-1}^{\text{real}} + A_{0+1}^{\text{real}} & A_{0+1}^{\text{im}} - A_{0-1}^{\text{im}} & A_{0-2}^{\text{real}} + A_{0+2}^{\text{real}} & A_{0+2}^{\text{im}} - A_{0-2}^{\text{im}} & \cdots & \cdots \\ A_1^{\text{real}} & A_{1-1}^{\text{real}} + A_{1+1}^{\text{real}} & A_{1+1}^{\text{im}} - A_{1-1}^{\text{im}} + 1\omega I_n & A_{1-2}^{\text{real}} + A_{1+2}^{\text{real}} & A_{1+2}^{\text{im}} - A_{1-2}^{\text{im}} & \cdots & \cdots \\ A_1^{\text{im}} & A_{1+1}^{\text{im}} + A_{1-1}^{\text{im}} - 1\omega I_n & A_{1-1}^{\text{real}} - A_{1+1}^{\text{real}} & A_{1+2}^{\text{im}} + A_{1-2}^{\text{im}} & A_{1-2}^{\text{real}} - A_{1+2}^{\text{real}} & \cdots & \cdots \\ A_2^{\text{real}} & A_{2-1}^{\text{real}} + A_{2+1}^{\text{real}} & A_{2+1}^{\text{im}} - A_{2-1}^{\text{im}} & A_{2-2}^{\text{real}} + A_{2+2}^{\text{real}} & A_{2+2}^{\text{im}} - A_{2-2}^{\text{im}} + 2\omega I_n & \cdots & \cdots \\ A_2^{\text{im}} & A_{2+1}^{\text{im}} + A_{2-1}^{\text{im}} & A_{2-1}^{\text{real}} - A_{2+1}^{\text{real}} & A_{2+2}^{\text{im}} + A_{2-2}^{\text{im}} - 2\omega I_n & A_{2-2}^{\text{real}} - A_{2+2}^{\text{real}} & \cdots & \cdots \\ \vdots & \vdots & \vdots & \vdots & \vdots & \cdots [\widetilde{\mathbf{A}}]_{kl} & \cdots \\ \vdots & \vdots & \vdots & \vdots & \vdots & \vdots & \ddots \end{bmatrix}, \tag{B.20}$$

where

$$[\widetilde{\mathbf{A}}]_{kl} = \begin{bmatrix} A_{k-l}^{\text{real}} + A_{k+l}^{\text{real}} & A_{k+l}^{\text{im}} - A_{k-l}^{\text{im}} + k\omega I_n \delta_{kl} \\ A_{k+l}^{\text{im}} + A_{k-l}^{\text{im}} - k\omega I_n \delta_{kl} & A_{k-l}^{\text{real}} - A_{k+l}^{\text{real}} \end{bmatrix} \tag{B.21}$$

is the block matrix in row $k$ and column $l$ that multiplies the element $\begin{bmatrix} P_l^{\text{real}} \\ P_l^{\text{im}} \end{bmatrix}$, and $\delta_{kl}$ is the Kronecker delta function.

# APPENDIX C. Representing LPTV Systems by 2×2 Real Blocks

## C.1 Representing Complex LPTV System by 2×2 Real Blocks

In general, an LPTV system may be complex, which means that the system matrix $A(t)$ is complex, so at least one of the matrices $P(t)$ and $R$ is complex. In this case, the real-imaginary decomposition equation and the properties discussed in Section B.2 are irrelevant because the assumptions that $A(t)$, $P(t)$, and $R$ are real are not valid. We consider two approaches.

First, we search the complex plane for a solution to the LPTV system as described in Section B.1. Alternatively, we convert the complex LPTV system into an LPTV system with real and imaginary parts separated and rearrange the system into 2×2 real block matrices. This approach is based on representing every complex number $a + ib$ by a 2×2 real block matrix $\begin{bmatrix} a & -b \\ b & a \end{bmatrix}$, as described below.

We use the following notation:

- $x \mapsto x + iy$;
- $A(t) \mapsto A(t) + iB(t)$;
- $P(t) \mapsto P(t) + iQ(t)$;
- $R \mapsto R + iS$.

The new notation is defined such that $x, y \in \mathrm{R}^n$ and $A(t),\ B(t),\ P(t),\ Q(t),\ R,$ $S \in \mathrm{R}^{n\times n}$. Using the new notation in Eq. (1.14) to obtain the new complex LPTV system in terms of $x + iy$ gives:

$$\begin{aligned}\dot{x} + i\dot{y} &= [A(t) + iB(t)](x + iy) \\ &= [A(t)x - B(t)y] + i[B(t)x + A(t)y].\end{aligned} \tag{C.1}$$

For Eq. (1.17), we use *Floquet theory* to obtain the complex linear differential equation in terms of the matrices $P(t) + iQ(t)$ and $R + iS$. The result is:

$$\begin{aligned}&[A(t) + iB(t)][P(t) + iQ(t)] = [\dot{P}(t) + i\dot{Q}(t)] + [P(t) + iQ(t)](R + iS) \\ &\Rightarrow (AP - BQ) + i(BP + AQ) = \dot{P} + i\dot{Q} + (PR - QS) + i(QR + PS).\end{aligned} \tag{C.2}$$

By separating the real and imaginary parts and rearranging the results into 2×2 real block matrices, we obtain a real LPTV system equivalent to that given in Eq. (C.1):

$$\begin{bmatrix}\dot{x} \\ \dot{y}\end{bmatrix} = \begin{bmatrix}A(t) & -B(t) \\ B(t) & A(t)\end{bmatrix}\begin{bmatrix}x \\ y\end{bmatrix}. \tag{C.3}$$

As per *Floquet theory*, this is solved by the following real linear differential equation equivalent to Eq. (C.2):

$$\begin{bmatrix}A(t) & -B(t) \\ B(t) & A(t)\end{bmatrix}\begin{bmatrix}P(t) & -Q(t) \\ Q(t) & P(t)\end{bmatrix} = \begin{bmatrix}\dot{P}(t) & -\dot{Q}(t) \\ \dot{Q}(t) & \dot{P}(t)\end{bmatrix} + \begin{bmatrix}P(t) & -Q(t) \\ Q(t) & P(t)\end{bmatrix}\begin{bmatrix}R & -S \\ S & R\end{bmatrix}. \tag{C.4}$$

We denote the state as $\tilde{x} = \begin{bmatrix}x \\ y\end{bmatrix}$, the real LPTV system as $\tilde{A}(t) = \begin{bmatrix}A(t) & -B(t) \\ B(t) & A(t)\end{bmatrix}$, and the *Floquet theory* real matrices as $\tilde{P}(t) = \begin{bmatrix}P(t) & -Q(t) \\ Q(t) & P(t)\end{bmatrix}$ and $\tilde{R} = \begin{bmatrix}R & -S \\ S & R\end{bmatrix}$. To fit the dimensions of the matrices and solve for $\tilde{P}(t)$ and $\tilde{R}$, we construct every block-matrix from Eq. (C.2) to (C.4) through isomorphic representations of every complex number by a 2×2 real block matrix ($a + ib \mapsto \begin{bmatrix}a & -b \\ b & a\end{bmatrix}$). However, to solve for the state $\tilde{x}$, it suffices to arrange it as a column vector $\begin{bmatrix}x \\ y\end{bmatrix}$. Note that, if a real LPTV system matrix has the form $\tilde{A}(t) =$ $\begin{bmatrix} A(t) & -B(t) \\ B(t) & A(t) \end{bmatrix}$, then, by considering the equation backwards, we conclude that its *Floquet theory* real matrices have the same form: $\tilde{P}(t) = \begin{bmatrix} P(t) & -Q(t) \\ Q(t) & P(t) \end{bmatrix}$ and $\tilde{R} = \begin{bmatrix} R & -S \\ S & R \end{bmatrix}$, so we reform the LPTV system into the complex plane as described in Eq. (C.2).

## C.2 Representing Even-Odd Decomposition of LPTV Systems by 2×2 Blocks

This section follows the ideas of Section 2.2.2 and uses an even-odd decomposition, but with the state vector separated into even and odd parts. Furthermore, the matrix of the LPTV system is separated according to blocks. We rewrite Eqs. (2.12) and (2.13) in the following block matrix form:

$$\begin{bmatrix} A^{\text{odd}}(t) & A^{\text{even}}(t) \\ A^{\text{even}}(t) & A^{\text{odd}}(t) \end{bmatrix} \begin{bmatrix} P^{\text{even}}(t) & P^{\text{odd}}(t) \\ P^{\text{odd}}(t) & P^{\text{even}}(t) \end{bmatrix} = \begin{bmatrix} \dot{P}^{\text{even}}(t) & \dot{P}^{\text{odd}}(t) \\ \dot{P}^{\text{odd}}(t) & \dot{P}^{\text{even}}(t) \end{bmatrix} + \begin{bmatrix} P^{\text{even}}(t) & P^{\text{odd}}(t) \\ P^{\text{odd}}(t) & P^{\text{even}}(t) \end{bmatrix} \begin{bmatrix} 0 & R \\ R & 0 \end{bmatrix}, \tag{C.5}$$

which, by *Floquet theory*, is related to the following LPTV system:

$$\begin{bmatrix} \dot{x}^{\text{even}} \\ \dot{x}^{\text{odd}} \end{bmatrix} = \begin{bmatrix} A^{\text{odd}}(t) & A^{\text{even}}(t) \\ A^{\text{even}}(t) & A^{\text{odd}}(t) \end{bmatrix} \begin{bmatrix} x^{\text{even}} \\ x^{\text{odd}} \end{bmatrix}. \tag{C.6}$$

The form in Eq. (C.6) can also be derived from an even-odd decomposition of Eq.1.14:

$$\dot{x}^{\text{even}} + \dot{x}^{\text{odd}} = \left[A^{\text{even}}(t) + A^{\text{odd}}(t)\right]\left[x^{\text{even}} + x^{\text{odd}}\right]. \tag{C.7}$$

As in the previous section C.1, we denote the state as $\tilde{x} = \begin{bmatrix} x^{\text{even}} \\ x^{\text{odd}} \end{bmatrix}$, the real LPTV system as $\tilde{A}(t) = \begin{bmatrix} A^{\text{odd}}(t) & A^{\text{even}}(t) \\ A^{\text{even}}(t) & A^{\text{odd}}(t) \end{bmatrix}$, and the *Floquet theory* real matrices as $\tilde{P}(t) = \begin{bmatrix} P^{\text{even}}(t) & P^{\text{odd}}(t) \\ P^{\text{odd}}(t) & P^{\text{even}}(t) \end{bmatrix}$ and $\tilde{R} = \begin{bmatrix} 0 & R \\ R & 0 \end{bmatrix}$. In addition, we observe that, if an LPTV system matrix has the form $\tilde{A}(t) = \begin{bmatrix} A^{\text{odd}}(t) & A^{\text{even}}(t) \\ A^{\text{even}}(t) & A^{\text{odd}}(t) \end{bmatrix}$, then, by looking at the equation backwards, we conclude that its *Floquet theory* real matrices have the same form: $\tilde{P}(t) = \begin{bmatrix} P^{\text{even}}(t) & P^{\text{odd}}(t) \\ P^{\text{odd}}(t) & P^{\text{even}}(t) \end{bmatrix}$ and $\tilde{R} = \begin{bmatrix} 0 & R \\ R & 0 \end{bmatrix}$. We recast this LPTV system into a more compact form, as described by Eq. (2.11), so we obtain some of the properties previously described in this work.

For example, referring to Eqs. (1.29) and (1.30), we note the average and trace properties for $A(t) = A^{\text{even}}(t) + A^{\text{odd}}(t)$ and $R$ : $\tilde{A}(t) = \begin{bmatrix} A^{\text{odd}}(t) & A^{\text{even}}(t) \\ A^{\text{even}}(t) & A^{\text{odd}}(t) \end{bmatrix}$ and $\tilde{R} = \begin{bmatrix} 0 & R \\ R & 0 \end{bmatrix}$: Average $A^o = \frac{A_0^{\text{even}}}{2}$, $\tilde{A}^o = \begin{bmatrix} 0 & \frac{A_0^{\text{even}}}{2} \\ \frac{A_0^{\text{even}}}{2} & 0 \end{bmatrix}$, Trace: $\text{trace}\{A^o\} = \text{trace}\{R\}$, $\text{trace}\{\tilde{A}^o\} = \text{trace}\{\tilde{R}\} = 0$.

## C.3 Generalizing Even-Odd Decomposition via Split Complex Numbers and their Representation as 2×2 Blocks

We generalize the structure of the even-odd decomposition by using, e.g., split complex numbers (also known as hyperbolic complex numbers) $\mathbb{H}$, which are defined as follows:

$$\mathbb{H} = \{a + jb : a, b \in \mathbb{R}, j^2 = +1, j \neq \pm 1\}. \tag{C.8}$$

If we replace the imaginary unit $i$ (where $i^2 = -1$) by the split imaginary unit $j$ (with $j^2 = 1$) in Eqs. (C.1) and (C.2), we obtain:

$$\begin{aligned}\dot{x} + j\dot{y} &= [A(t) + jB(t)](x + jy) \\ &= [A(t)x + B(t)y] + j[B(t)x + A(t)y],\end{aligned} \tag{C.9}$$

$$\begin{aligned}&[A(t) + jB(t)]\big(P(t) + jQ(t)\big) = \big[\dot{P}(t) + j\dot{Q}(t)\big] + [P(t) + jQ(t)](R + jS) \\ &\Rightarrow (AP + BQ) + j(BP + AQ) = \dot{P} + j\dot{Q} + (PR + QS) + j(QR + PS).\end{aligned} \tag{C.10}$$

That can be represented block-wise by:

$$\begin{bmatrix}\dot{x} \\ \dot{y}\end{bmatrix} = \begin{bmatrix}A(t) & B(t) \\ B(t) & A(t)\end{bmatrix}\begin{bmatrix}x \\ y\end{bmatrix}, \tag{C.11}$$

$$\begin{bmatrix}A(t) & B(t) \\ B(t) & A(t)\end{bmatrix}\begin{bmatrix}P(t) & Q(t) \\ Q(t) & P(t)\end{bmatrix} = \begin{bmatrix}\dot{P}(t) & \dot{Q}(t) \\ \dot{Q}(t) & \dot{P}(t)\end{bmatrix} + \begin{bmatrix}P(t) & Q(t) \\ Q(t) & P(t)\end{bmatrix}\begin{bmatrix}R & S \\ S & R\end{bmatrix}, \tag{C.12}$$

from which, with a suitable change of notation, we find that the even-odd decomposition in Eqs. (C.6) and (C.5) is a special case of Eqs. (C.11) and (C.12), respectively.

# APPENDIX D. Dynamic Eigen Decomposition of LTV Systems

This appendix outlines the main results of (Wang, 2017), who generalizes the notion of eigen decomposition of LTV systems [the related proofs are given in (Wang, 2017)]. Consider the following general LTV system defined by the matrix $\mathbf{A}(t) \in \mathbb{R}^{n\times n}$:

$$\dot{x} = \mathbf{A}(t)x\,. \tag{D.1}$$

According to (Wang, 2017), an LTV system matrix $\mathbf{A}(t) \in \mathbb{R}^{n\times n}$ can be decomposed as follows:

$$\mathbf{A}(t) = \sum_{k=1}^{n}\left[\lambda_k(t)\mathbf{e}_k(t)\mathbf{r}_k^T(t) + \dot{\mathbf{e}}_k(t)\mathbf{r}_k^T(t)\right], \tag{D.2}$$

where $\lambda_k(t)$ is the $k$th dynamic ($t$-variant) eigenvalue of $A(t)$, $\mathbf{e}_k(t) \not\equiv 0$ is the corresponding dynamic (column) eigenvector, $\mathbf{r}_k^T(t)$ is the row eigenvector, and the reciprocal of $\mathbf{e}_k(t)$ such that $\mathbf{r}_j^T(t)\mathbf{e}_k(t) = \delta_{kj}$, where $\delta_{kj}$ is the Kronecker delta function. The pair $\{\lambda_k(t), \mathbf{e}_k(t)\}$ is called the eigenpair of $\mathbf{A}(t)$. The transition matrix $\Phi_{\mathbf{A}}(t, t_0)$ is constructed from the eigenpairs of $\mathbf{A}(t)$ as follows:

$$\Phi_{\mathbf{A}}(t, t_0) = \sum_{k=1}^{n} \exp\left\{\int_{t_0}^{t} \lambda_k(\tau)d\tau\right\} \mathbf{e}_k(t)\mathbf{r}_k^T(t_0)\,. \tag{D.3}$$

Each eigenpair $\{\lambda_k(t), \mathbf{e}_k(t)\}$ is a solution to obtain the pair $\{\lambda(t), \mathbf{e}(t)\}$ in the following equation:

$$\left\{\mathbf{A}(t) - \mathbf{I}_n \frac{d}{dt}\right\}\mathbf{e}(t) = \lambda(t)\mathbf{e}(t). \tag{D.4}$$

The procedure suggested in (Wang, 2017) is to use an *auxiliary equation*; that is, to define a new LTV system with an arbitrary matrix $\mathbf{G}(t)$ such that:

$$\dot{\boldsymbol{e}}(\mathrm{t}) = \mathbf{G}(t)\mathbf{e}(t)\ . \tag{D.5}$$

Equation (D.4) can be rearranged as:

$$\{\lambda(t)\mathbf{I}_n - \mathbf{A}(t) + \mathbf{G}(t)\}\mathbf{e}(t) = 0, \tag{D.6}$$

which has a nontrivial (nonzero) solution for $\mathbf{e}(t)$ if and only if:

$$\det\{\lambda(t)\mathbf{I}_n - \mathbf{A}(t) + \mathbf{G}(t)\} = 0\ . \tag{D.7}$$

The procedure consists of the following steps: Select an arbitrary matrix $\mathbf{G}(t)$; solve Eq. (D.7) to find all eigenvalues $\lambda_k(t)$; solve Eq. (D.6) to find all eigenvectors $\mathbf{e}_k(t)$; if Eq. (D.5) is *not* satisfied, select a new arbitrary matrix $\mathbf{G}(t)$ and repeat the process; if Eq. (D.5) *is* satisfied, construct $\Phi_{\mathbf{A}}(t, t_0)$ [see Eq. (D.3)].

**Remarks:**

1. The set of all eigenpairs are not a unique solution [depends on the selection of $\mathbf{G}(t)$].

2. If $\mathbf{A}(t)$ satisfies the commutative property, $\mathbf{A}(t_1)\mathbf{A}(t_2) = \mathbf{A}(t_2)\mathbf{A}(t_1)$ $\forall\, t_1, t_2$, then we can use $\mathbf{G}(t) \equiv 0$, which implies that each eigenvector $\mathbf{e}_k(t)$ is a constant vector.

3. If $\mathbf{A}(t) = \begin{bmatrix} 0 & 1 \\ -a_0(t) & -a_1(t) \end{bmatrix}$, then the eigenvalues are the result of the Riccati equation $\dot{\lambda}(t) + \lambda^2(t) + a_1(t)\lambda(t) + a_0(t) = 0$, with the corresponding eigenvectors $\mathbf{e}_{1,2}(t) = \left[1, \lambda_{1,2}(t)\right]^T$. A more general procedure for using the Riccati equation is given by van der Kloet and Neerhoff (van der Kloet & Neerhoff, 2004a), (van der Kloet & Neerhoff, 2004b).

# APPENDIX E. Generalization of LTI-System Tools for LPTV Systems

This appendix outlines the main results of (Wereley, 1991), which suggest generalizing tools for LPTV systems to analyze and control LTI systems (e.g., transfer functions, zeros and poles in $s$ and $z$ domains). In this appendix, we focus only on the $s$-plane generalizations. Consider the following state space for an LPTV system:

$$\dot{x} = A(t)x + B(t)u,$$
$$y = C(t)x + D(t)u, \tag{E.1}$$

where each matrix set $\{A(t), B(t), C(t), D(t)\}$ is $\frac{2\pi}{\omega}$-periodic with exponential Fourier series representation:

$$A(t) = \sum_{l=-\infty}^{\infty} A_l e^{i\omega l t}. \tag{E.2}$$

Consider $s_l = s + il\omega$ for some $s \in \mathbb{C}$ and let $u(t)$ be the following input:

$$u(t) = \sum_{l=-\infty}^{\infty} u_l e^{s_l t}\,. \tag{E.3}$$

According to (Wereley, 1991), theorem 3.14 for zero initial condition and the steady state, $x(t)$, $\dot{x}(t)$, and $y(t)$ have the same form as $u(t)$:

$$x(t) = \sum_{l=-\infty}^{\infty} x_l e^{s_l t}\,,$$
$$\dot{x}(t) = \sum_{l=-\infty}^{\infty} s_k x_k e^{s_l t}, \tag{E.4}$$

$$y(t) = \sum_{l=-\infty}^{\infty} y_l e^{s_l t}.$$

Upon inserting Eqs. (E.2)–(E.4) into Eq. (E.1), re-indexing the $\frac{2\pi}{\omega}$-periodic matrix set $\{A(t), B(t), C(t), D(t)\}$ by, e.g., $A_l \longmapsto A_{k-l}$, comparing the coefficients of $e^{s_k t}$, and using $s_k = s + ik\omega \; and \; ik\omega = \sum_{l=-\infty}^{\infty} \delta_{kl} il\omega$, where $\delta_{kl}$ is the Kronecker delta function, we obtain:

$$\begin{aligned} sx_k &= \sum_{l=-\infty}^{\infty} [A_{k-l} - \delta_{kl} il\omega I_n] x_l + \sum_{l=-\infty}^{\infty} B_{k-l} u_l, \\ y_k &= \sum_{l=-\infty}^{\infty} C_{k-l} x_l + \sum_{l=-\infty}^{\infty} D_{k-l} u_l. \end{aligned} \tag{E.5}$$

Based on (Wereley, 1991), theorem 3.14 and on the *Lyapunov reducibility theorem* [see (Yakubovich & Starzhinskii, 1975), Chapter 2, Section 2.4], and assuming that we know the matrices $P(t) \; and \; R$ [or at least $P(t)$][10] to obtain the transition matrix $\Phi_A(t, t_0)$, we use $\tilde{x} = P^{-1}(t)x$ to transform Eq. (E.1) into a state-space model with time-invariant dynamics in the following sense:

$$\begin{aligned} \dot{\tilde{x}} &= \widetilde{A}\tilde{x} + \widetilde{B}(t)u, \\ y &= \tilde{C}(t)\tilde{x} + \widetilde{D}(t)u, \end{aligned} \tag{E.6}$$

where $\tilde{A} = R$ (see footnote[10]), $\tilde{B}(t) = P^{-1}(t)B(t)$, $\tilde{C}(t) = C(t)P(t)$, $\widetilde{D}(t) = D(t)$.

[10] If $P(t)$ is correct, then $R$ is obtained by $R = P^{-1}(t)\left[A(t)P(t) - \dot{P}(t)\right]$ so the RHS is constant matrix. If $P(t)$ is incorrect, then $\tilde{A} = P^{-1}(t)\left[A(t)P(t) - \dot{P}(t)\right]$ is some $T$-periodic matrix.

**Definition E.1**: The *harmonic state-space model* is defined by the system of equations (E.5) represented by:

$$s\boldsymbol{x} = (\boldsymbol{\mathcal{A}} - \boldsymbol{\mathcal{N}})\boldsymbol{\mathcal{X}} + \boldsymbol{\mathcal{B}}\boldsymbol{u},$$
$$\boldsymbol{y} = \boldsymbol{\mathcal{C}}\boldsymbol{x} + \boldsymbol{\mathcal{D}}\boldsymbol{u}, \tag{E.7}$$

where $\boldsymbol{\mathcal{A}}$, $\boldsymbol{\mathcal{B}}$, $\boldsymbol{\mathcal{C}}$, $\boldsymbol{\mathcal{D}}$ are the Toeplitz transforms of $A(t), B(t), C(t), D(t)$, respectively [see Eq. (B.9)], $\boldsymbol{\mathcal{N}} = \text{blkdiag}(\{il\omega I_n, l \in \mathbb{Z}\})$, and $\boldsymbol{u}, \boldsymbol{x}, \boldsymbol{y}$ are the doubly infinite vector representations of $u, x, y$, respectively (e.g., $\boldsymbol{u}^T = [\dots u_{-1}^T, u_0^T, u_1^T \dots]$). The set $\{\boldsymbol{\mathcal{A}} - \boldsymbol{\mathcal{N}}, \boldsymbol{\mathcal{B}}, \boldsymbol{\mathcal{C}}, \boldsymbol{\mathcal{D}}\}$ is used to denote the harmonic state-space model. If the state-space model is transformed so that it has time-invariant dynamics [Eq. (E.6)], then the set of corresponding Toeplitz-transformed matrices $\{\boldsymbol{\mathcal{R}} - \boldsymbol{\mathcal{N}}, \widetilde{\boldsymbol{\mathcal{B}}}, \widetilde{\boldsymbol{\mathcal{C}}}, \widetilde{\boldsymbol{\mathcal{D}}}\}$ is used, where $\boldsymbol{\mathcal{R}} = \text{blkdiag}(\dots, R, R, \dots)$, $\boldsymbol{\mathcal{P}}^{-1} = [\boldsymbol{\mathcal{T}}\{P^{-1}(t)\}] = [\boldsymbol{\mathcal{T}}\{P(t)\}]^{-1}$, $\widetilde{\boldsymbol{\mathcal{B}}} = \boldsymbol{\mathcal{P}}^{-1}\boldsymbol{\mathcal{B}}$, $\widetilde{\boldsymbol{\mathcal{C}}} = \boldsymbol{\mathcal{C}}\boldsymbol{\mathcal{P}}$, and $\widetilde{\boldsymbol{\mathcal{D}}} = \boldsymbol{\mathcal{D}}$.

□

**In the s plane, the form of an LPTV *harmonic state space* is similar to that of an LTI *state space*.**

**Definition E.2**: The *harmonic transfer function* $\widehat{\boldsymbol{\mathcal{G}}}(s)$ is an infinite-dimensional matrix that describes how the input $u$ is related to the output $y$ such that:

$$\boldsymbol{y} = \widehat{\boldsymbol{\mathcal{G}}}(s)\boldsymbol{u},$$
$$\widehat{\boldsymbol{\mathcal{G}}}(s) = \boldsymbol{\mathcal{C}}[s\boldsymbol{\mathcal{I}} - (\boldsymbol{\mathcal{A}} - \boldsymbol{\mathcal{N}})]^{-1}\boldsymbol{\mathcal{B}} + \boldsymbol{\mathcal{D}}, \tag{E.8}$$

where $\boldsymbol{\mathcal{I}}$ is the infinite identity matrix, if the inverse exists. If the state-space model is transformed to have a time-invariant dynamic [Eq. (E.6)], $\widehat{\boldsymbol{\mathcal{G}}}(s)$ is identically obtained by:

$$\widehat{\boldsymbol{\mathcal{G}}}(s) = \widetilde{\boldsymbol{\mathcal{C}}}[s\boldsymbol{\mathcal{I}} - (\boldsymbol{\mathcal{R}} - \boldsymbol{\mathcal{N}})]^{-1}\widetilde{\boldsymbol{\mathcal{B}}} + \widetilde{\boldsymbol{\mathcal{D}}}. \tag{E.9}$$

In the s plane, the form of an LPTV *harmonic transfer function* is similar to that of an LTI *transfer function*.

□

**Definition E.3**: The *LPTV poles in the $s$ plane* mark the location in the complex $s$ plane where the harmonic transfer function is not analytic.

□

The poles of an LPTV system are obtained by solving the eigenvalue problem:

$$[s\boldsymbol{\mathcal{I}} - (\boldsymbol{\mathcal{A}} - \boldsymbol{\mathcal{N}})]\boldsymbol{v} = 0, \tag{E.10}$$

where, for each eigenvalue, $s$ is a pole of the LPTV system with a corresponding infinite eigenvector $\boldsymbol{v}$. Equivalently, if the LPTV state space is transformed to be time invariant [Eq. (E.6)], the poles can be obtained from the union of all the eigenvalues from the eigenvalue problem:

$$\forall\ l \in \mathbb{Z}, [sI_n - (R - il\omega I_n)]\mathbf{v} = 0. \tag{E.11}$$

In the s plane, the form of an LPTV pole definition is similar to that of an LTI pole definition.

**Definition E.4**: The *LPTV transmission zeroes in the $s$ plane* mark the location $s_z$ in the complex $s$ plane, along with the corresponding input:

$$u(t)\big|_{s=s_z} = \sum_{l=-\infty}^{\infty} u_l e^{(s_z+il\omega)t}. \tag{E.12}$$

The initial condition for state $x_0$ implies that the output is identically zero, i.e., $y(t) \equiv 0$.

□

These transition zeroes can be obtained by solving the following generalized eigenvalue problem:

$$\begin{bmatrix} s_z\boldsymbol{\mathcal{I}} - (\boldsymbol{\mathcal{A}} - \boldsymbol{\mathcal{N}}) & -\boldsymbol{\mathcal{B}} \\ -\boldsymbol{\mathcal{C}} & -\boldsymbol{\mathcal{D}} \end{bmatrix} \begin{Bmatrix} \boldsymbol{x}_0 \\ \boldsymbol{u}_0 \end{Bmatrix} = , \quad \begin{Bmatrix} \boldsymbol{x}_0 \\ \boldsymbol{u}_0 \end{Bmatrix} \neq 0, \tag{E.13}$$

where $\boldsymbol{u}_0$ contains the harmonics determined for $u(t)|_{s=s_z}$, and $\boldsymbol{x}_0$ contains the harmonics for the initial condition of $x(t)$ determined by $\boldsymbol{u}_0$. Equivalently, if the LPTV state space is transformed to be time invariant [Eq. (E.6)], the transmission zeroes are obtained by solving the eigenvalue problem

$$\begin{bmatrix} s_z\boldsymbol{\mathcal{I}} - (\boldsymbol{\mathcal{R}} - \boldsymbol{\mathcal{N}}) & -\widetilde{\boldsymbol{\mathcal{B}}} \\ -\widetilde{\boldsymbol{\mathcal{C}}} & -\widetilde{\boldsymbol{\mathcal{D}}} \end{bmatrix} \begin{Bmatrix} \widetilde{\boldsymbol{x}}_0 \\ \boldsymbol{u}_0 \end{Bmatrix} = 0, \begin{Bmatrix} \widetilde{\boldsymbol{x}}_0 \\ \boldsymbol{u}_0 \end{Bmatrix} \neq 0. \tag{E.14}$$

In the s plane, the definition of the transmission zeros for an LPTV system is similar to that of an LTI system.

יולי 2021

# תקציר

עבודה זו מציגה ניתוח של מערכות לינאריות מחזוריות המיוצגות על ידי מערכת משוואות דיפרנציאליות לינאריות רגילות בארגומנט הזמן $t$ עם מטריצת מקדמים מחזורית עם זמן מחזור $T$ כלשהו (או באופן שקול בעלת תדירות $\omega = \frac{2\pi}{T}$ כלשהי) המבוססת על תיאוריית Floquet. לפי תיאוריית Floquet עבור מערכת לינארית מחזוריות המיוצגת ע"י מטריצה ריבועית $A(t) = A(t+T)$, מטריצת מעבר של $A(t)$ ניתנת לייצוג ע"י מכפלה של מטריצה ריבועית מחזורית $P(t) = P(t+T)$ באקספוננט של מטריצה מהצורה $Rt$ כאשר $R$ הינה מטריצה קבועה ביחס ל-$t$. למרות נכונות תאוריה זו, קשה מאוד למצוא בצורה אנליטית ביטויים סגורים ל-$P(t)$ ו-$R$ במקרים בהם מטריצת מעבר של $A(t)$ לא יודעה מראש, או במילים אחרות קשה מאוד למצוא בצורה אנליטית ביטוי סגור לפתרון של מערכת מחזורית.

המטרות של העבודה הינה לבחון את ההשפעה של שינוי פרמטר התדר $\omega$ במערכות LPTV ופתרונן (לדוגמא: ניתוח יציבות), ולנסות לאפיין משפחה (רצוי גדולה) של מטריצות מחזוריות $A(t)$ עם מספר הרמוניות סופי, אשר מתכן מתקבל כי החלק המחזורי של הפתרון $P(t)$ הינו בעל מספר הרמוניות סופי. בנוסף, עבור משפחה זו נדרש למצוא דרך יעילה שתניב את $P(t)$ ו-$R$ הדרושים לפתרון. שיטת המחקר הינה לבחון דוגמאות למטריצות מחזוריות $A(t)$ אשר התדירות שלה $\omega$ הינו פרמטר חופשי (כלומר אינו נתון ע"י מספר קבוע), על מנת לקבל משפחה כמה שיותר גדולה של מערכות מחזוריות אשר למעשה מגדירות את מטריצת המעבר שלה כתלות ב-$\omega$ . בעבודה זו נתמקד במשפחה של מטריצות מחזוריות $A(t)$ אשר יש להן פירוק פורייה סופי עם מקדמי פורייה שהינם פולינום סופי ב-$\omega$, כאשר החלק המחזורי של הפתרון $P(t)$ בעל פירוק פורייה סופי עם מקדמי פורייה בלתי תלויים בתדר $\omega$ והחלק הקבוע -$R$ הינו פולינום סופי ב-$\omega$.

תוצאות המחקר מראות כי עבור המשפחה הנ"ל ניתן לבצע השוואת מקדמים לפי חזקות של $\omega$ בנוסף להשוואות הרמוניות (מקדמים של cos\sin או של exp מרוכב) על מנת לקבוע את הצמד את $P(t)$ ו-$R$ . בנוסף על כך, מעבודה זו ניתן להסיק על הקשר בין מערכת לינארית מחזורית עם תדר $\omega$ כלשהו לבין מערכת לינארית בלתי תלויה בזמן המתקבלת כתוצאה מהצבה $\omega = 0$. בהנחת מבנה -$R$ כפולינום ב-$\omega$ ,ניתן לבחור את איבר החופשי בפולינום להיות כל מטריצה הדומה למטריצה הקבועה של $A(t)$ המחושבת ב- $\omega = 0$. חשיבות לשימוש ברעיון בו התדר $\omega$ הינו פרמטר חופשי הינה לבחינת היציבות של הפתרון כתלות בפרמטר $\omega$ אשר נובעת מתלות הערכים העצמיים של החלק הקבוע $R$ בפרמטר $\omega$ (כאשר אכן ישנה תלות כזו).

יולי 2021

**אוניברסיטת בן-גוריון בנגב**

**הפקולטה למדעי ההנדסה**

**בית הספר להנדסת חשמל ומחשבים**

# ניתוח של מערכות לינאריות, משתנות בזמן ומחזוריות

חיבור זה מהווה חלק מהדרישות לקבלת תואר מגיסטר בהנדסה

מאת: אורן פייבל

מנחה:

פרופ' יצחק לבקוביץ'

המחבר: אורן פייבל ............................ תאריך: 15 יולי 2021

מנחה: פרופ' יצחק לבקוביץ' ............................ תאריך: 15 יולי 2021

יו"ר ועדת הוראה לתואר שני:

שם: **ד"ר קובי כהן** ............................ ............................ תאריך: 02/08/2021

יולי 2021

**אוניברסיטת בן-גוריון בנגב**

**הפקולטה למדעי ההנדסה**

**בית הספר להנדסת חשמל ומחשבים**

# ניתוח של מערכות לינאריות, משתנות בזמן ומחזוריות

חיבור זה מהווה חלק מהדרישות לקבלת תואר מגיסטר בהנדסה

מאת: אורן פייבל

מנחה:

פרופ' יצחק לבקוביץ'

יולי 2021